\documentclass{aa}

\usepackage{xcolor}
\usepackage{graphicx}
\usepackage{txfonts}
\usepackage{float}
\usepackage{amsmath}
\usepackage{hyperref}
\usepackage{bm}

\usepackage{hyperref}	
\hypersetup{colorlinks=true,linkcolor=blue,citecolor=blue,filecolor=blue,urlcolor=blue}

\begin{document}
\title{Dust settling and rings in the outer regions of protoplanetary discs subject to ambipolar diffusion}
\titlerunning{Dust dynamics in outer regions of protoplanetary discs}

\author{A. Riols\inst{1}\and  G. Lesur\inst{1}}
\institute{%
$^1$ Univ. Grenoble Alpes, CNRS, Institut de Planétologie et d’Astrophysique de Grenoble (IPAG), F-38000, Grenoble, France
}

\date{\today}

\abstract{Magnetohydrodynamic (MHD) turbulence plays a crucial  role in the dust dynamics of protoplanetary discs. It affects planet formation, vertical settling and is one possible origin of  the large scale axisymmetric structures, such as rings, recently imaged by ALMA and SPHERE. Among the variety of MHD processes in discs, the magnetorotational instability (MRI) has raised particular interest since it provides a source of turbulence and potentially organizes the flow into large scale structures. However, the weak ionization of discs prevents the MRI from being excited beyond 1 AU.  Moreover, the low velocity dispersion observed in CO and strong sedimentation of millimetre dust measured in T-Tauri discs are in contradiction with predictions based on ideal MRI turbulence.}{In this paper, we study the effects of non-ideal MHD and magnetised winds on the dynamics and sedimentation of dust grains. We consider a weakly ionized plasma subject to ambipolar diffusion characterizing the disc outer regions ($\gg 1$ AU).} {To compute the dust and gas motions,  we perform numerical MHD simulations in the stratified shearing box, using a modifed version of the PLUTO code. We explore different grain sizes from micrometer to few centimetres and different disc vertical magnetizations with plasma beta ranging from $10^3$ to $10^5$. }{Our simulations show that the mm-cm dust is contained vertically in a very thin layer, with typical heightscale $\lesssim 0.4$ AU at $R=30$ AU, compatible with recent ALMA observations.  Horizontally, the grains are trapped within the pressure maxima (or zonal flows) induced by ambipolar diffusion, leading to the formation of dust rings.  For micrometer grains and strong magnetization, we find that the dust layer has a size comparable to the disc heightscale $H$. In this regime, dust settling cannot be explained by a simple 1D diffusion theory but results from a large scale 2D circulation induced by both MHD winds and zonal flows.}{ Our results suggest that non-ideal MHD effects and MHD winds associated with zonal flows play a major role in shaping the radial and vertical distribution of dust in protoplanetary discs. Leading to effective accretion efficiency $\alpha \simeq 10^{-3}- 10^{-1}$, non-ideal MHD models are also a promising avenue to reconcile the low turbulent activity measured in discs with their relatively high accretion rates.}


\maketitle


\section{Introduction}

One of the most challenging problem in astrophysics is to understand how protoplanetary (PP) discs accrete and form planets. For more than forty years, accretion has been thought to originate from turbulent motions,  acting like an effective viscosity \citep{shakura73}. The magneto-rotational instability \citep[MRI, ][]{balbus91,hawley95}  has long been considered as the main source of turbulence, among various other instabilities (of gravitational or thermodynamics nature) whose role is still debated. Local and global MRI simulations in the ideal limit suggest in particular that the rate of angular momentum transport is compatible with that inferred from observations \citep{flock11,flock13,bai13}. However the MRI has been shown to survive only in the very inner part of the disc, below 0.1 -1 AU  where the ionization sources are strong enough to couple the gas with the ambient magnetic field \citep{gammie96}.  Further away, non-ideal MHD effects, such as ohmic diffusion, Hall effect and ambipolar diffusion prevail and tend to suppress any form of MRI-driven turbulence \citep{fleming00, sano02,wardle12,bai13b,lesur14, bai15}. \\

Observationally, the prevalence of a sustained  and vigorous MRI-turbulence in the regions $\gtrsim 1$ AU is also largely questioned.  The high resolution imaging at optical, near-infrared (VLT/SPHERE) and sub-millimetric wavelenght  (ALMA) have provided precious information about the disc structure and turbulence, essentially by probing the spatial distribution of micrometer and millimetre dust grains.  Few observations have been able to map the dust heightscale, by measuring the rings/gaps contrast \citep{pinte16} or directly by imaging edge-on discs \citep{wolff17}.  In particular, the survey of \citet{pinte16} in  HL Tau suggests that the millimeter dust forms a very thin layer around the midplane with typical scaleheight  $\lesssim$ 2 AU at a radius of 100 AU. This result, together with the measurements of gas velocity dispersion in CO  \citep[e. g.][]{Flaherty17}  indicate that the turbulence in protoplanetary discs is weak, contrary to the current theoretical predictions based on the ideal MRI. In particular the accretion efficiency that one would deduce from  such low turbulent levels  is smaller by one or two orders of magnitude than those inferred from direct observations \citep[see Table 4 of ][]{venuti14}. These elements suggest that accretion is happening through an essentially
laminar process which is yet to be identified\\

In parallel to that, ALMA and SPHERE have unveiled a lot of structures in the scattered light emission or the dust continuum, such as rings, spiral arms, gaps or vortices \citep{garufi17}.  Numerous effort has been deployed to explain the formation of concentric rings, a feature observed in various protoplanetary discs such as HL tau \citep{alma15}, TW Hydra \citep{andrews16} or the disc around Herbig Ae star HD 163296 \citep{isella16}.  If the presence of planets is often invoked to explain these structures, there is today no consensus on their origin. Many other scenarii have been proposed such as {dust-drift-driven viscous ring instability \citep{wunsch05,dullemond18}}, snow lines \citep{okuzumi16}, MHD turbulence with dead zones \citep{flock15}, zonal flows \citep{johansen09}  or  secular gravitational instabilities in the dust \citep{takahashi14}.  \\

A possible solution to account for all these observations  is to consider the atypical dynamics induced by non-ideal MHD effects in protoplanetary discs, in particular the Hall effect and ambipolar diffusion. Indeed, if the latter tend to weaken considerably the disc turbulence,  both are able to produce large scale horizontal magnetic fields within the disc, producing strong currents. These currents allow the development of vigorous outflows at the disc surface and a laminar Maxwell stress in the midplane \citep{lesur14}, both transporting significant angular momentum \citep{bai13b,simon15,gressel15}.  Moreover, these non-ideal effects lead in the nonlinear regime to the formation of self-organized axisymmetric structures, such as zonal flows \citep{kunz13,bai15}, corresponding to concentric rings in a  global disc view \citep{bethune16}. These rings, which  are very stable and coincide with gas density maxima, are likely to be ideal locations for dust accumulation and planet formation.   

Whether the Hall effect and ambipolar diffusion are key physical processes accounting for the observed disc quiescence (in terms of turbulent motions) and ring-like structures remains an open question. Since current sub-millimeter or near-infrared observations
are sensitive to dust, direct comparison with them requires to go one step forward and determine the dust dynamics in such non-ideal MHD flows. Ultimately it requires to compute its scattered and direct emission, but this remains out of the scope of the paper.  Although the dust behaviour has been widely studied in local and global MRI disc simulations \citep{johansen05,carbadillo06,fromang06b,balsara09,zhu15}, it remains relatively unexplored in non-ideal MHD simulations characterizing the disc outer regions $(r\gtrsim 0.1 -1$AU). \citet{ruge16} studied the case of zero net flux MRI with ohmic resistivity but did not include other relevant non-ideal effects.  Several simulations by \citet{zhu15} have combined the dust dynamics with ambipolar diffusion but focused into a regime of large particle size  ($\gtrsim$ mm),  in relatively small domains compared to the characteristic length of the zonal MHD flows.

Our aim is to generalize the study to smaller particles sizes and more extended domains. A first step is to characterize the degree of dust sedimentation,  and the ability of non-ideal MHD effect to concentrate the grains into rings. The goal is to make comparison with recent ALMA observations for millimeter size particles.  Our ambition is also to provide a theoretical model that predicts the dust scaleheight and its vertical distribution, as a function of the disc magnetization and particle size. Current models of dust settling in turbulent discs are based on a 1D diffusion theory \citep{dubrulle95,dullemond04} which is not necessarily appropriate to describe non-ideal MHD flows characterised by nearly laminar midplane and windy corona.  \\

To address the problem, we performed 3D MHD shearing
box simulations of stratified discs with an imposed vertical magnetic field,   using a modified version of the  PLUTO code.  To make the interpretation easier, we restrict our study to a large radius where the Hall effect can be neglected. The dust population is approximated as a pressure-less fluid made of different particle sizes. We explore different grain sizes from few micrometers to few centimetres and different disc magnetization with plasma beta ranging from $10^3$ to $10^5$. We enable the back reaction of the dust on the gas. As a preliminary step, we do not include radiative transfer and neglect coagulation or fragmentation processes. 
The paper is organized as follows: in Section \ref{framework}, we describe the model and review the main characteristics of non-ideal physics and dust/gas interaction.  We also present the numerical methods  used to simulate the dust/gas dynamics. In Section \ref{sec_withoutdust}, we perform numerical simulations without dust in order to characterise the properties of the flow subject to ambipolar diffusion (transport, turbulent levels, winds). In Section \ref{sec_withdust}, we add the dust in the simulations and determine its vertical and radial distribution. We show in particular that the sub-millimetre dust is highly sedimented and form ring structures with high density contrast. We also explore the effect of a mean radial pressure gradient in the disc. In Section \ref{sec_modele}, we compare different settling models and show that the classical 1D model fails to predict the vertical dust distribution for small particle sizes or large magnetization. We then propose a 2D model including the effect of coronal winds and radial gas density structures (zonal flows).  Combined each other, these two features induce a large scale circulation of the dust in the corona, which affects the sedimentation process. Finally, in Section \ref{sec_obs} we compare the vertical and radial dust distributions with those inferred from ALMA observations \citep{pinte16}. We conclude in Section \ref{sec_conclusions} by discussing the applications of our work on protoplanetary discs evolution and planet formation.

\section{Model and numerical framework}
\label{framework}
\subsection{MHD and dust equations in the shearing sheet}
\label{gas_equations}
To study the gas and dust dynamics in the outer part of protoplanetary discs, we use the local shearing sheet model \citep{goldreich65}. This corresponds to a Cartesian patch of the disc, centred at $r=R_0$, where the Keplerian rotation is approximated locally by a linear shear flow plus a uniform rotation rate $\boldsymbol{\Omega}=\Omega \, \mathbf{e}_z$. We note  $(x,y,z)$ respectively the radial, azimuthal and vertical directions.  We adopt a multi-fluid approximation in which the ionized gas and the dust interact and exchange momentum through drag forces.  

The gas is coupled to a magnetic field $\mathbf{B}$ and is assumed to be inviscid and isothermal, its pressure $P$ and density $\rho$ related by $P=\rho c_s^2$, with $c_s$ the uniform sound speed.  The evolution of its density ${\rho}$, and total velocity {field} $\mathbf{v}$ follows:
\begin{equation}
\dfrac{\partial \rho}{\partial t}+\nabla\cdot \left(\rho \mathbf{v}\right)=0, 
\label{mass_eq}
\end{equation}
\begin{equation}
\rho\left(\frac{\partial{\mathbf{v}}}{\partial{t}}+\mathbf{v}\cdot\mathbf{\nabla
  v} +2\boldsymbol{\Omega}\times\mathbf{v}\right) =\rho \mathbf{g}
  -\mathbf{\nabla}{P}+(\nabla\times \mathbf{B})\times\mathbf{B}+\rho\,{\bm{\gamma}_{d->g}}, 
\label{ns_eq}
\end{equation}
where the total velocity field can be decomposed into a mean shear and a perturbation $\mathbf{u}$: 
\begin{equation}
\mathbf{v}=-q \Omega x\,  \mathbf{e}_y+\mathbf{u}.
\end{equation}
with $q=3/2$. The term $\mathbf{g}=-\Omega^2 z \,\mathbf{e}_z+2q\Omega^2\,x \, \mathbf{e}_x$ corresponds to the tidal gravity field in the local frame. The last term in the momentum equation (2) contains the acceleration $\bm{\gamma}_{d->g}$ exerted by the dust drag force on a gas parcel (detailed below). The magnetic field $\mathbf{B}$, which appears in the Lorentz force  ${(\nabla\times \mathbf{B})\times\mathbf{B}}$, is governed by the non-ideal induction equation, 
\begin{equation}
\frac{\partial{\mathbf{B}}}{\partial{t}} =\nabla\times(\mathbf{v}\times\mathbf{B})+\mathbf{\nabla} \times \left[\eta_A (\mathbf{J} \times \mathbf{e_b}) \times \mathbf{e_b}\right].
\label{magnetic_eq} 
\end{equation}
The term $\mathbf{\nabla} \times \left[\eta_A (\mathbf{J} \times \mathbf{e_b}) \times \mathbf{e_b}\right]$  corresponds to ambipolar diffusion, with $\mathbf{J}=\mathbf{\nabla} \times \mathbf{B}$  the current density, $\mathbf{e_b}=\mathbf{B}/ \Vert \mathbf{B} \Vert$  the unit vector parallel to the field line and $\eta_A$ the ambipolar diffusivity. This diffusion is due to ions-neutral collisions and is assumed to be the dominant non-ideal effect in the regions considered in this paper, i.e  $R_0=30 $ AU \citep[see justifications in][]{simon15}. Therefore, the Hall effect and ohmic diffusion are neglected. Due to our assumption of ionization, $\eta_A$ is not uniform and its vertical profile is detailed in Section \ref{ionization_profile}. \\

Finally, the dust is composed of a mixture of different species,  characterizing different grain sizes. Each specie is described by a pressure-less fluid,  with a given density $\rho_{d}$ and velocity $\mathbf{v}_{d}$. We assume in this paper that dust grains remain electrically neutral so that they do not feel the magnetic field. The equations of motion for each specie are: 
\begin{equation}
\dfrac{\partial \rho_d}{\partial t}+\nabla\cdot \left(\rho_d \mathbf{v_d}\right)=0,
\label{eq_mass_dust}
\end{equation}
\begin{equation}
\rho_d\left(\dfrac{\partial \mathbf{v_d}}{\partial{t}}+\mathbf{v_d}\cdot\mathbf{\nabla  v_d} +2\boldsymbol{\Omega}\times\mathbf{v_d}\right) =\rho_d \,(\mathbf{g}+ \bm{\gamma}_{g->d}),
\label{ns_eq_dust}
\end{equation}
with $\bm{\gamma}_{g->d}$ the drag acceleration induced by the the gas on a dust particle.  If we label each dust specie by a subscript $k$, we obtain by conservation of total angular momentum:
\begin{equation}
\bm{\gamma}_{d->g}= -\dfrac{1}{\rho} \sum_k \rho_{d_k} \,\bm{\gamma}_{g->{d_k}}.
\end{equation}
In the next section, we specify the form of the drag term $\bm{\gamma}_{g->d}$ and its dependence on the particle size. 
\subsection{Drag and stopping times}
In this study we consider that dust particles are spherical and small enough that they are in the Epstein regime \citep{weiden77}. The acceleration associated with the drag and acting on a single particle is given by:
\begin{equation}
 \bm{\gamma}_{g->d}=\dfrac{1}{\tau_s}  (\mathbf{v}-\mathbf{v_d}).
\end{equation}
For a spherical particle of size $a$ and internal density $\rho_s$ (should not be confused with the fluid density $\rho_d$), the stopping time $\tau_s$ is
\begin{equation}
\label{eq_stoptime}
\tau_s  = \dfrac{\rho_s a}{\rho c_s}.
\end{equation}
This corresponds to the timescale on which frictional drag will cause an order-of-unity change  in the momentum of the dust grain. This is a direct  measure of the coupling between dust particles and gas. A useful dimensionless quantity to parametrize this coupling is the Stokes number 
\begin{equation}
\text{St}=\Omega \tau_s.
\end{equation}
If not precised, St denotes the Stokes number in the midplane. Note that the effective Stokes number in the disc atmosphere is larger than St, since it is inversely proportional to the density. 

\subsection{Converting Stokes to particle size}
\label{conversion}
In this paper, we preferentially use the Stokes number rather than particle size to describe the dust dynamics. Indeed St  is a dimensionless quantity and does not depend on the disc properties and geometry.  However, to allow comparison between our simulations and real systems, it is helpful to associate the Stokes number to a grain size. 

In the limit of small magnetization, the disc is in hydrostatic equilibrium and its column (or surface) density is  $\Sigma = \rho_0  H \sqrt{2\pi}$ where $\rho_0$ is the midplane density and $H= c_s/\Omega$ is the disc scaleheight. Thus, combining these different relations, we obtain:
\begin{equation}
\text{St}=a_d \left( \dfrac{\rho_s \sqrt{2\pi}}{\Sigma}\right).
\end{equation}

To evaluate $\Sigma$,  a first possibility is to assume that the  disc can be described through the standard MMSN model \citep{hayashi81}, with surface density 
\begin{equation}
\label{eq_MMSN}
\Sigma = 1700 \, (R_0 / \text{1 AU})^{-3/2} \text{g}\,\text{cm}^{-2}. 
\end{equation}
This gives $\Sigma = 10.34$ g/cm$^{2}$at $R_0=30$ AU.  However, this model is probably not realistic since recent observations suggest flatter profiles with a power law between -0.5 and -0.2 \citep{williams16}, and in some cases $\Sigma$  varying up to 20 $\text{g}\,\text{cm}^{-2}$ at $R_0=30$ AU \citep[see][in the HL Tau disc for instance]{pinte16}. 
By assuming $\rho_s=2.5$ g/cm$^{-3}$,  we obtain the following conversion 
\begin{equation}
\text{St} \simeq 0.3-0.6 \,\,  \left(\dfrac{a}{\text{1 cm}}\right) \, \left(\dfrac{R_0}{ \text{30 AU}}\right)^{3/2}.
\end{equation}
\subsection{Ionization profile and ambipolar diffusivity  $\eta_A$}
\label{ionization_profile}
If the only charged particles are the ions and electrons, the ambipolar diffusivity $\eta_A$ is given by:
\begin{equation}
\eta_A=\dfrac{\mathbf{B}^2}{\gamma_i \rho_n  \rho_i}
\end{equation}
\citep{wardle07}, where $\gamma_i=\langle \sigma v \rangle_i /(m_n+m_i) $ and $\langle \sigma v \rangle_i=1.3\times 10^{-9} \text{cm}^3 \text{s}^{-1}$ is the ion-neutral collision rate. $\rho_n$ and $\rho_i$ are respectively the neutral and ion mass density; $m_n$ and  $m_i$ their individual mass. By introducing the Alfv\'en speed $v_A=B/\sqrt{\rho}$, we also define the dimensionless ambipolar Elsasser number
\begin{equation}
\text{Am}=\dfrac{v_A^2}{\Omega \eta_A} = \dfrac{\gamma_i \rho_i}{\Omega}.
\end{equation}
To calculate $\rho_i$,  one generally needs to model the various ionization sources of the disc (X rays, cosmic rays, radioactive decay and FUV) and compare them to the dissociative recombination mechanisms. However, this requires to compute the complex chemistry occurring  in the gas phase and on grain surfaces. To simplify the problem, we use the same model as \citet{simon15} which captures the essential physics of disc ionization, i.e the effect of FUV in the disc corona.  In this model,  Am is constant and of order unity in the midplane, and increases abruptly up to a certain height $z_{io}$.  The height $z_{io}$ is free to vary during the simulation and corresponds to the base of the ionization layer, above which the FUV can penetrate.  If we note $\Sigma_i(z)$ the horizontally-averaged density, integrated from the vertical boundary, we assume that FUV are blocked for $\Sigma_i(z)>\Sigma_{ic}=0.1$ g/cm$^{2}$. Given that carbon and sulfur atoms are fully ionized by FUV and immune to recombination on grains, the ionization fraction $x_i=\rho_i m_n/(\rho m_i)$ above $z_{io}$ is $10^{-5}$ \citep{perez11}.  To avoid any discontinuity of this quantity  at $z=z_{io}$, we use a smooth function to make the transition between the midplane regions and the ionized layer, so that the ionization profile is
\begin{equation}
x_i(z)=10^{-5}\exp\left[-\left(\dfrac{\Sigma_i(z)}{\Sigma_{ic}}\right)^4\right].
\end{equation}
The profile of Am is then given by:
\begin{equation}
\text{Am} = \text{Am}_{0} +3.3\times 10^{12}x_i(z)\,\dfrac{\rho}{\rho_0} \,\,(R_0/1 \text{AU})^{-5./4}, 
\end{equation}
where $\text{Am}_{0}$  is the  constant midplane value.  Note that we have supposed an MMSN disc model (see Eq.~\ref{eq_MMSN}) to convert numerical $\Sigma_i$  into g/cm$^{2}$.
\subsection{Numerical methods}
\label{numerics}
We use a modified version of the Godunov-based PLUTO code \citep{mignone07} to simulate the gas and dust motions.  This code provides a conservative finite-volume scheme that solves the approximate Riemann problem at each inter-cell
boundary. Physical quantities are  evolved in time with  a second  order Runge-Kutta scheme, and the orbital-advection algorithm FARGO is used to reduce numerical dissipation. The implementation of ambipolar diffusion is described in \citet{lesur14} (Appendix B) and has been used in various studies \citep{lesur14,bethune16,bethune17}.  For the dust, we implemented our own version which is described and tested in Appendix \ref{appendixA}. 
Our simulations are  computed in a shearing box
of size $L_x=8H ,L_y=8H$ and $L_z=12H$  with resolution $N_X=128, N_Y=64$,  $N_Z=192$.  The inverse of the orbital frequency $ \Omega^{-1}=1$ defines our unit of time and $H=c_s/\Omega$ the disc scaleheight, our unit of length. 
Boundary conditions are periodic in $y$ and shear-periodic in $x$. In
the vertical direction, we use a standard outflow condition for both the
gas and dust velocities but compute a hydrostatic balance in the ghost cells
for the gas density.  For the magnetic field, we adopt the so called "vertical field" boundary with $B_x=B_y=0$ at $z=\pm L_z/2$. The presence of magnetic fields produces outflows, which depletes the mass contained in the box. Then, at each time step, we artificially multiply $\rho$ in each cell by a uniform factor such that the total mass in the box is maintained constant.  We checked that the mass injected at each orbital period is negligible compared to the total mass ($\approx 1 \%$) and does not affect the results. Note that the mean $B_z$ is conserved in the box but not the mean horizontal magnetic field, whose evolution depends on the electromotive forces (EMFs) at the boundary.  
\begin{figure}
\centering
\includegraphics[width=1.01\columnwidth]{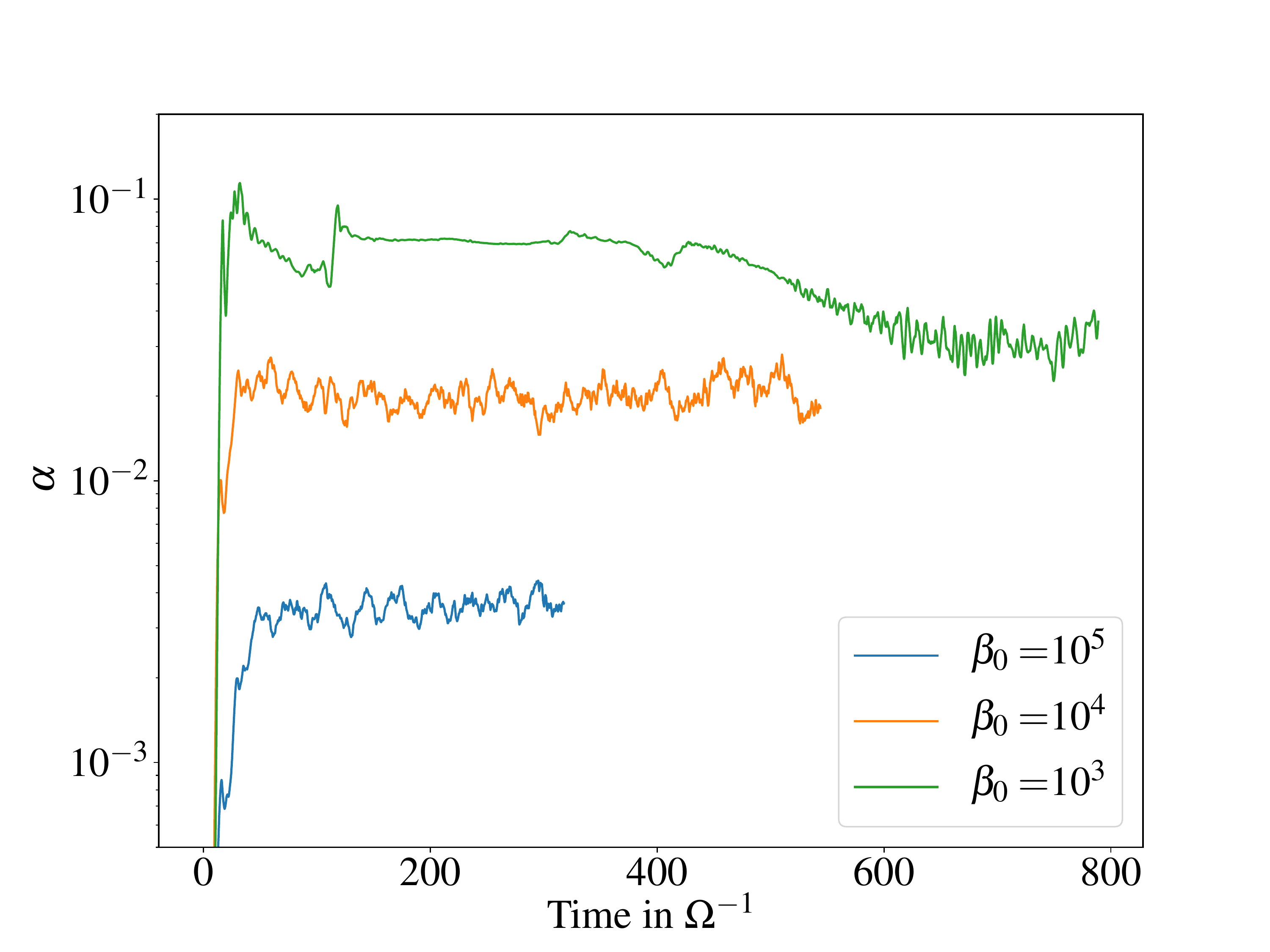}
 \caption{Transport efficiency $\alpha$ as a function of time for different vertical magnetization and $Am_0=1$.}
\label{fig_transport}
 \end{figure} 
 \subsection{Initialization and main parameters}
We  fix $R_0=30\, \text{AU}$ in all runs. Non ideal MHD simulations without dust are initialized with hydrostatic equilibrium  and weak random motions. The box is threaded by a net vertical field $B_z$, which allows us to define
\begin{equation}
\beta=\dfrac{2 c_s^2 \rho_0}{B_z^2}
\end{equation}
the beta-plasma parameter with $\rho_0=1$ the midplane density.  
The midplane ambipolar Elsasser number is fixed to Am$_0=1$ in all cases.  Simulations with dust are systematically initialized from developed non-ideal MHD states. Their detailed setup and the range of Stokes numbers considered  are given  in Section \ref{dust_init}. 
\subsection{Diagnostics}
To analyse the statistical properties of the MHD flow, we define the box average:
\begin{equation}
\overline{X}^B=\frac{1}{L_xL_yL_z}\int_V X\,\, dV, 
\end{equation}
where $L_x$, $L_y$ and $L_z$ are the dimensions of a Cartesian portion of volume $V$.  We also define the horizontally averaged vertical profile of a physical quantity: 
\begin{equation}
\overline{X} =\overline{X}^Z=\dfrac{1}{L_xL_y} \int\int X\,\, dxdy, 
\end{equation}
and its mean radial profile, averaged in $y$ and $z$ (over a size $l_z\leq L_z$): 
\begin{equation}
\overline{X}^R =\dfrac{1}{l_z L_y} \int_{-l_z/2}^{l_z/2}\int X\,\, dydz.
\end{equation}
Finally we note $\langle X \rangle$ the time average of a quantity over the simulation time. \\

An important quantity that characterizes the turbulent dynamics is the coefficient $\alpha$ which measures the radial angular momentum transport efficiency. This quantity is the sum of the total stress, which includes Reynolds $H_{xy}=\rho u_xu_y$  and Maxwell stresses $M_{xy}=-B_xB_y$, divided by the box average pressure. 
\begin{align}
\label{def_alpha}
\alpha=\dfrac{
\overline{H}_{xy}^B+\overline{M}_{xy}^B }{\overline{\rho}^B c_s^2}.
\end{align} 
\section{Ambipolar-MHD simulations without dust}
\label{sec_withoutdust}
A first and necessary step before computing the dust/gas interaction is to characterize the properties of the flow  in outer regions of PP discs. For that purpose, we start with non-ideal MHD simulations without dust that include ambipolar diffusion. We examine the turbulent transport, velocities dispersion and flow structures  at $R=30$ AU for different vertical magnetization. The turbulent states obtained will then be used in Section \ref{sec_withdust} to initialize simulations with dust. 
\begin{figure*}
\centering
\includegraphics[width=0.93\textwidth]{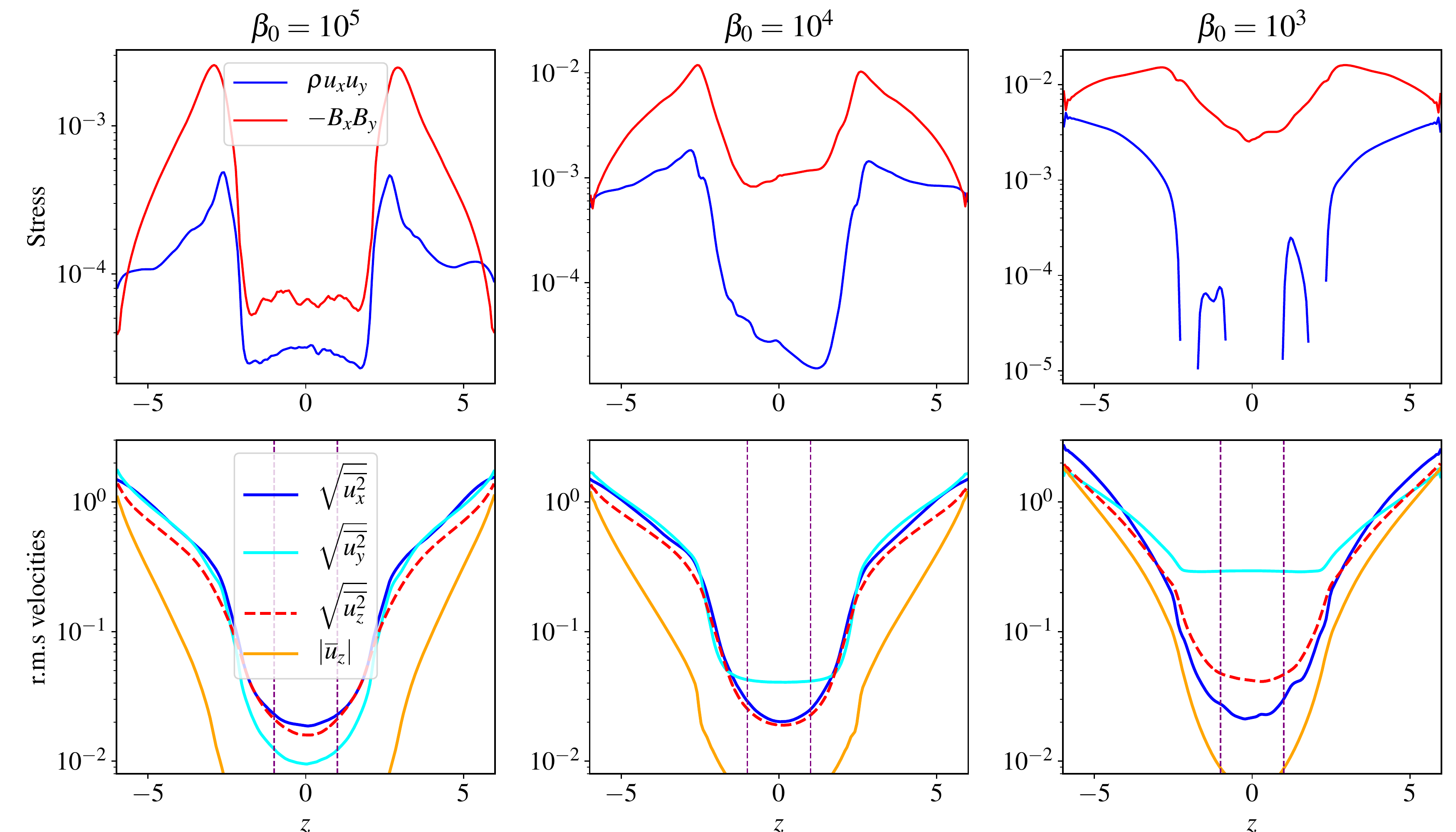}
 \caption{Top: vertical profiles of horizontally-averaged Reynolds and Maxwell stresses. Bottom: vertical profiles of horizontally-averaged r.m.s  velocity fluctuations, computed in the three directions. The orange lines denote the absolute value of the mean wind $\overline{u}_z$ component. To help with reading, we add vertical dashed lines in the bottom panels which indicate the disc heightscale $H$. From left to right, $\beta= 10^5$, $10^4$ and $10^3$. }
\label{fig_zaverages}
 \end{figure*} 
 \subsection{Transport and turbulence properties}
 \label{turbulence_properties}
To characterise the properties of non-ideal MHD flows with ambipolar diffusion, we perform three different simulations (assuming $\text{Am}_0=1$) with $\beta=10^5$, $10^4$ and $10^3$ respectively, which corresponds to $B_z=0.0045$,  0.014 and 0.045 in our code units.  Figure \ref{fig_transport} shows the evolution of the box averaged radial transport efficiency $\alpha$  for the three different cases. For all runs, a quasi-steady state that transports a significant amount of angular momentum is obtained. The transport is an increasing function of $B_z$,  with values in agreement with those found in the literature.  For instance, we obtain $\alpha\simeq 0.0034$ for $\beta_0=10^5$ and $\alpha\simeq 0.02$ for $\beta_0=10^4$, which is identical to what \citet{simon13b} found for a similar setup with the ATHENA code (see their Table 1). 

Figure \ref{fig_zaverages} (top panels) shows the vertical profiles of horizontally-averaged Maxwell and Reynolds stress. Each of them  contributes positively to the radial transport efficiency $\alpha$ (except $H_{xy}$ in the midplane for $\beta=10^3$), although the Maxwell stress always dominates over the Reynolds stress.  For all $\beta$, both quantities peak around  $z\simeq2H$ where the ionization is sufficiently large to excite the MRI. Their profiles show a dip in the midplane where the ionization is weak and the MRI absent. 

However, as the vertical field increases, a laminar stress due to the large scale $B_x$ and $B_y$ starts to be important in the midplane and induces a shallower dip in Maxwell stress. In particular, the ratio of Maxwell to Reynolds stress sharply increases with magnetization,  from 2.5 at $\beta=10^5$ to $\sim 100$ at $\beta=10^3$ , and thus  deviates significantly from that of ideal MRI. We will see in Section \ref{sec_obs} that such result has consequences on the so-called Schmidt number, ratio between the gas effective viscosity and dust turbulent mixing. Note that a transition to a magnetic and quasi-laminar stress occuring around $\beta\simeq10^3$ has been already pointed out by \citet{simon13b} and also exists when the Hall effect is included \citep{lesur14}.   We emphasize also that the radial stress $W_{xy}$ is not the only source of angular momentum transport (and thus accretion) in these type of flows. The vertical magnetic field  also removes angular momentum from the disc via strong winds powered by the MRI in the active layers. This leads to a $zy$ component of the stress tensor which is smaller than $W_{xy}$ but can influence the accretion rates \citep[due to a factor $R/H$ appearing in the equation for $\dot{M}$ , see ][]{fromang13,simon13b} . 

Finally, the bottom panels of Fig.~\ref{fig_zaverages} show the vertical profiles of the horizontally averaged r.m.s velocity  $\mathbf{u}_{\text{rms}}(z)$=$(\overline{\mathbf{u}^2})^{1/2}$ in all directions. These quantities are important to quantify the diffusion of dust particles (see Section \ref{sec_modele}).  We checked that their dependence in $z$ is similar to that of \citet{bai15}. For weak magnetization $\beta=10^5$ and $10^4$, radial and vertical velocity dispersions in the midplane are comparable in both cases ($\approx 2\times 10^{-2}$)  while they reach unity near the vertical boundaries. In the midplane region, they are mainly associated with the fluctuating part of the velocity field $\mathbf{u}-\overline{\mathbf{u}}(z)$. The mean  component $\overline{\mathbf{u}}(z)$ (orange dashed lines in Fig.~\ref{fig_zaverages})  associated with a large scale wind remains negligible compared to the fluctuations up to $z \simeq 4H$.  However, for higher magnetization ($\beta=10^3)$ ,  this wind component is stronger and becomes similar in amplitude to the fluctuating part for $z \gtrsim 1.5 H$. 
Note that the azimuthal velocity dispersion increases significantly with  $B_z$ in the midplane. This is a consequence of the emergence of large scale radial zonal flows whose amplitude is particularly enhanced at large $B_z$ (see next section). 
 \begin{figure*}
\centering
\includegraphics[width=\textwidth]{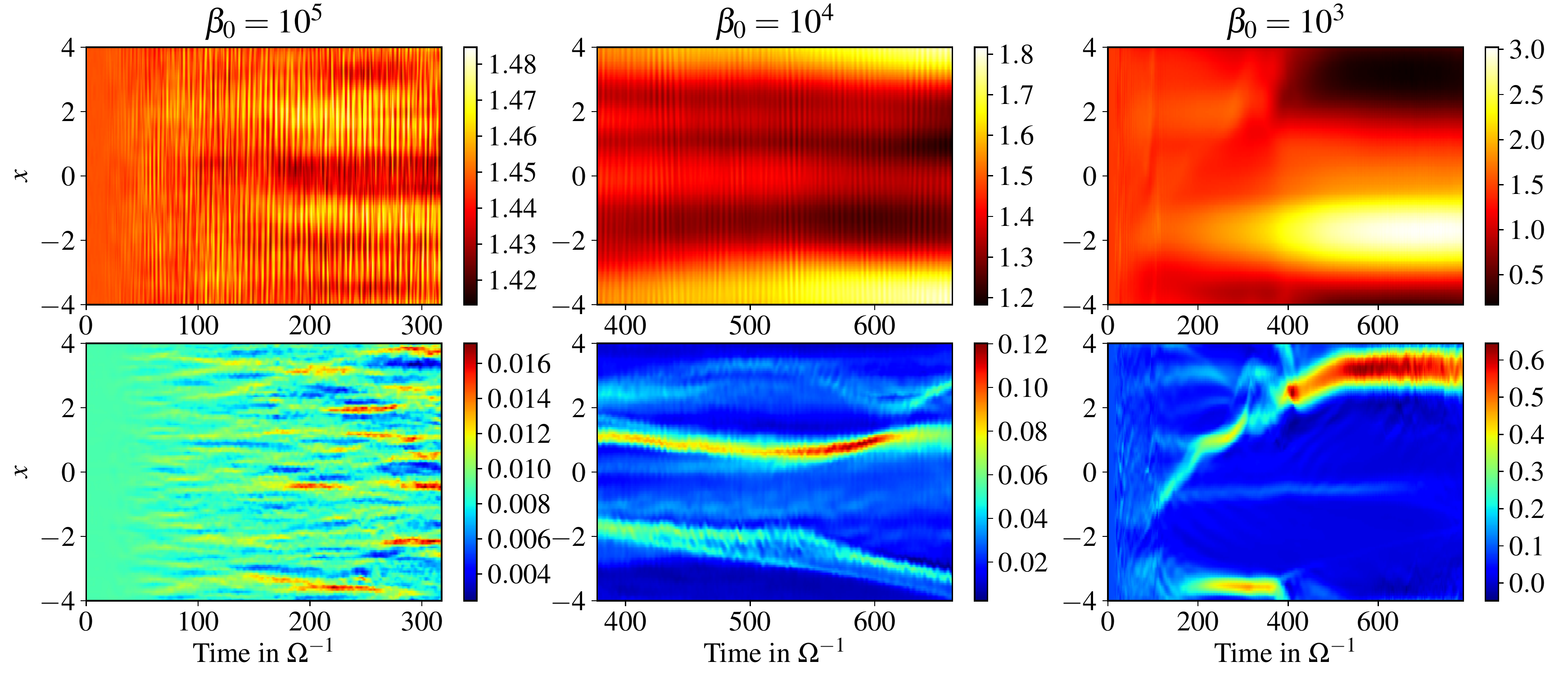}
 \caption{Time evolution of the radial profiles of gas density $\overline{\rho}^R$ (top panels)  and vertical magnetic field $\overline{B}^R_z$ (lower panels) averaged over the $y-z$ plane within $z = \pm 1.5H$. From left to right, $\beta= 10^5$, $10^4$ and $10^3$.}
\label{fig_xaverages}
 \end{figure*} 
\subsection{Zonal flows, pressure maximas and sound  waves}
Self-organized axisymmetric structures, or commonly called "zonal flows" are found in a variety of non-ideal MHD flows threaded by a net vertical field:  Hall-dominated turbulence \citep{kunz13}, ambipolar-dominated flows \citep[with or without ohmic diffusion, see][]{simon14,bai15} and flows combining all non-ideal effects \citep{lesur14, bai15}.  It has been pointed out that ambipolar diffusion enhances these features \citep{bai15} and can lead to zonal flows with density amplitude of 10–20\%  the background disc density \citep{simon14}. {These zonal flows become large scale rings in the global configuration, as revealed by the simulations of \citet{bethune17}}.

To check whether these axisymmetric structures exist in our stratified simulations, we show in Fig.~\ref{fig_xaverages} the spacetime-diagram ($t,x$) of the mean gas density (top) and $B_z$ (bottom), averaged in $y$ and $z$, for our three different $\beta$.  Note that the average is done between $z=-1.5 H_0$ and $z=1.5 H_0$ to exclude the ionized layers where coherent structures are mixed up by the strong MHD turbulence.  The top panels show that the disc develops radial density variations associated with zonal flows whose amplitude increases with mean $B_z$ (decreases with $\beta$). In case $\beta=10^5$, the density structures are very faint and do not exceed $5\%$ of the mean background density. For $\beta=10^4$ and $10^3$, the zonal flows are much more developed and reach amplitudes respectively  $30-40\%$  and  $170\%$ of the mean background density. Within the simulations time ($\sim 100$ orbits), these structures remain almost steady, although variations on longer timescale occur \citep[see][]{bai15}. Note that for $\beta=10^4$ and $10^5$, the density plots of Fig.~\ref{fig_xaverages} reveal the presence of large scale acoustic waves propagating radially in the disc. By computing the Fourier transform of $\rho$ in time and space (along the $x$ direction), we checked that  the signal frequencies match the dispersion relation of p-modes (sound waves). These waves might be generated via the turbulent stress, through a process called "aerodynamic noise" \citep{lighthill52,heinemann09}. \\

The bottom panels show that for $\beta=10^4$ and $10^3$, the vertical magnetic flux is concentrated into very thin shells (or filaments), while outside these shells, the net vertical flux is close to zero. These filaments are located within the gas density minima. Their  existence  have already been pointed out by \citet{bai15} but their origin is still unclear. One possibility suggested by \citet{bai14} is that the flux concentration is driven by the anisotropic turbulent diffusivity associated with the MRI turbulence.  However ambipolar diffusion seems to reinforce the magnetic shells and  might also play an important role. It is likely that the disc stratification is also essential to the formation of the structures. We explore in next section the effect of zonal density and $B_z$  structures on the dust sedimentation and radial concentration. 
\section{Simulations with dust and ambipolar diffusion}
\label{sec_withdust}
We now study the dust dynamics in the MHD flows described in Section \ref{sec_withoutdust}. For each $\beta$, we simulate the evolution of different grain sizes, corresponding to different midplane Stokes numbers St. We first study the settling of the dust and its evolution toward a quasi-steady state. We then characterize their vertical distributions and typical scaleheight with respect to  St and $\beta$. We analyse the effect of zonal flows (pressure bumps) on their radial structures and vertical dynamics. Finally, the presence of a mean radial pressure gradient in the disc is investigated. 
\begin{figure*}
\centering
\includegraphics[width=\textwidth]{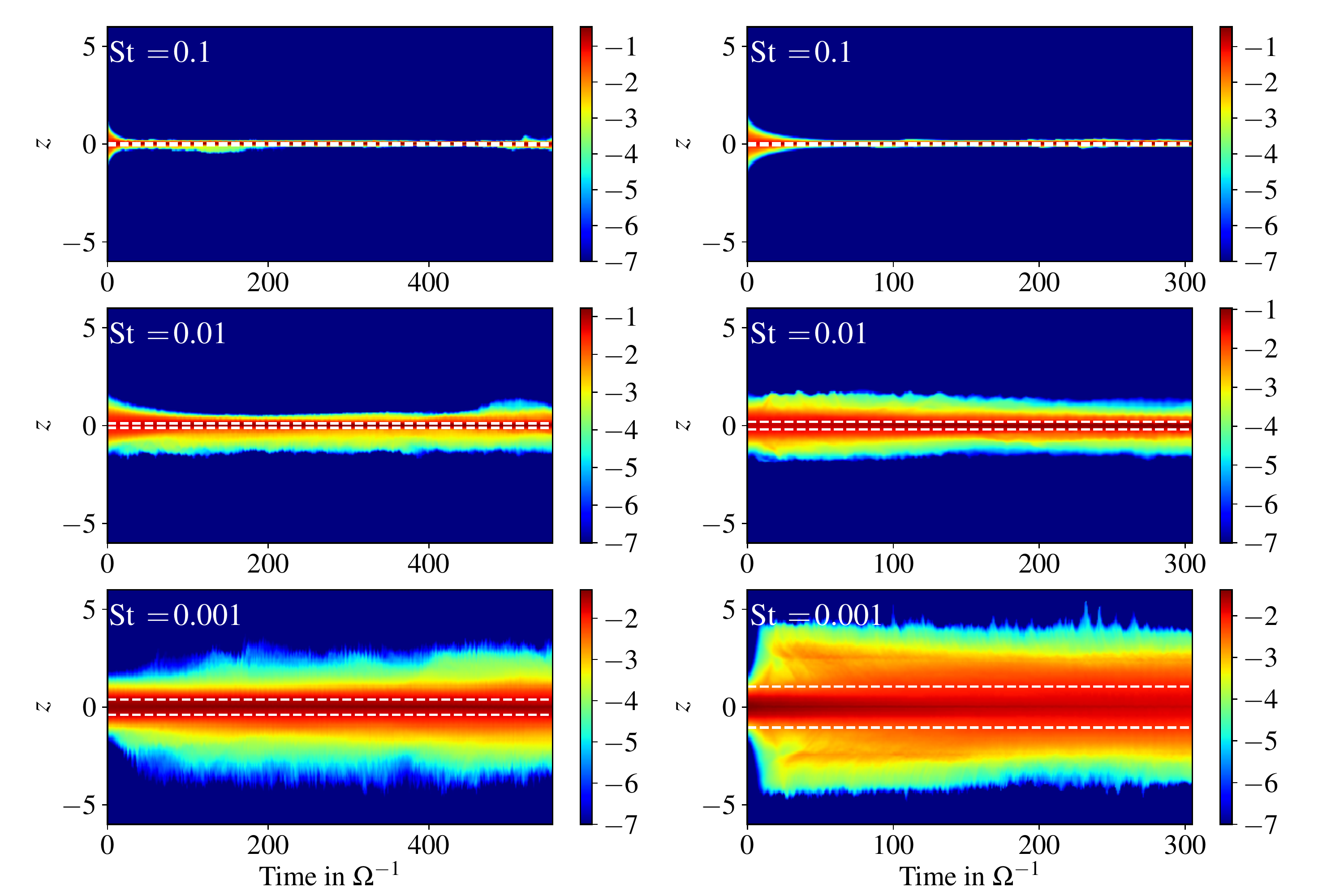}
 \caption{Time-evolution of the vertical profiles of dust density $\overline{\rho_d }(z,t)$ (in log space), averaged in $x$ and $y$, for three different Stokes number in the case $\beta=10^4$ (left) and  $\beta=10^3$ (right).  The white dashed lines indicate the typical dust heightscale $H_d$ measured using the relation \ref{eq_defHd}.}
\label{fig_spacetime}
 \end{figure*} 

\subsection{Initial condition, settling times and convergence}
\label{dust_init}

\begin{figure}
\centering
\includegraphics[width=\columnwidth]{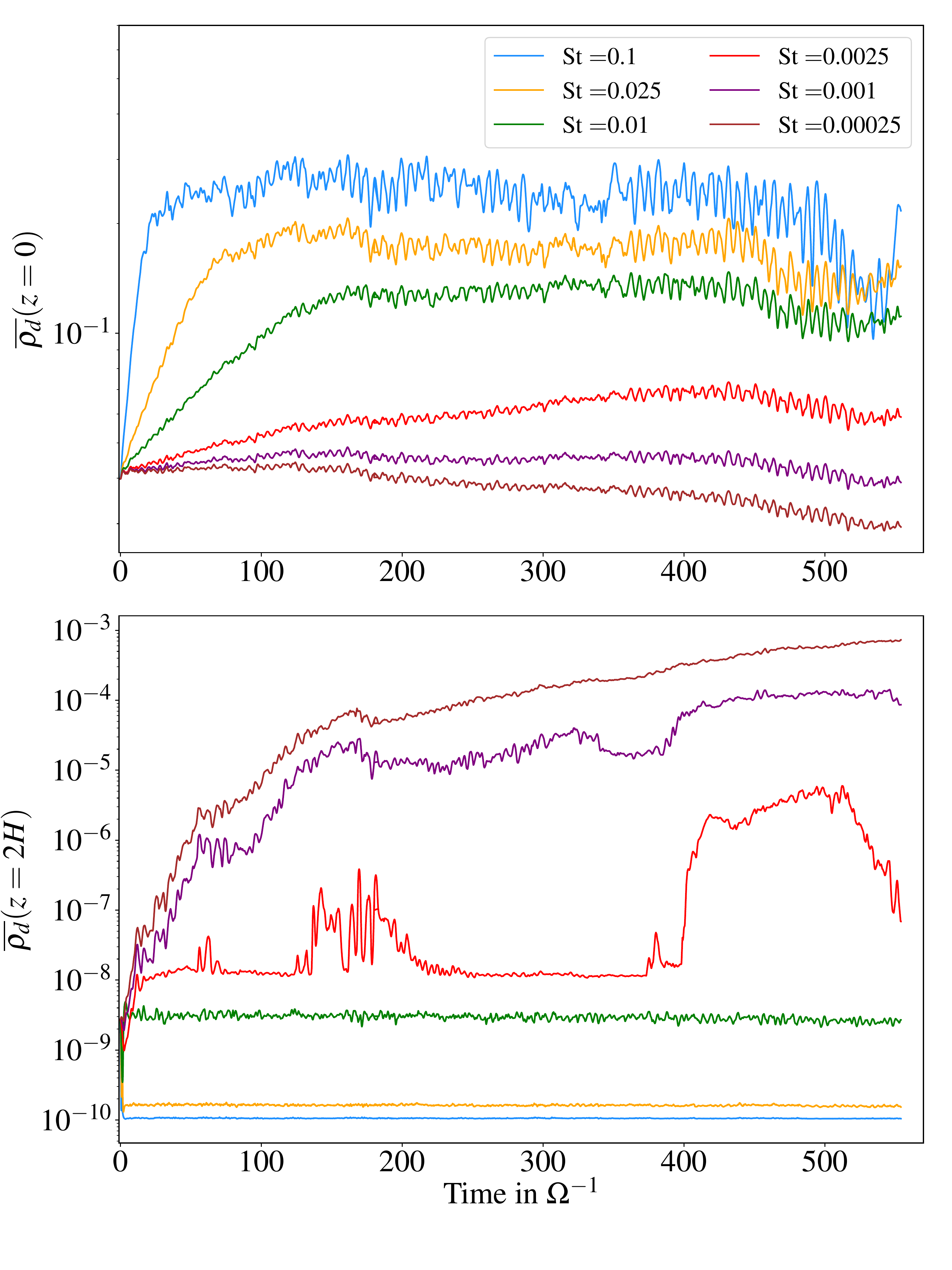}
 \caption{Time-evolution of the horizontally-averaged dust density for different Stokes numbers in the case $\beta=10^4$. The top panel is for $z=0$ (midplane) while the bottom panel is for $z=2 H$}
\label{fig_dustevol}
 \end{figure} 
 \begin{figure}
\centering
\includegraphics[width=\columnwidth]{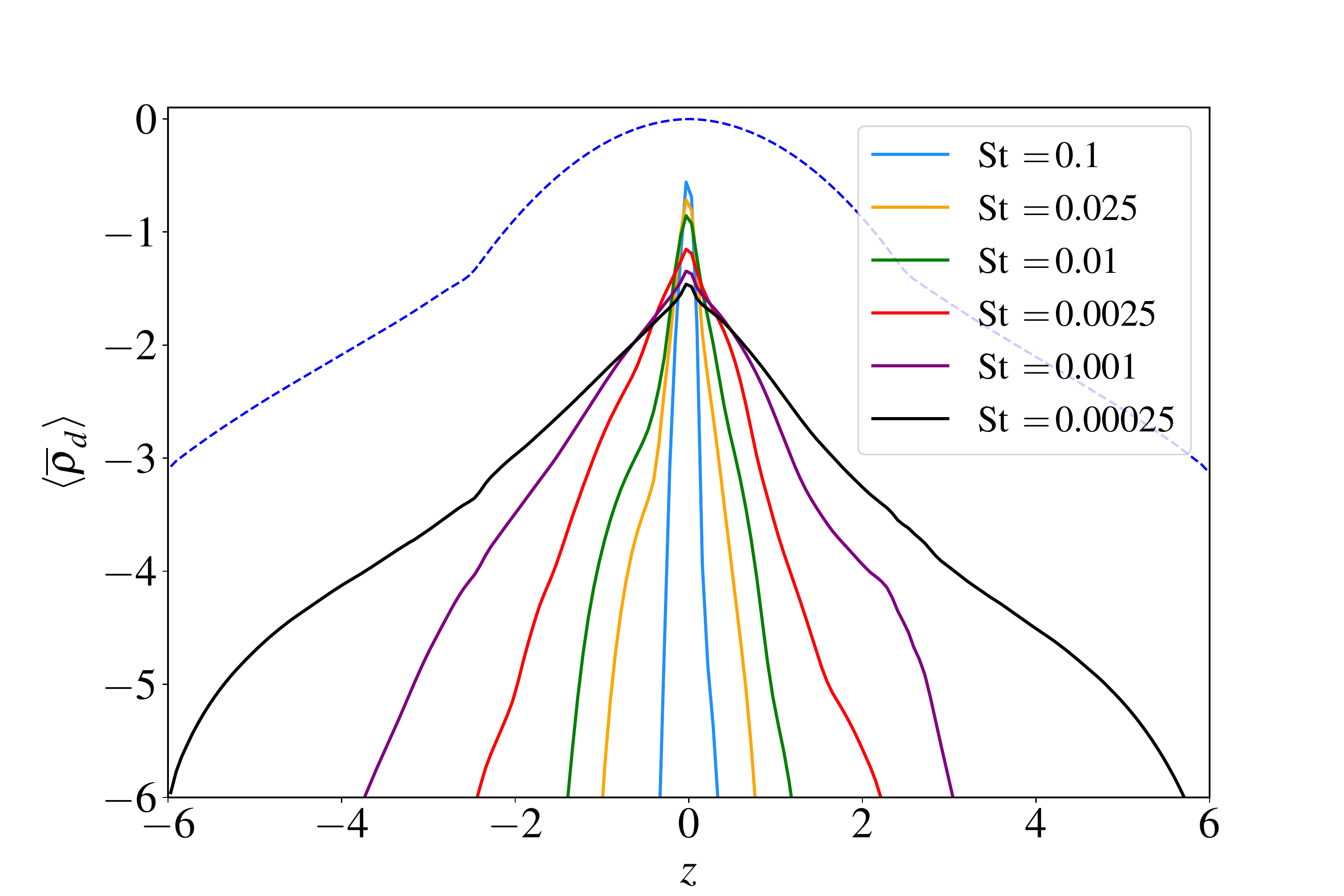}
 \caption{Vertical density profiles of dust grains for different Stokes number (in log scale) in the case $\beta=10^4$. For comparison, the dashed blue line corresponds to the gas profile. }
\label{fig_dustprofiles}
 \end{figure} 
The initial conditions for the dust are similar for all $\beta$ and Stokes number. The distribution at $t=0$ is
\begin{equation}
\rho_d(t=0)=\rho_0\exp\left(-\dfrac{z^2}{2 H_{d_0}^2}\right), 
\end{equation}
with $H_{d_0}=0.35 \,H$ and $\rho_0$ a constant evaluated  so that the ratio of surface densities $\Sigma_d/\Sigma$ is $0.014$. The dust velocity is initially Keplerian with zero perturbation.  For $\beta=10^5$, we integrate simultaneously, in a same simulation,  the motion of four different grain sizes with Stokes numbers 0.1, 0.025, 0.01 and  0.001. For $\beta=10^4$, we compute simultaneously six different grain sizes with Stokes numbers 0.1, 0.025, 0.01, 0.0025, 0.001 and 0.00025  ($\simeq 5$ mm to $10\, \mu$m) . Finally, for $\beta=10^3$,  we ran two different simulations, one with four dust species ($\text{St} = 0.1, 0.025, 0.01$ and 0.001) and the second with two dust species ($\text{St}=0.0025$ and 0.00025). 
Note that for simplicity the dust mass distribution is initially independent of the particle size, which is far from being the case in real protoplanetary discs. However, since the gas to dust ratio remains small when summed over all sizes,  this approximation has little effect on the results. We checked in particular that the back reaction of the dust onto the gas does not produce substantial  effects (at least within the simulation time).  Moreover, we found that the dust evolution and distribution are independent of the initialization (provided that $\sum_k \rho_{d_k}/\rho \ll 1$), and whether the dust components are simulated altogether or  separately.\\

Figure \ref{fig_spacetime} shows the space-time diagrams of $\overline{\rho}_d(z,t)$ (horizontally averaged vertical density distribution)  obtained from the simulations,  for $\beta=10^4$ (left) and $\beta=10^3$ (right), and for different Stokes numbers.  The density profiles seem to converge to a steady state depending on the Stokes number after a certain period of time.  To quantify more precisely this timescale, we plot in Fig.~\ref{fig_dustevol} the time-evolution of $\overline{\rho}_d$ at $z=0$ (midplane) and $z=2H$, for $\beta=10^4$.  If $\text{St}\gtrsim 0.001$, the initial evolution is purely exponential and dominated by gravitational settling.  The dust particles fall toward the midplane within a time proportional to $ \Omega^{-1}/\text{St}$, which is characteristic of such process \citep{dullemond04}. Ultimately, when turbulent diffusion and mixing start to be important,  dust particles stop settling toward the midplane and their mean vertical distribution stabilizes. Note that at $z=2H$, the gas density is  smaller and dust particles have an effective Stokes number greater than in the midplane. Therefore, the settling and convergence toward a steady state are more rapid at this location. Nevertheless, Fig.~\ref{fig_dustevol} (bottom) shows  that the dust density at $z=2H$ undergoes brutal variations  (especially for $\text{St} = 0.0025$ between $t=400$ and $500\,\Omega^{-1}$)  probably associated with stochastic events, like jets or wind plumes.  The case $\text{St}= 0.00025$ (our smallest particle size) is a bit more delicate since the typical settling time can be, in theory, much larger than the simulation time. In this case, an equilibrium is not necessarily reached and the vertical dust distribution keeps evolving slowly in time. Note that for $\beta=10^3$, the settling rates are similar but equilibria are reached sooner. The reason is that turbulence is more vigorous and the winds more efficient  when the magnetization is higher (see Fig.\ref{fig_zaverages}). 

 \begin{figure}
\centering
\includegraphics[width=\columnwidth]{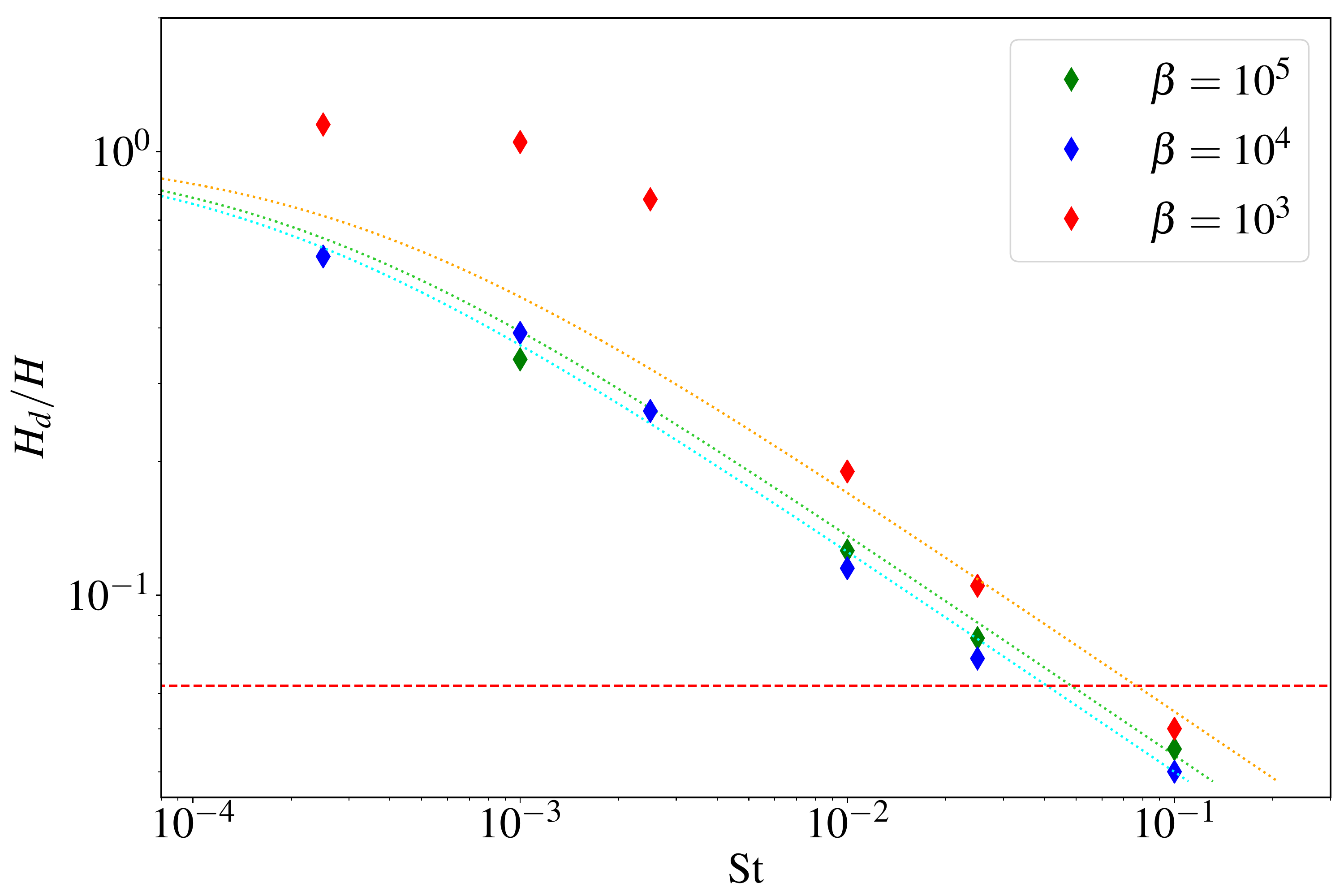}
 \caption{Dust heightscale $H_d$ as a function of the Stokes number for three different plasma $\beta$. The diamond markers correspond to direct measurement from numerical simulations, with the definition given by Eq.~\ref{eq_Hd1}. The dotted curves are derived from a simple diffusion model assuming that gravitational settling is balanced by homogenous turbulent diffusion \citep[see][and section \ref{sec_modele}]{dubrulle95, fromang06b}. The dashed/red horizontal line denotes the size of the numerical cells.}
\label{fig_Hd}
 \end{figure}  
 \begin{figure*}
\centering
\includegraphics[width=\textwidth]{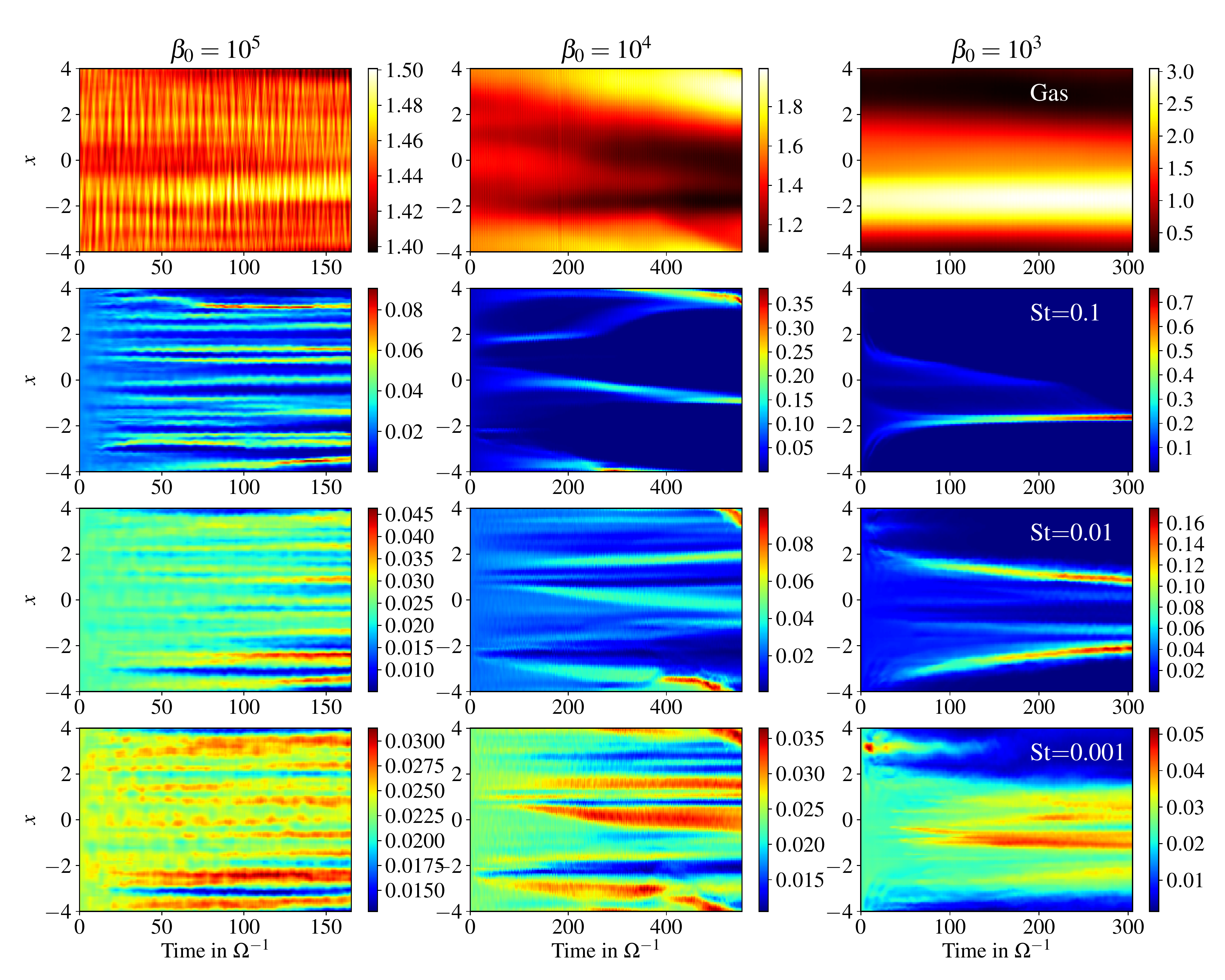}
 \caption{Time-evolution of the radial profiles of gas density (top panels) and dust densities (other panels). The second, third and fourth rows correspond to Stokes numbers of 0.1, 0.01 and 0.001. The densities are averaged over the $y-z$ plane within $z = \pm 1.5H$.   From left to right, $\beta= 10^5$, $10^4$ and $10^3$. }
\label{fig_rings}
 \end{figure*} 
\subsection{Vertical profiles and dust scale height with respect to St}
\label{vertical_profiles}
We now characterise in more detail the shape of the vertical equilibrium and estimate the typical dust scaleheight as a function of the Stokes number. A quick inspection of Fig.~\ref{fig_spacetime} shows that the size of the dust layer increases with decreasing Stokes number. This is physically expected, as small dust particles are less sensitive to gravitational settling and tend to follow the turbulent gas motion. To be more quantitative, we show in Fig.~\ref{fig_dustprofiles} the vertical density profiles, averaged in time,  for $\beta=10^4$ and different dust sizes.  For comparison we superimposed the gas vertical density profile (dashed blue line)  fitting perfectly a gaussian of width $H=1$ in the domain $\vert z \vert <2H$. For large Stokes numbers $0.0025 \lesssim \text{St} \lesssim 0.1$, the dust density profiles fit also quite well with gaussians but with a width much smaller than $H$. However, for $ \text{St}\lesssim 0.0025$, the profile turns out to be more flared  and follows an exponential distribution of the form $\rho_d = \rho_0 \exp(-z/\lambda_d)$ with $\lambda_d \simeq H$ for the minimum grain size. \\

To be more quantitative, we define the dust scaleheight $H_d\,(\text{St},\beta)$ as the altitude $z$ such that
\begin{equation}
\label{eq_defHd}
\rho_d (z=H_d) = \rho_d (z=0) \, e^{-\frac{1}{2}} \simeq 0.6\, \rho_{d0}.
\end{equation}
In case of large St, this corresponds to the standard height of the Gaussian distribution. For each $\beta$ and St, we calculate $H_d$ from the simulations data and plot their values in Fig.~\ref{fig_Hd}. For the largest $\text{St}=0.1$, almost all dust material is contained within two grid cells around the midplane, so that our estimation of $H_d$  is probably incorrect. We remind also that for $\text{St}=2.5\times 10^{-4}$, the distribution is probably not fully converged and one might expect that $H_d$ is slightly underestimated. \\

By examining the scaling relations of Fig.~\ref{fig_Hd}, we found that 
\begin{align}
\label{eq_Hd1}
H_d/H \simeq 0.013 \,\, \text{St}^{-\frac{1}{2}}   \quad  \text{for} \quad \beta = 10^{4-5}\\\simeq 0.019 \,\, \text{St}^{-\frac{1}{2}}     \quad  \text{for} \quad \beta = 10^3
\end{align}
for intermediate particle sizes, but saturates to $H_d \approx H$ in the limit of small $\text{St}$. The dependence on St$^{-1/2}$ has been obtained in many other disc turbulence simulations coupling dust and gas dynamics \citep{fromang06b,zhu15} and can be understood, in a first and crude approach,  within the framework of the classical diffusion theory \citep{morfill85,dubrulle95}. In this theory, the equilibrium distribution in the vertical direction is simply achieved by the balance between the gravitational settling and the diffusion due to the turbulence. The latter is assumed to be homogeneous and isotropic, and can be merely described through a uniform anomalous diffusion coefficient. A simple 1D advection-diffusion equation is solved in the vertical direction which gives the relation (\ref{eq_Hd1}) in the limit of $H_d \ll H$ and a flatter dependency in the limit of small St (see Sect \ref{dubrulle_model}). For a comparison with the numerical data, we superimposed in Fig.~\ref{fig_Hd} the predicted scaleheight from this model for the three different magnetizations  (see section \ref{sec_modele} for details about their calculation).\\

Although the turbulent diffusion theory seems to reproduce quite well the numerical dependency for $\beta = 10^5$ and for $\beta = 10^4$,  it does not explain the flared shape of the density profiles at small St  (Fig.~\ref{fig_dustprofiles}). In the case $\beta=10^3$, it is even worst;  the dust scaleheight is 2 or 3 times larger than the predicted value in the regime of small St ($\ll  0.01$). 
There are actually many caveats to this simple  approach. First it assumes that the turbulence is homogeneous and isotropic and neglects the presence of a wind (mean $u_z$). Indeed,  for small St, the typical timescale associated with MHD outflows in the simulations becomes smaller than the stopping time, which suggests that the wind might lift small dust particles and modify their vertical equilibria. For $\beta=10^3$, the association of a powerful wind with a zonal flow drastically changes the settling process (see next section) and breaks the assumptions usually made in the classical model.  All these effects are studied quantitatively and in detail in Section \ref{sec_modele}. 

\subsection{Effect of zonal flows}
\label{zonal_flows}
As we showed in Section \ref{sec_withoutdust},  MHD flows dominated by ambipolar diffusion develop quasi-steady zonal density structures (see Fig.~\ref{fig_xaverages}).  What are their impact on the radial distribution of grains and, more indirectly, on their vertical settling?
\subsubsection{Dust trapped in rings}
First, zonal flows or annular structures are expected to have an effect on the dust radial distribution. Indeed, they correspond  to gas pressure maxima in isothermal discs, which are known to collect dust particles and play a fundamental role in the formation of planetesimals. To concentrate dust in the radial direction, the drift associated to the pressure maxima has to be stronger than the radial turbulent diffusion of grains. Since the drift velocity is directly proportional to St (in the limit $\text{St}\ll1$), we expect that dust will be efficiently concentrated into radial bands for sufficiently large values of the Stokes number. The dependence of the process on $\beta$ is however less clear. If the radial zonal density structures are less prominent in the low magnetization regime, the turbulent diffusion might be also weaker.  Assuming that the dust rings form, another unknown is the stability of these structures. One might ask whether they reach a steady configuration before the dust to gas ratio becomes excessively high and leads to instabilities (e.g. the streaming instability). 

To address these questions, we show in Fig.~\ref{fig_rings} (second, third and fourth rows)  the radial density distribution of the dust as a function of time, for $\text{St}=0.1$, 0.01 and 0.001 and different $\beta$. For $\beta=10^5$ (left column), the grains concentrate into very thin radial bands, spaced out by less than a scaleheight $H$ in the radial direction. As expected, the concentration is stronger for the largest particles (high St). For $\text{St}=0.1$, the density contrast can be quite high, of order 1000, while for $\text{St}=0.001$, it is never greater than a factor 2. Although the density contrasts are pretty high, the dust to gas ratio remains bounded and smaller than 0.06.   For higher magnetization, ($\beta=10^4$ and $\beta=10^3$), dust rings also form, with density contrast and concentration much higher than in the case $\beta=10^5$ but with similar dependence on the Stokes number. Moreover, the spacing between the rings is larger than $H$ and comparable to the box size. In particular for the largest magnetization ($\beta=10^3$) and $\text{St}=0.1$, all the dust material accumulates into one single narrow band with averaged dust to gas ratio $\sim 0.3$.  In this particular case,  the stability of the ring is not guaranteed as its density keeps growing in time, even after 300 $\Omega^{-1}$. The evolution of such structure with high dust concentration is hard to predict from our simulations, since the bi-fluid assumption is not valid anymore when $\rho_d/\rho \simeq 1$.\\

We check that in most cases, the location where the dust is trapped corresponds  to a gas pressure maximum (for comparison, we plotted the radial density profiles of the gas in the top panels of Fig.~\ref{fig_rings}). However, for small \text{St}, some rings develop outside gas pressure maxima, like the upper one in the right panels of  Fig.~\ref{fig_rings},   for $\text{St}=0.01$ ($\beta=10^3$). Actually, they correspond to locations nearby magnetic shells where $B_z$ is strong. At these radii, the radial velocity associated with the steady zonal MHD structure is strong  enough to advect and concentrate the small dust particles. This occurs because the gas velocity exceeds  the radial drift velocity due to the local pressure gradients.

\begin{figure}
\centering
\includegraphics[width=\columnwidth]{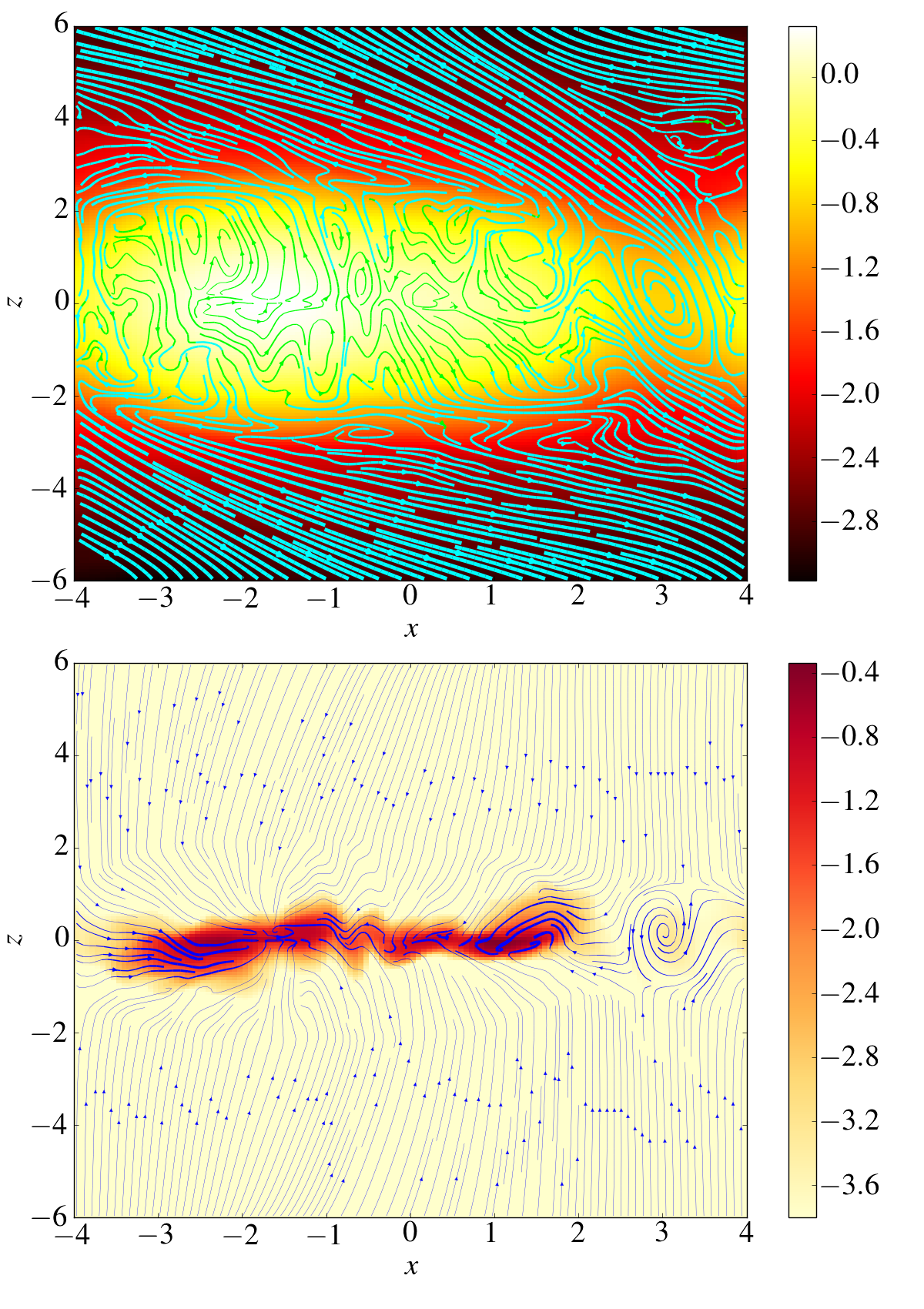}
\includegraphics[width=\columnwidth,trim=0cm 0cm 0cm 21.5cm, clip=true]{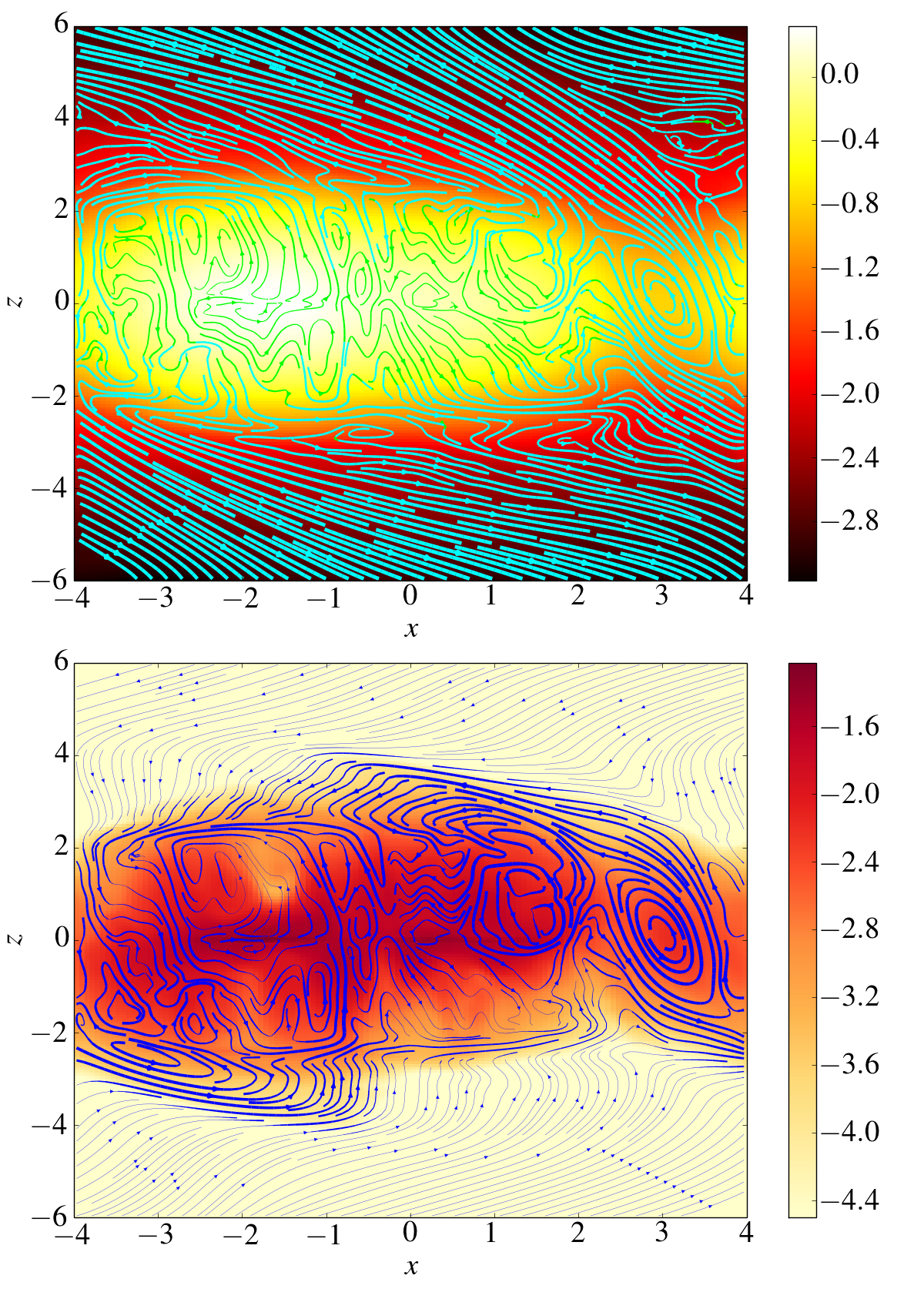}
 \caption{Top panel: gas density (colormap) and poloidal streamlines (cyan lines) showing the outflow topology. The line thickness accounts for the norm of the velocity. Center panel:  dust density (colormap) and poloidal streamlines for $\text{St}=0.01$. The thickness of the arrows accounts for the norm of the mass flux $\rho_d \mathbf{u_d}$. Lower panel: same for $\text{St}=0.001$.}
\label{fig_circulation}
 \end{figure}  
\subsubsection{Dust vertical circulation}
\label{circulation}
In the previous paragraph, we showed that zonal flows directly impact the radial dust distribution by efficiently concentrating the grains. But do they also impact vertical settling? 
First, the stopping time in the density (or pressure) maxima is reduced so that the effect of gravitational settling is weaker. We checked that the dust layer is puffed up in these regions (and shrinked in the density minima by using similar arguments). Note however that this effect, when averaged in $x$, should not change significantly the mean vertical dust profile. Another and more subtle effect originates from the wind topology and vertical structure of the zonal MHD flow. The top panel of  Fig.~\ref{fig_circulation} shows the gas density and poloidal streamlines averaged in time and $y$, for $\beta=10^3$. We see that the wind flow is not distributed uniformly along $x$. Instead,  a strong and coherent windy plume emanates from the right part of the box where $\rho$ is minimum and $B_z$ is maximum (see Fig.~\ref{fig_xaverages} for comparison). This localized  structure is maintained during the entire simulation time $\simeq 300 \,\Omega^{-1}$. It is inclined and connected to a large scale roll in the midplane.  {Such asymmetric outflow with odd parity is actually reminiscent of global simulations \citep[Fig.~25 of][]{bethune17,gressel15b} and does not appear as an artefact of the shearing box. There is locally, near the midplane,  a spontaneous breaking of vertical symmetry of the flow and magnetic fields. Our local model however does not capture the outflow behaviour at large $z$ and its connection to the global disc geometry. Global models predict that at a certain height, the stream that flows toward the star  (on one side of the disc) strongly bends and ends up flowing away from the star (see Fig.~25 of \citet{bethune17}).}
Note finally that the rest of the disc is characterized by fainter and smaller scale axisymmetric rolls (or eddies) with smaller correlation time. Some of these features are probably residuals of our averaging procedure, which contains a finite number of datasets (typically 50 in time).  

The second and third panels of  Fig.~\ref{fig_circulation} show the dust density and mass flux streamlines in the poloidal plane. For intermediate Stokes number ($\text{St}=0.01$), the dust remains confined in the pressure maxima,  out of the windy plume structure. A tiny fraction of the material is advected in the large scale roll but the gravitational settling is too strong for the dust to be lifted away by the plume. The effect of the wind is then partially impeded.  However for $\text{St}=0.001$, more material enters the large scale roll (which also corresponds to gas density minimum) and get lifted by the plume. The flux streamlines of Fig.~\ref{fig_circulation} (bottom) shows that the dust is then dragged horizontally toward the pressure maxima, because of the plume inclination, and finally falls back onto the disc. Therefore a large scale circulation of small dust grains develops between the midplane regions and the disc corona.  This circulation, driven by the wind plume associated with the zonal flow,  allows grains to reach altitudes higher than those allowed by a simple turbulent diffusion (see Section \ref{sec_modele} for a more quantitative evidence of this result). 
\subsection{Simulations with an imposed mean pressure gradient}
\label{gradP}
\begin{figure}
\centering
\includegraphics[width=\columnwidth]{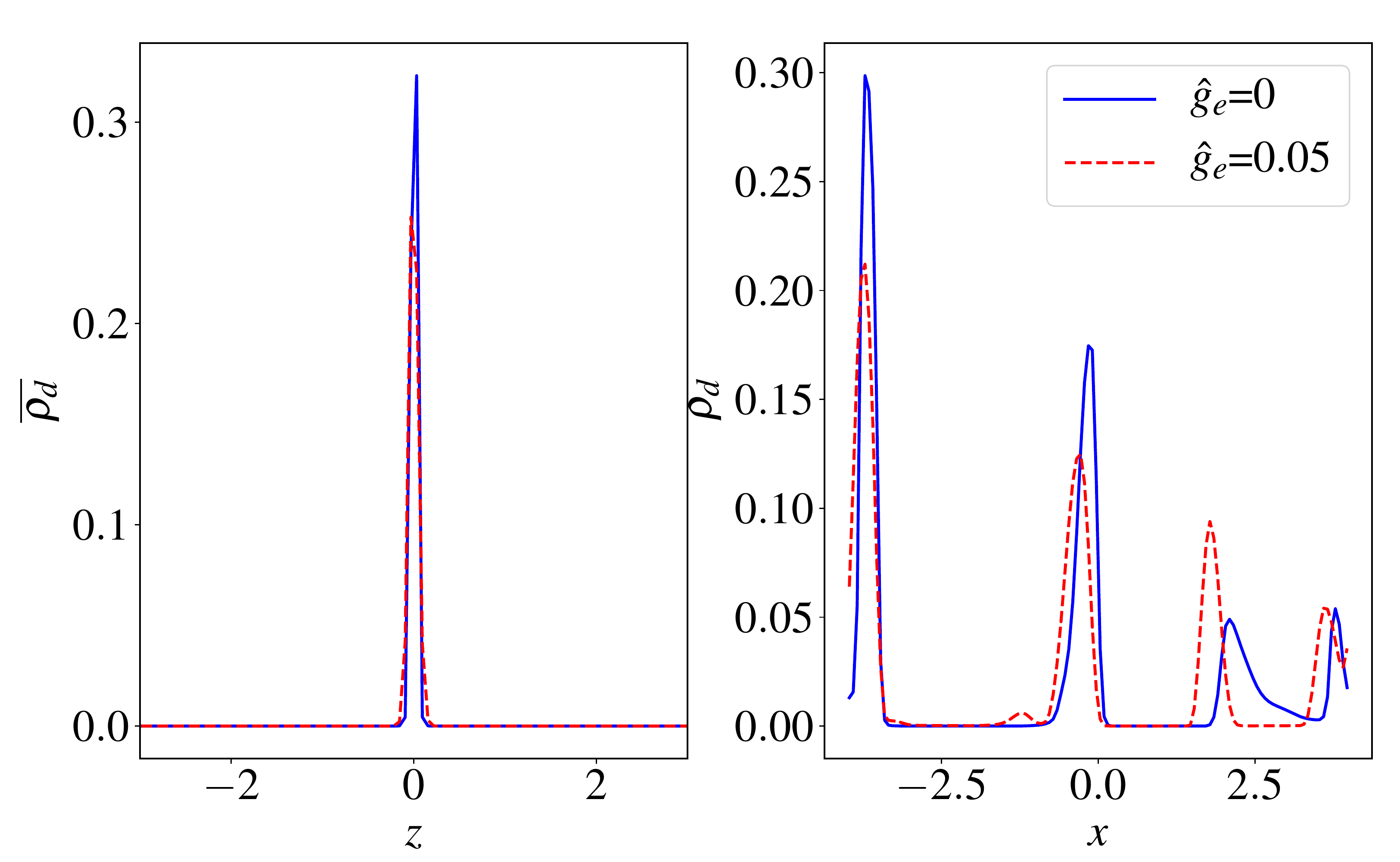}
 \caption{Right: dust vertical density profiles (averaged in $x$ and $y$) at $t=220 \, \Omega^{-1}$ for the case without mean radial pressure gradient (blue/plain line) and the case with $\hat{g_e}=0.05$ (red/dashed line). Left: Surface density profile in $x$ for both cases ($\rho_d$ averaged over $y$ and $z$).}
\label{fig_gradP}
 \end{figure}  
In protoplanetary discs, gas pressure decreases with radius.  In our previous simulations, we neglected the effect of a mean radial pressure gradient, which is known to cause a mean drift of dust particles \citep{birnstiel16}, and potentially lead to streaming instabilities  \citep{youdin05,youdin07,jacquet11}. The aim of this paragraph is to understand whether the mean drift changes the radial and vertical distribution of solids. 

We introduce a dimensionless parameter $\hat{g}_e$ that characterises the acceleration induced by the pressure gradient:
\begin{equation}
\hat{g}_e = -\dfrac{1}{H \Omega^2} \dfrac{1}{\rho}\left(\dfrac{ \partial P}{\partial R}\right) \approx  \left( \dfrac{H}{R} \right)
\end{equation}
with R the radius. In the Minimum Mass Solar nebula model  (MMSN) of \citet{chiang10},  the aspect ratio is  $H/R \approx 2.8\times 10^{-2}  (R/\text{AU})^{2/7}$, which gives $g_e \simeq 0.07$ at 30 AU. Observations of T-Tauri discs suggest a disc aspect ratio $H/R= 0.08$ at  $R=30$ A.U \citep[see Fig.~8 of ][]{pinte16}.  We recall that in this configuration, the mean drift velocity $\mathbf{\Delta v}=\mathbf{v_d}-\mathbf{v}$ is: 
\begin{equation}
\label{eq_radialdrift}
\mathbf{\Delta v}= -c_s  \hat{g}_e   \left[\dfrac{ \text{St}}{1+(f_g \text{St})^2}\mathbf{e}_x +\dfrac{f_g \text{St}^2}{2+2(f_g \text{St})^2}\mathbf{e}_y \right], 
\end{equation}
where $f_g=\rho/(\rho+\rho_d)$ \citep{youdin05}. The above relation shows that the drift increases with St for $\text{St}\lesssim1$ and reaches a maximum for $\text{St}\approx 1$. Therefore, to maximise the effect of the radial pressure gradient, it is suitable to consider the largest size simulated $\text{St}=0.1$.   We ran two different  simulations, one with $\hat{g}_e=0$ (no pressure gradient) and the other with $\hat{g}_e=0.05$. The latter corresponds to a radial drift of 0.005 $c_s$ in the midplane.  To allow comparison, we integrate a single size ($\text{St}=0.1$) and use identical initial conditions in both cases. 
Note that the radial pressure gradient in PLUTO is implemented by adding a force term $S_{\text{drift}}=\rho  H \Omega^2  \hat{g}_e $ per unit volume in the gas momentum equation, with uniform and fixed $\hat{g}_e$. 

Figure \ref{fig_gradP} shows the vertical and radial density profiles for both cases $\hat{g}_e=0$ and $\hat{g}_e=0.05$, at a given time $t=220 \Omega^{-1}$. When a radial drift is present, the dust layer is slightly wider in $z$, although the difference remains very small. Axisymmetric rings are still formed but the dust concentration inside  is weaker. We checked that the gas radial density structures  as well as the dust rings remain at the same location as in the the case $\hat{g}_e=0$.  

In the limit $\text{St} \ll 1$ and for perturbations wavelength $\lambda \gg 2\pi \hat{g}_e \text{St}^2$, \citet{jacquet11} showed that the pressure gradient induces a streaming instability with growth rate $\sigma$:
\begin{equation}
\sigma/\Omega \simeq f_p \text{St}^3 \hat{g}_e^2 \left(\dfrac{k k_x H}{k_z} \right)^2,
\end{equation}
where $k_x$ and $k_z$ are the radial and vertical perturbations wavenumbers and $f_p=\rho_d/(\rho+\rho_d$). This instability is only relevant for Stokes numbers $\gtrsim 0.1$  as $\sigma$ depends on the cube of St.   However it is not detected during the course of the simulations even for our largest St; the reason is that our resolution is probably too low. Indeed, the dust settles in a very thin layer, with a size not exceeding 2 or 3 grid points. We do not exclude that the streaming instability might occur on very short wavelength ($\ll H$) and destabilize the rings, but this requires to increase the number of grid points in $z$ and probably in  $x$ as well.
\section{Settling models}
\label{sec_modele}
The aim of this section is to elaborate a settling model  that accounts for the vertical density profiles obtained in Section \ref{vertical_profiles}.  We start with the historical 1D diffusion model of \citet{dubrulle95} and attempt to apply it to our non-ideal MHD flows. We examine in particular the influence of a mean wind component in the model. In a second part, we propose a 2D model particularly adapted to the case $\beta=10^3$,  including the effect of  zonal flows and vertical circulation induced by the MHD wind plumes.
\begin{figure*}
\centering
\includegraphics[width=\textwidth]{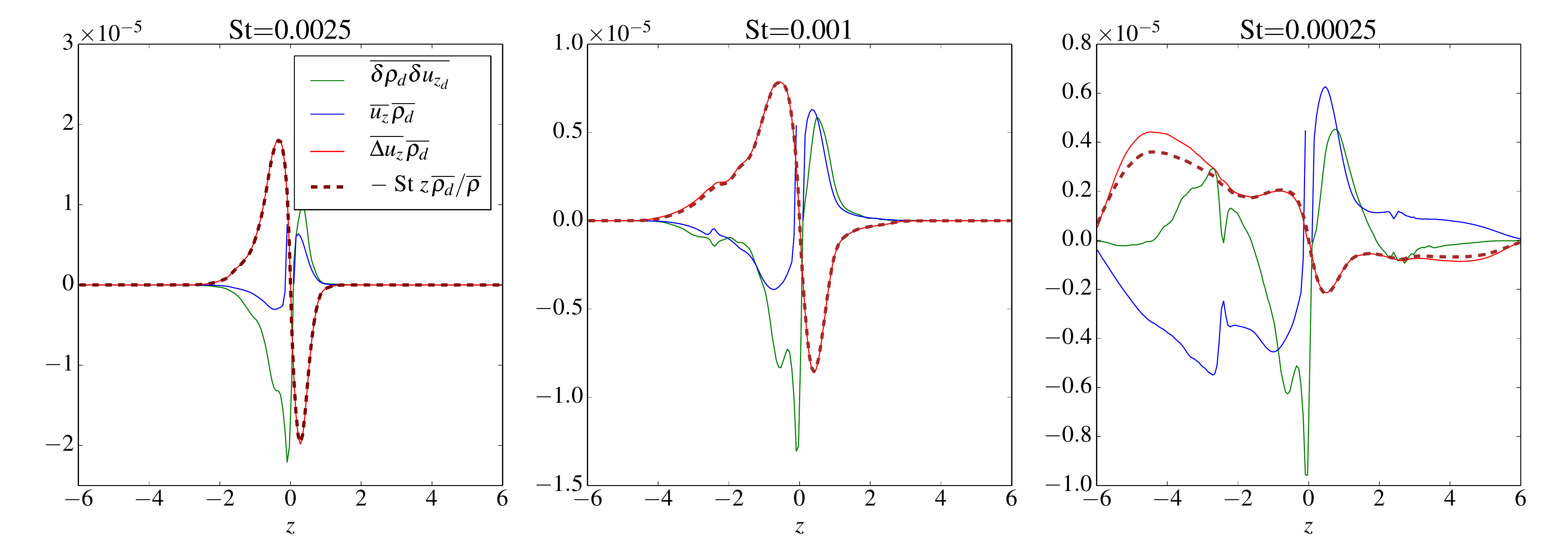}
 \caption{Vertical profiles of the flux terms appearing in the right-hand side of Eq.~(\ref{mass_eq_dust2}), calculated directly form simulations and averaged in time,  for $\beta=10^4$ and three different Stokes numbers. From left to right, $\text{St}=0.0025$, 0.001 and 0.00025. Green, blue and red lines account respectively  for the turbulent correlation term, mean wind and mean drift. The brown dashed line is the mean drift due to gravitational settling estimated in the terminal velocity approximation and assuming that $\overline{\rho^{-1}}=\overline{\rho}^{-1}$ (small  gas density fluctuations). }
\label{fig_flux}
 \end{figure*}  
\begin{figure}
\centering
\includegraphics[width=\columnwidth]{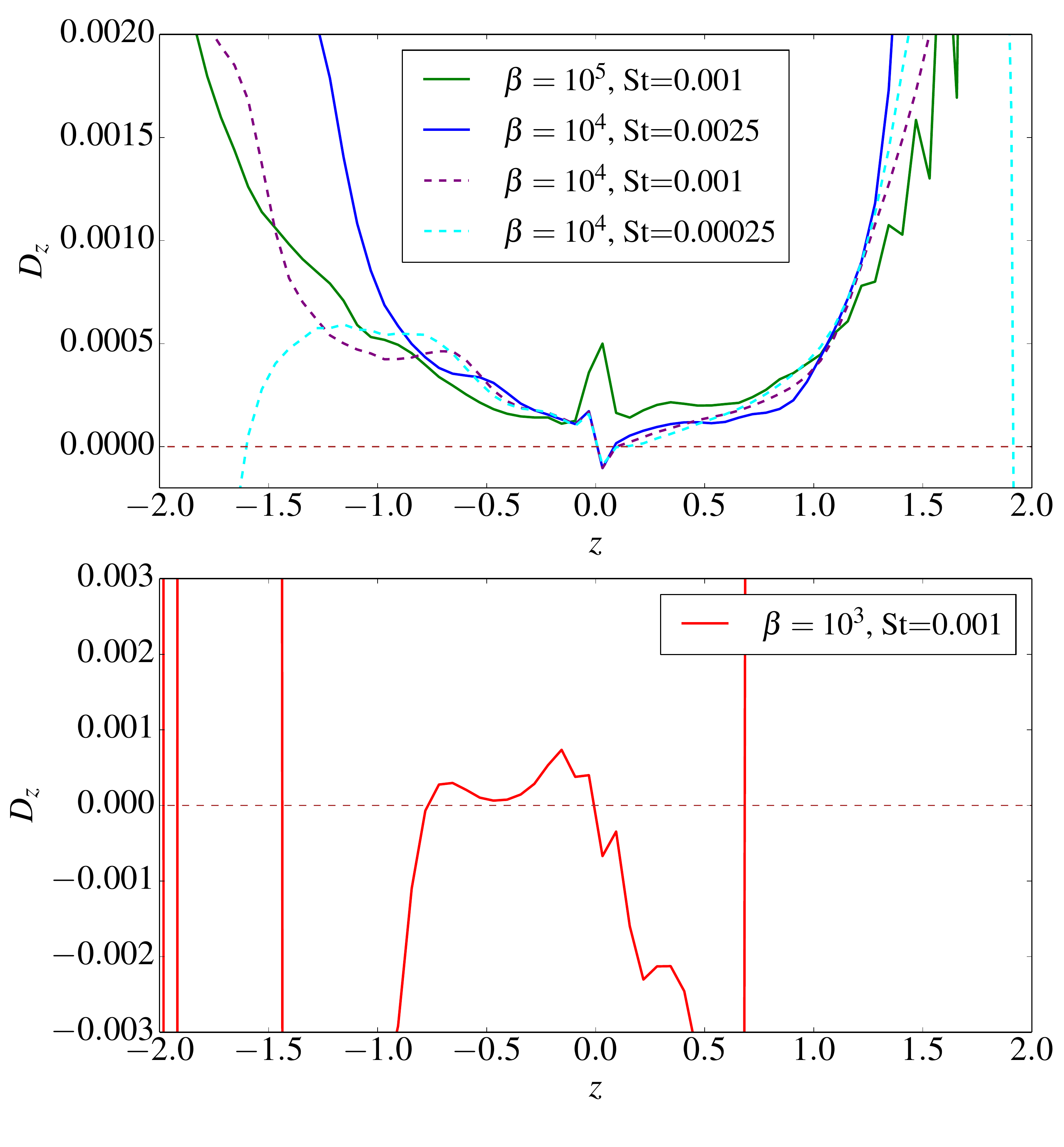}
 \caption{Turbulent diffusion coefficient computed numerically for different $\beta$ and Stokes numbers (estimated from relation \ref{def_Dz}). }
\label{fig_Dz}
 \end{figure}  
 \subsection{1D models}
\subsubsection{The turbulent diffusion model of \citet{dubrulle95} without winds}
\label{dubrulle_model}
The classical 1D diffusion theory \citep{dubrulle95} assumes that all quantities can be decomposed into a mean component (that depends only on $z$) and a small fluctuating part:
\begin{equation}
\label{eq_decompo2}
\rho_d = \overline{\rho_d}+ \delta \rho_d; \quad   \mathbf{v} = \overline {\mathbf{v}}+ \delta \mathbf{v};  \quad \mathbf{v_d} = \overline {\mathbf{v_d}}+ \delta \mathbf{v_d},
\end{equation}
By using such decomposition, the mass conservation equation (\ref{eq_mass_dust}), averaged in $x$ and $y$, becomes: 
\begin{equation}
\dfrac{\partial \overline{\rho_d}}{\partial t}+\dfrac{\partial}{\partial z} \left(\overline{\rho_d}\, \overline{ v_z} + \overline{\delta \rho_d \,\delta v_{z_d}}+ \overline{\rho_d} \, \overline{\Delta v_z}\right)=0,
\label{mass_eq_dust2}
\end{equation}
where $\mathbf{\Delta v}=\mathbf{v}_d-\mathbf{v}$ is the drift velocity between dust an gas. The first term in the $z$-derivative corresponds to the advection/stretching of dust by the mean vertical gas flow (wind) that we leave behind here. The second term is due to the correlation of turbulent fluctuations which is reduced to a diffusion operator in Dubrulle's theory. The third term accounts for the mean vertical drift motion of dust due to gravitational settling.  By making some assumptions, listed in Section \ref{validity_assumptions}  ,  it is possible to show that Eq.~(\ref{mass_eq_dust2}) takes  the form of an advection-diffusion equation:
\begin{equation}
\label{eq_dubrulle}
\dfrac{\partial\overline{\rho_d} }{\partial t} = \dfrac{\partial}{\partial z} \left(z \Omega^2\, \overline{\tau_s} \,\overline{\rho_d} \right) +\dfrac{\partial}{\partial z}  \left[ D_z \, \overline{\rho}  \dfrac{\partial}{\partial z} \left(\dfrac{\overline{\rho_d}}{\overline{\rho}}\right) \right]
\end{equation}
\citep{morfill85,dubrulle95,Schrapler04},  with $D_z \simeq  \langle \overline{v_z^2} \rangle $ $\tau_{\text{corr}}>0$ the diffusion coefficient and $\tau_{\text{corr}}$ the correlation time of the turbulent eddies. This equation and the relation liking $D_z$ to the vertical r.m.s fluctuations are  derived in appendix \ref{appendixB}. 
To find equilibrium solutions of this equation, we assume the ergodicity of the system and $\left \langle {\partial \overline{\rho_d}}/{\partial t} \right\rangle \simeq 0$. If the  gas is in  hydrostatic equilibrium $\rho=\rho_0 e^{-\frac{z^2}{2H}}$, an analytical solution is 
\begin{equation}
\label{sol_dubrulle}
\rho_d(z)=\rho_{d_0} \exp\left(-\dfrac{z^2}{2H^2}\right)\exp\left(-\int \dfrac{\text{St}\,\Omega \,z\,e^\frac{z^2}{2H^2}}{D_z(z)} dz \right)
\end{equation}
For uniform diffusion coefficient $D_z$ and $z \ll H$,  this gives: 
\begin{equation}
\label{eq_Hdgauss}
\rho_d(z)\simeq \rho_{d_0} \exp\left(- \dfrac{z^2}{2 H_d^2} \right)  \quad \text{with } \quad \dfrac{H_d}{H} =\dfrac{1}{\sqrt{1+\dfrac{\text{St}\,\Omega \,H^2}{D_z }}}
\end{equation}
The distribution is Gaussian and the dust heightscale has a dependence on $\text{St}^{-1/2}$ for  $\text{St} \gg D_z/(\Omega H^2)$. It tends toward unity in the limit of small St. 
\subsubsection{Check of the assumptions}
\label{validity_assumptions}
To obtain Eq.~(\ref{eq_dubrulle}) and the relation (\ref{eq_Hdgauss}), one needs to make several assumptions (see appendix \ref{appendixB}): 
\begin{enumerate}
\item Gas density fluctuations  small compared to the background density.
\item Vertical drift $\Delta u_z=v_{z_d}-v_{z}$ proportional to $z \,\text{St} / \overline{\rho}$. This is called the "terminal velocity approximation". 
\item Wind advection $\overline{u}_z \overline{\rho_d}$ negligible compared to the turbulent transport $\overline{\delta \rho_d \,\delta v_{z_d}}$.
\item Turbulent transport in the form of a diffusion operator $-D_z \,\overline{\rho}  \dfrac{\partial}{\partial z} \left(\dfrac{\overline{\rho_d}}{\overline{\rho}}\right)$ with $D_z$ uniform diffusion  coefficient (homogeneous and isotropic turbulence)
\end{enumerate}
The aim of this paragraph is to discuss the validity of these assumptions in the different regimes probed by our simulations. \\

The first assumption (1) is verified for $\beta=10^5$ and $\beta=10^4$. Indeed, in both cases, the density fluctuations never reach more than $30\%$ of the background density. However it is not valid for $\beta=10^3$ for which the density contrast can be higher than 3. 
To check assumptions (2) and (3), we plot in Fig.~\ref{fig_flux} the three different flux terms appearing in Eq.~(\ref{mass_eq_dust2}) for $\beta=10^4$. {The turbulent flux  $\overline{\delta \rho_d \,\delta v_{z_d}}$ is computed directly from the 3D numerical data, with perturbations
$\delta \rho,\, \delta \mathbf{u}$ obtained by subtracting to each quantity its mean value}. We superimposed in the same figures the approximated settling term $z \text{St} \overline{\rho_d}/ \overline{\rho}$. We show that there is an excellent agreement between the actual vertical drift and its value in the terminal velocity approximation. A tiny difference is however obtained at $z\gtrsim 2H$ for $ \text{St}=2.5\times 10^{-4}$. 
Regarding hypothesis (3),  Fig.~\ref{fig_flux} shows that the flux associated with wind advection (blue/plain lines) remains small compared to gravitational settling and turbulent transport, provided that $\text{St}\gtrsim 0.0025$,. However for $\text{St}<0.001$, it is comparable to other terms and carries a large fraction of the dust vertically. By neglecting the wind component, the 1D model of \citet{dubrulle95} is probably incomplete for small particles with $\text{St}\lesssim 0.001$. \\
Finally, to check assumption (4), we plot in Fig.~\ref{fig_Dz} the ratio
\begin{equation}
\label{def_Dz}
D_z = -\dfrac{\langle\overline{\delta \rho_d \,\delta v_{z_d}}\rangle}{\left \langle \overline{\rho} \dfrac{\partial}{\partial z} \left(\dfrac{\overline{\rho_d}}{\overline{\rho}}\right)\right \rangle },
\end{equation}
for different $\beta$ and  St. The diffusion coefficient $D_z$ is computed by averaging the numerical 3D datasets in time with a sampling period of $5 \Omega^{-1}$. For $\beta=10^5$ and $\beta=10^4$ (top panel),  $D_z$ is positive and well-defined. In the midplane, within $z\pm 0.5 H$, we found that $D_{z_0} \simeq 1.9 \times 10^{-4}$ for $\beta=10^5$ and $D_{z_0} \simeq 1.6 \times 10^{-4}$ for $\beta=10^4$. 
However $D_z$ is not uniform in $z$ and is multiplied almost by a factor 10 at $z=1.5\,H$.  The diffusion coefficient  increases rapidly with altitude,  fitting quite-well with an exponential function $D_z(z) \simeq D_{z_0} \exp(1.6 \,\vert z \vert /H)$.  These results are consistent with the theoretical prediction that $D_z \simeq  \langle \overline{v_z^2} \rangle  \tau_{\text{corr}}$ with  $\tau_{\text{corr}} \simeq  0.5 \Omega^{-1}$.  In particular, we showed in Section \ref{turbulence_properties} that the vertical r.m.s fluctuations increase quasi-exponentially with altitude and have similar values for both $\beta=10^5$ and $\beta=10^4$. Note that the diffusion coefficient does not depend too much on the particle size. \\

To summarize, as long as the size of the dust layer remains much smaller than $H$, the approximation of uniform diffusion coefficient holds and leads to the simple solution given in (\ref{eq_Hdgauss}). However for $\text{St} \lesssim 0.001$, this assumption does not hold; winds and non-uniform $D_z$   have to be included in the model to obtain acceptable solutions. 
For $\beta=10^3$ (bottom panel of Fig.~\ref{fig_Dz}), we found however that the 1D diffusion coefficient is not well-defined, as it can be either positive or negative, and undergoes brutal variation in  $z$.  We checked in particular that the flux associated with the turbulent fluctuations is negative for $z>0$, carrying mass downward. This result suggests that the simple 1D diffusion theory is inappropriate to describe dust settling in this regime. 
\subsubsection{Numerical vs theoretical density profiles}
\begin{figure}
\centering
\includegraphics[width=\columnwidth]{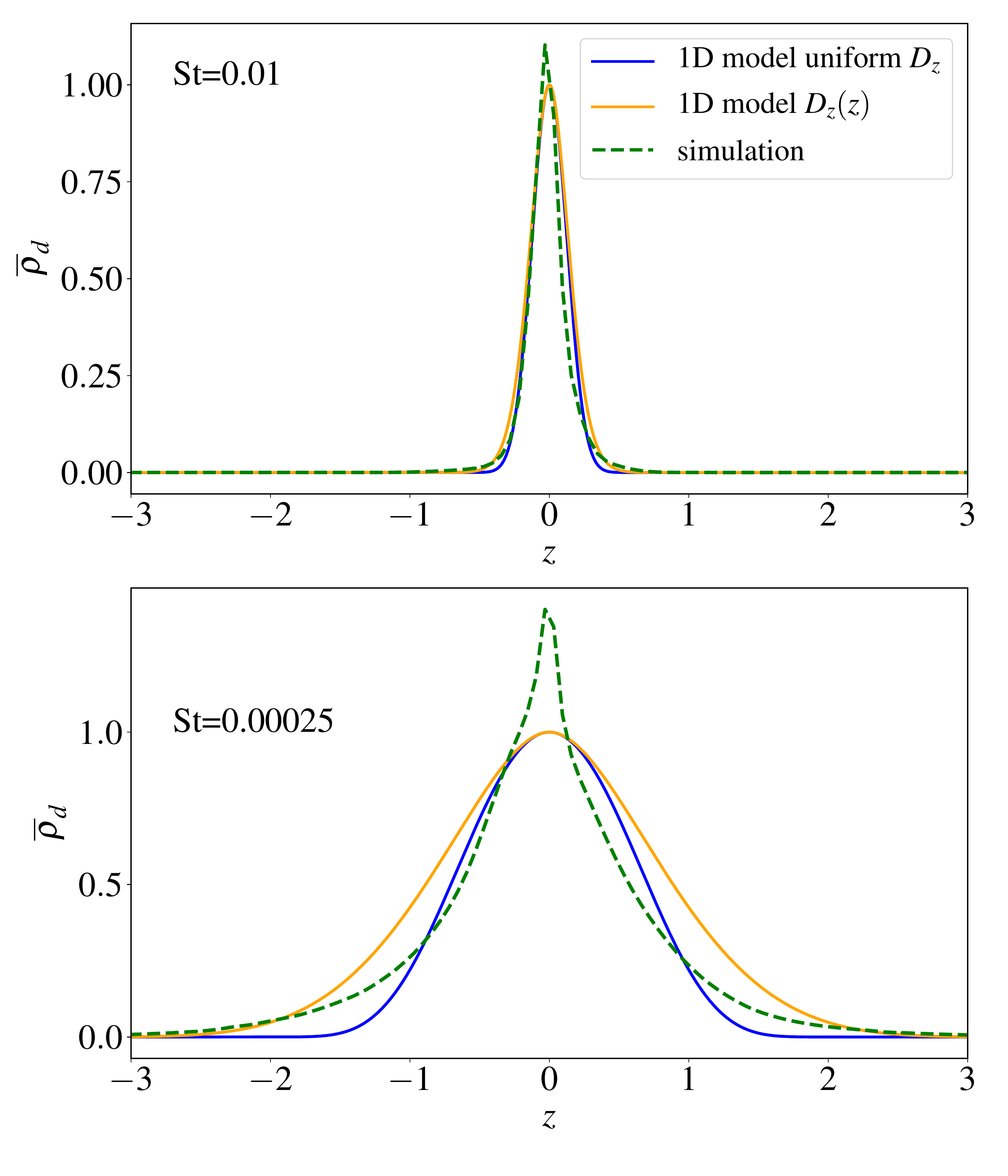}
 \caption{Comparison between numerical and theoretical density profiles for $\text{St}=0.01$ and $\text{St}=0.00025$ with Dubrulle's diffusion prescription ($\beta=10^4$). The blue line is derived from the simple model with uniform $D_z$, while the orange line is calculated with $D_z = D_{z_0} \exp(\vert \tilde{z} \vert /h_i)$, $hi=0.6$.}
\label{fig_profile1D}
 \end{figure}  
We focus here particularly on simulations with $\beta=10^4$ (the case $\beta=10^3$ requires to go beyond the 1D model and is treated further in Section \ref{2Dmodel}). Figure \ref{fig_profile1D} shows a comparison between the numerical density profiles in $z$ and the theoretical prediction of the 1D model, in case of a uniform diffusion coefficient $D_z = D_{z_0}=1.6 \times 10^{-4}$ (blue lines) and a varying $D_z = D_{z_0} \exp(\vert \tilde{z} \vert /h_i)$ (orange lines) with $h_i=0.6$ and $\tilde{z}=z/H$. The theoretical profile in case of a non-uniform $ D_z$ is obtained analytically by integrating (\ref{sol_dubrulle}): 
\begin{equation}
\overline{\rho_d}= \overline{\rho}_{d_0} \exp{\left(-\dfrac{z^2}{2H^2}\right)} \exp{\left(-\dfrac{\text{St}\,\Omega H^2 }{D_{z_0}}\xi(z)\right)}
\end{equation}
with 
\begin{equation}
\xi(z)=\dfrac{1}{h_i} \sqrt{\dfrac{\pi}{2}} e^{-\dfrac{1}{2 h_i^2}}\left[\text{erfi}\left(\dfrac{h_i \tilde{z}-1}{\sqrt{2} h_i}\right)-\text{erfi}\left(\dfrac{-1}{\sqrt{2}h_i}\right)\right]+e^{\dfrac{\tilde{z}^2}{2}-\dfrac{\tilde{z}}{h_i}}-1
\end{equation}
The solution with uniform diffusion coefficient (\ref{sol_dubrulle}) is recovered by setting $h_i=\infty$ and expanding the exponential in $\xi(z)$ to first order. 

For $\text{St}=0.01 \gg D_{z_0}/(\Omega H^2)$, Fig.~\ref{fig_profile1D} (top) shows that the analytical solution fits closely the numerical profile. As expected, the difference between uniform $D_z$ and varying $D_z$ solutions is very small.  However, for $\text{St}=2.5 \times 10^{-4} \simeq D_{z_0}/(\Omega H^2)$, both solutions fail to reproduce the numerical vertical distributions. If the heightscale calculated from the theoretical solution is not too far from the simulated one, the shape of the profile differs significantly. In particular, the 1D model does not account for the flaring at large $z$.  \\

In conclusion, we showed that the 1D diffusion model of \citet{dubrulle95} applied to a non-ideal MHD turbulent flow works pretty well for $\beta=10^5$ and $\beta=10^4$, provided that the dust particles are not too small.  For $\text{St} \lesssim 0.001$, the model does not reproduce very well the shape of the vertical profile, even when the assumption of uniform diffusion in $z$ is relaxed.  The wind starts to be important in this regime and has to be included in the model. This is investigated in the next sections. 

\subsubsection{1D models with a uniform and mean wind}
\label{wind1d}
\begin{figure}
\centering
\includegraphics[width=\columnwidth]{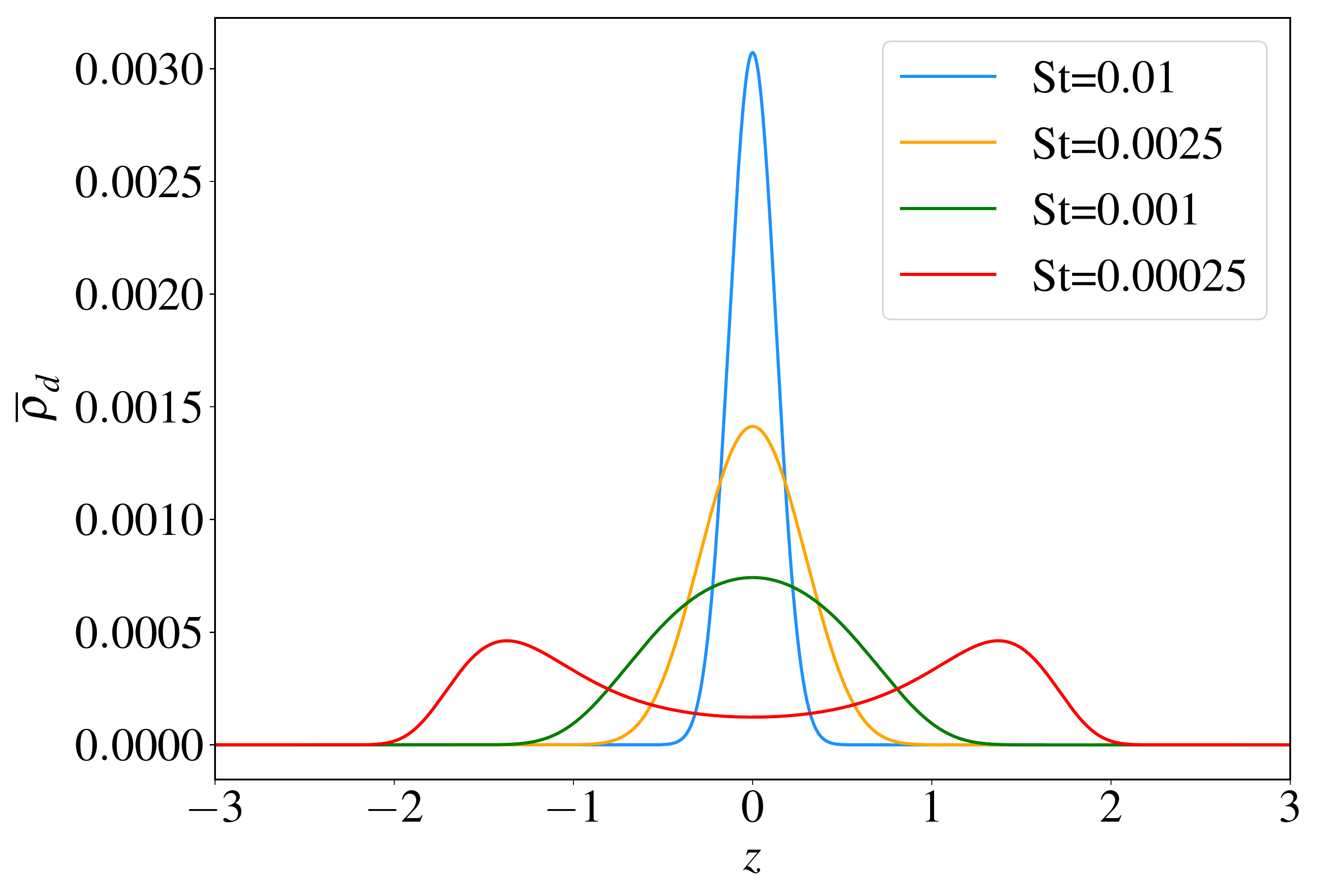}
\caption{Density profiles calculated in the 1D model of \citet{dubrulle95}, extended to the case with a mean wind $\overline{ v_z} = \Omega_w z$, for different St. They correspond to analytical solutions of Eq.~(\ref{sol_wind}) with $D_z(z)=cst=1.6\times 10^{-4}$ and $\Omega_w= 7\times 10^{-4}$.}
\label{fig_profile1D_wind}
 \end{figure}  
We showed that for small particles (typically $\text{St}<0.001$), the advection of the mean wind $\overline{\rho_d}\, \overline{ v_z}$ in Eq.~(\ref{mass_eq_dust2}) becomes comparable to the gravitational settling. Then, assumption (3) needs to be relaxed. A first and naive approach is to model the wind by a linear function of $z$ but uniform in $x$:  
\begin{equation}
\overline{ v_z} = \Omega_w z, 
\end{equation}
with a characteristic rate $\Omega_w$.  We checked that this simple dependence matches relatively well with the simulations data, at least for $z$ within $\pm H$.   We found respectively $\Omega_w=1.6\times 10^{-4}$, $7\times 10^{-4}$ and $6 \times 10^{-3} \Omega$  for $\beta=10^5$, $10^4$ and $10^3$. The modified equilibrium,  including the effect of the wind,  is: 
\begin{equation}
\label{sol_wind}
\rho_d(z)=\rho_{d_0} \exp\left(-\dfrac{z^2}{2H^2}\right)\exp\left(-\int \dfrac{(\text{St}\,\Omega e^\frac{z^2}{2H^2} -\Omega_w) \,z}{D_z(z)} dz \right)
\end{equation}
Assuming a uniform $D_z=D_{z_0}$, it is straightforward to show that for 
\begin{equation}
\text{St} \leq  \dfrac{\Omega_w}{\Omega} - \dfrac{D_{z_0}}{\Omega H^2}, 
\end{equation}
the dust density increases with altitude in the vicinity of $z=0$.  In  case $\beta=10^4$ with $\Omega_w=7 \times 10^{-4}$, this occurs for a critical $\text{St}=5.4\times 10^{-4}$. We plot in Fig.~\ref{fig_profile1D_wind} the solutions of Eq.~(\ref{sol_wind}) for different Stokes numbers and $\beta=10^4$. For $\text{St}= 2.5\times 10^{-4}$, the density profile exhibits two bumps at $z\simeq 1.5 H$ where the dust settles. This configuration corresponds to  "floating" or "levitating" dust layers. Particles are trapped between $z\simeq H$ where the wind dominates and regions of higher altitudes where gravitational settling dominates (the latter is an exponential function of $z$ while the wind advection is only proportional to $z$). We checked that assigning a $z$-dependence of $D_z$ leads to similar results. These floating layers are actually reminiscent of the simulations by \citet{miyake16}, who found that magnetorotational driven winds can concentrate dust grains with size $20-40 \,\mu$m  around four disc scaleheights by a similar mechanism. \\

In conclusion, the simple 1D wind model does not reproduce at all the profiles computed numerically. Indeed, none of them exhibit such distribution with floating particles.  The reason behind  that is intimately related to the baseline assumptions of the 1D model  and particularly to the decomposition (\ref{eq_decompo}). While the gas density fluctuations remain small compared to the background density, it is not necessarily true for the gas vertical velocity  (see  Fig.~\ref{fig_zaverages} for example). The term "fluctuations" here is actually misleading since it does include both random turbulent motions and the $x$-dependent self-organized and coherent axisymmetric flow induced by ambipolar diffusion. In particular, we showed in Section \ref{circulation} that the wind is distributed non-uniformly in $x$  and rather active in the gas density minima, where the dust concentration is low.  By averaging quantities in $x$, we thus overestimate the wind effect. Moreover, the presence of zonal flows in $x$ leads to a large scale meridional circulation (see Fig.~\ref{fig_circulation}) which is not taken into account in the 1D model.  To better understand the dust distribution in magnetized discs with zonal flows  (typically for $\beta=10^3$ and marginally for $\beta=10^4$), one needs to treat the advection/diffusion equation in 2D. This is the object of the next section. 

\begin{figure}
\centering
\includegraphics[width=\columnwidth]{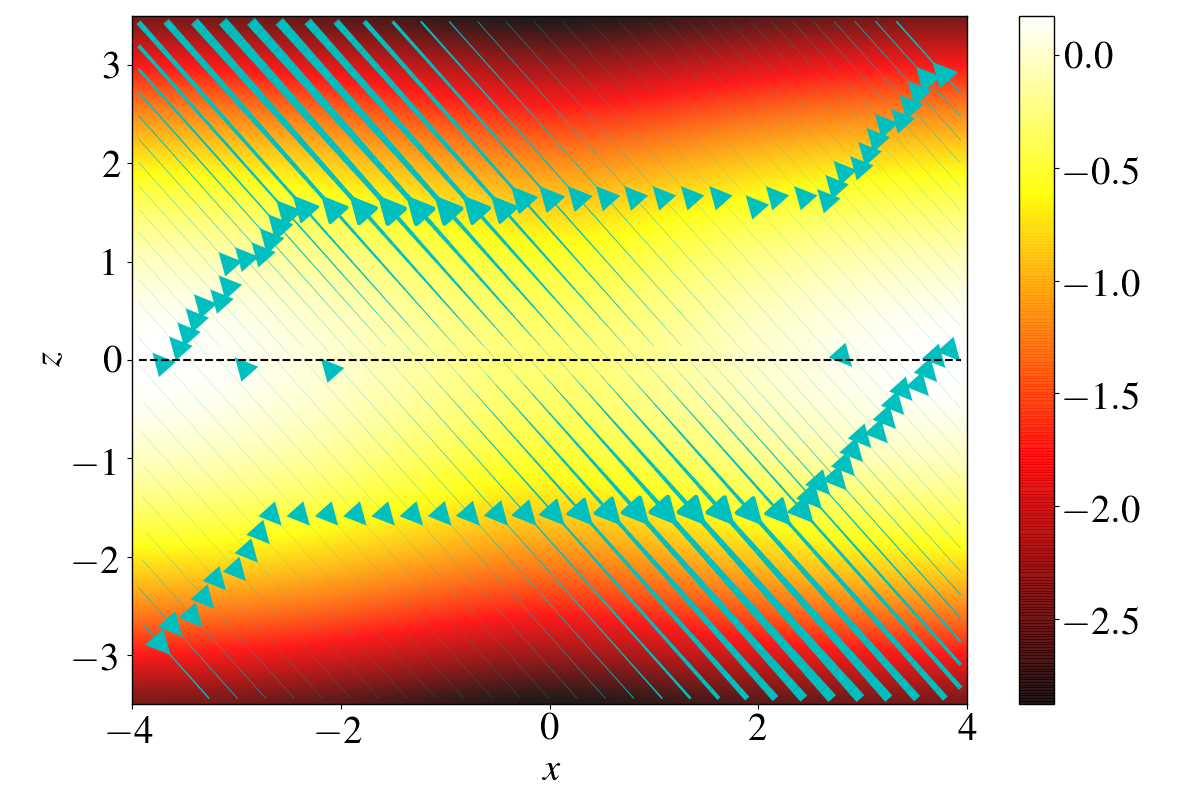}
 \caption{Gas density (colormap) and velocity streamlines (cyan arrows) of the 2D toy model. The lines thickness is proportional to the velocity $\Vert v \Vert$. The outflow inclination, width and averaged intensity are fixed and equal respectively to $i=40^\circ$, $\delta=0.7 H$, and $\Omega_w=6\times 10^{-3} \Omega$.  The density is Gaussian in $z$ and has a cosine perturbation in $x$ with amplitude $A_x=0.5$. The  outflow configuration can be directly compared with that of Fig.~\ref{fig_circulation} (here the plume has been shifted toward the center of the 2D domain)}
\label{fig_wind2D}
\end{figure}

\subsection{2D models with MHD zonal flows and wind plumes}
\label{2Dmodel}

In this section, we develop a 2D model that takes into account the wind geometry (in radius) and the zonal flow structures. We focus particularly on the case $\beta=10^3$ for which the settling process cannot be described by a 1D turbulent diffusion model (see \ref{validity_assumptions}), Note that the 2D model described below applies also for the case $\beta=10^4$ in the regime of small dust particles.  
 
\subsubsection{Formalism and toy model}
\label{toy_model}
To begin, we assume that the dust and gas motion is the sum of a steady axisymmetric background flow (physically representing the zonal structure) and turbulent fluctuations $\delta \mathbf{v}$:
\begin{equation}
\label{eq_decompo}
\rho_d = \langle \overline{\rho_d}^a \rangle+ \delta \rho_d; \quad   \mathbf{v} =\langle  \overline {\mathbf{v}}^a \rangle+ \delta \mathbf{v};  \quad \mathbf{v_d} =\langle  \overline {\mathbf{v_d}}^a \rangle+ \delta \mathbf{v_d},
\end{equation}
where $\langle\overline{X}^a \rangle(x,z)=\langle \int  X \, dy \rangle$ denotes the average in time and $y$ of a quantity $X$. Like in the 1D theory, we model the turbulent transport (cross-correlation of density and velocity fluctuations) through a diffusion operator: 
\begin{equation}
\langle \overline{\delta \rho_d \,\delta v_{z_d}}^a \rangle = -D_z \, \langle \overline{\rho}^a \rangle \dfrac{\partial}{\partial z} \left(\dfrac{\langle \overline{\rho_d}^a \rangle }{\langle \overline{\rho}^a \rangle }\right), 
\end{equation}
\begin{equation}
 \langle \overline{\delta \rho_d \,\delta v_{x_d}}^a \rangle=- D_x  \dfrac{\partial  \langle \overline{\rho_d}^a \rangle }{\partial x}, 
\end{equation}
with $D_x$ and $D_z$ the radial and vertical diffusion coefficients.  To simplify notation we drop the brackets and the overline denoting the time and $y$ averages. The 2D steady axisymmetric dust distribution is then governed by a second order partial differential equation:
\begin{equation}
\label{eq_model2D}
\dfrac{\partial (\rho_d v_{x_d})}{\partial x} + \dfrac{\partial (\rho_d v_{z_d})}{\partial z} = \dfrac{\partial}{dz} \left[ D_z \, \rho \dfrac{\partial}{\partial z} \left(\dfrac{\rho_d}{\rho}\right)\right] + \dfrac{\partial}{dx} \left[D_x  \dfrac{\partial \rho_d}{\partial x}\right]. 
\end{equation}
Before solving this equation for $\rho_d$, it is necessary to prescribe the background axisymmetric gas density and velocities. To model the zonal flow, we assume the following form for the gas density:
\begin{equation}
\hat{\rho} = \rho_0 \left[1-A_x\cos{\left(\dfrac{2\pi x}{L_x}\right)}\right]\exp{\left(-\dfrac{z^2}{2 H^2}\right)}, 
\end{equation}
with $A_x$ the amplitude of the radial density pattern.   Instead of a uniform wind $\overline{ v_z} = \Omega_w \, z$ prescribed in \ref{wind1d}, we consider here a $x$-dependent steady axisymmetric outflow modelled by: 
\begin{equation}
\label{eq_v2D}
\begin{cases}
v_z=\Omega_w z \, \mathcal{G}(x)   \quad  \text{with} \quad \mathcal{G}(x) = \sum_{n=-\infty}^{\infty} \exp{\left(-\dfrac{(x_i+n L_x \cos{i} )^2}{2 \delta^2}\right)} \\
 v_x= -v_z \tan{i} 
\end{cases}
\end{equation}
where $x_i = x\cos{i}+z\sin{i}$. This simple prescription mimics the inclined outflow obtained for $\beta=10^3$ and illustrated in the top panel of Fig~\ref{fig_circulation}. The inclination $i$ refers to the angle relative to the vertical axis. The plume is confined  within the gas density minimum, with a Guaussian profile in $x$ and a given width $\delta$. The infinite sum ensures the strict periodicity of the flow in the radial direction (with period $L_x$).
Figure \ref{fig_wind2D} shows the gas density and the streamlines associated with the 2D outflow for a given set of parameters $i$, $\delta$, $\Omega_w$ and $A_x$.  Given such prescription for the gas motion, it is possible to calculate the dust velocities. To simplify the problem as much as possible  we neglect the radial drift induced by the pressure maximum, so that $v_{x_d}=v_x$. This is correct in the limit of small Stokes numbers. In $z$, we use the terminal velocity approximation (see Appendix \ref{appendixB}) to estimate the vertical drift: 
\begin{equation}
v_{d_z} = v_{z} - \dfrac{\text{St}\,\Omega \,z}{\hat{\rho}}
\end{equation}
\begin{figure}
\centering
\includegraphics[width=1.02\columnwidth]{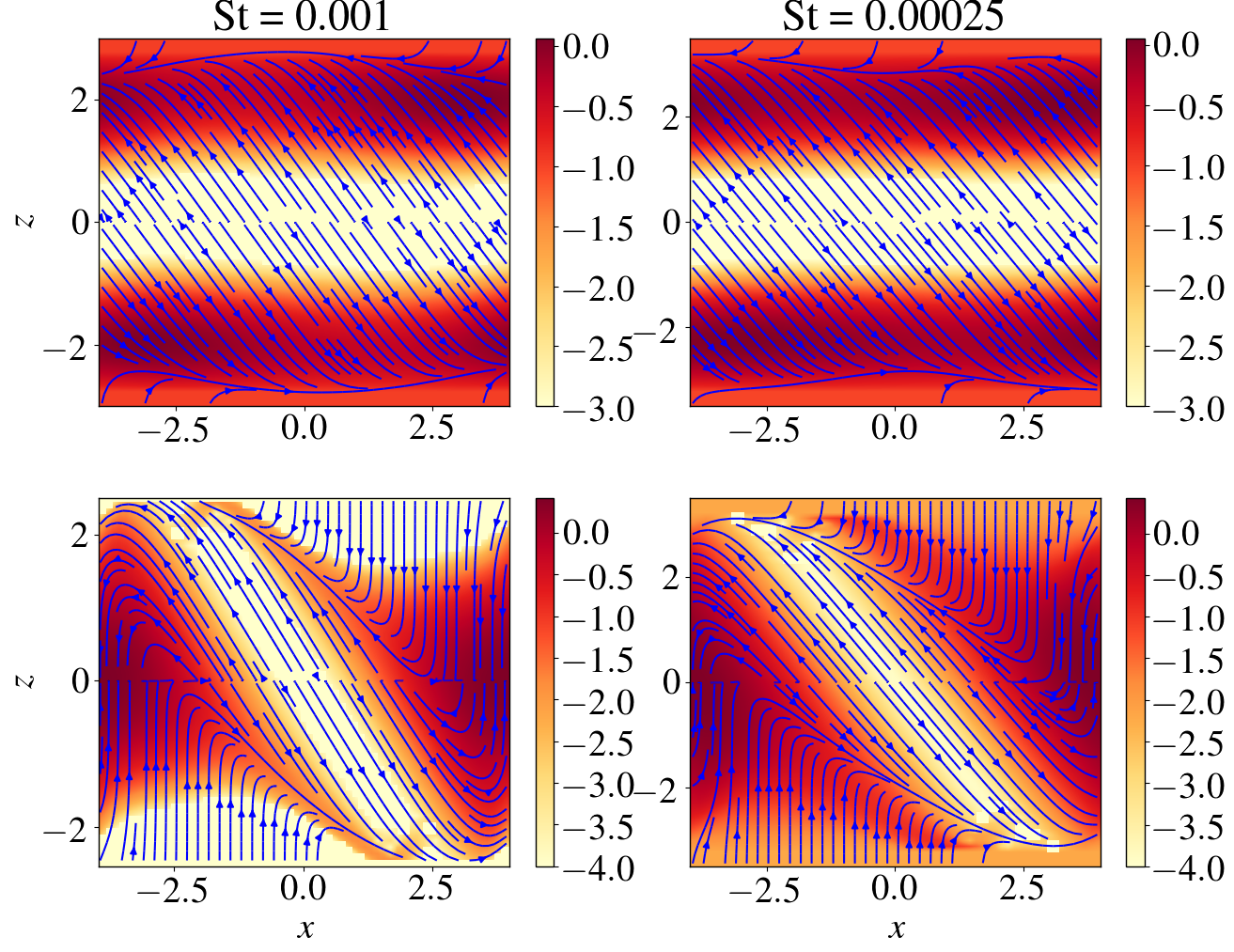}
 \caption{Equilibrium solutions of the 2D advection-diffusion equations (\ref{eq_model2D}) with prescribed $v_x$ and $v_z$ (Eq.~\ref{eq_v2D}) for two different grain sizes. The diffusion coefficients are uniform in $z$. The colormaps indicate the dust density distribution and the blue lines are the velocity streamlines in the poloidal plane. The top panels correspond to a uniform wind with $\delta=\infty$ and $\Omega_w=6 \times 10^{-3}$ while the bottom panels correspond to wind plume of width $\delta=0.7 H$. In both cases, we fixed the inclination of the wind  $i=40^{\circ}$ and the amplitude of the zonal flow $A_x=0.5$, as well as the two diffusion coefficients $D_x=2.5\times 10^{-4}$ and $D_z=3\times 10^{-4}$. }
\label{fig_rhod2D}
\end{figure}
\begin{figure}
\centering
\includegraphics[width=\columnwidth]{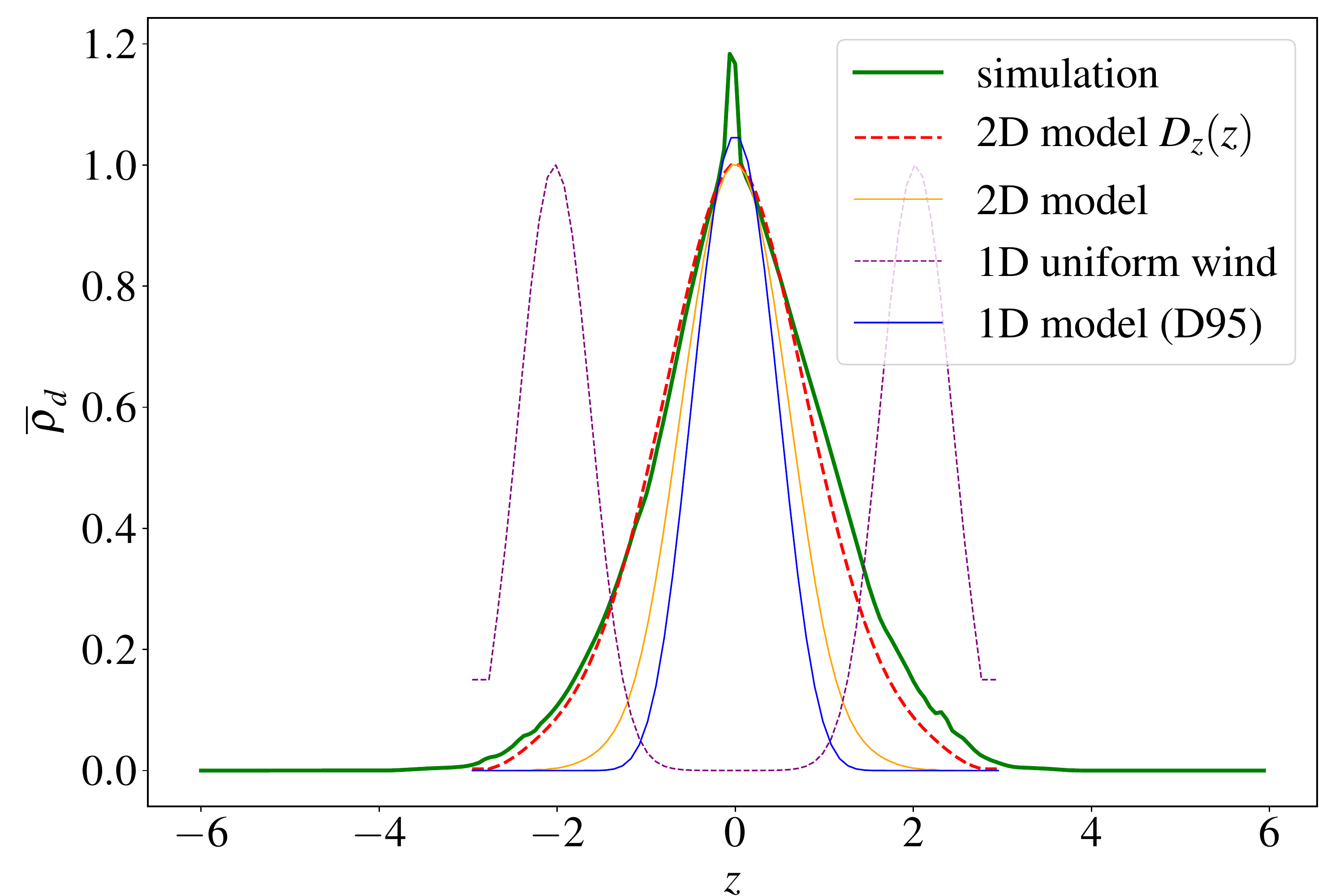}
\includegraphics[width=\columnwidth]{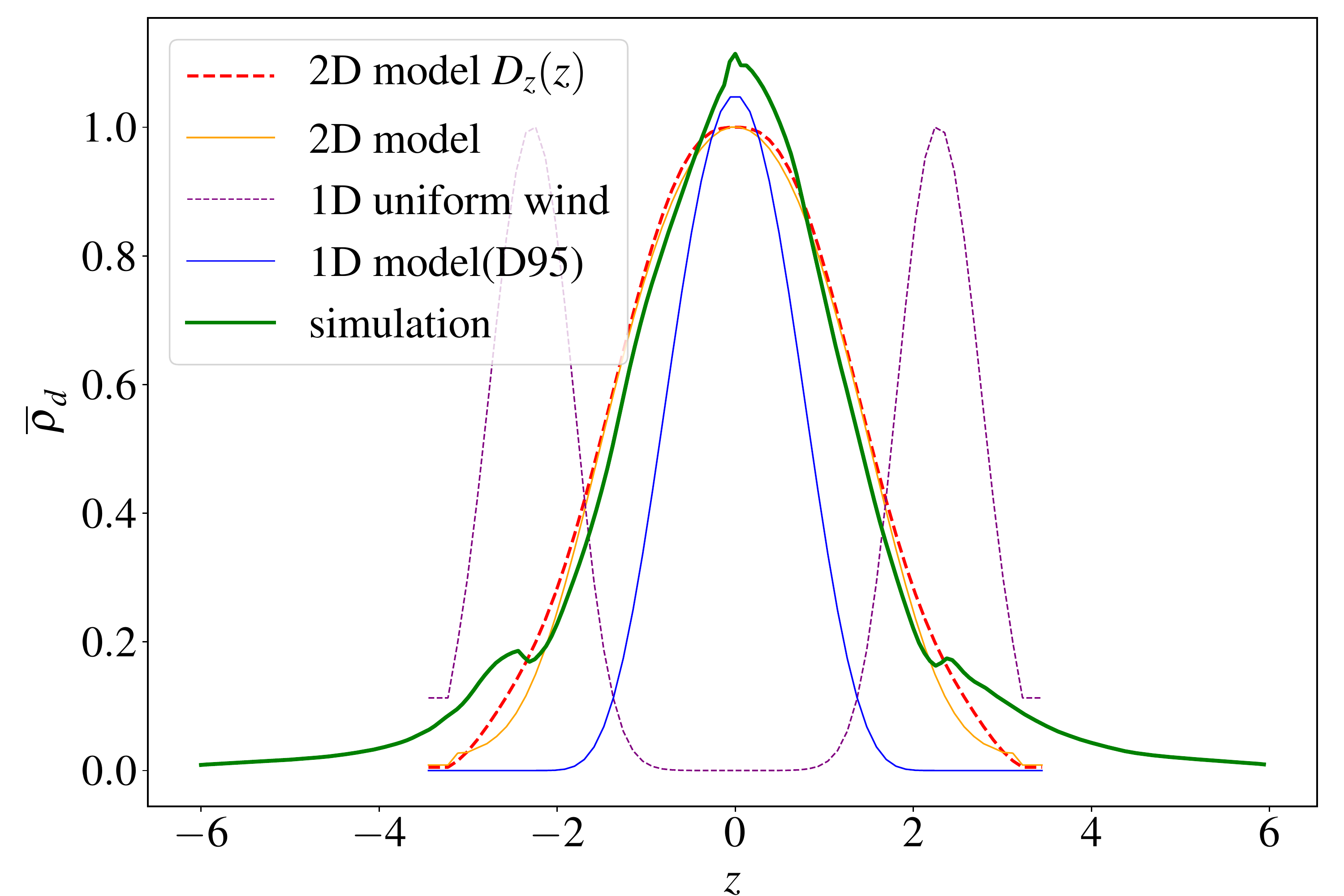}
 \caption{Comparison between different settling models for $\text{St}=0.001$ (top) and $\text{St}=0.00025$ (bottom) in the case $\beta=10^3$. The green line represents the vertical density profile obtained from the simulation. The red/dashed and orange lines are computed from the 2D semi-analytical models with parameters detailed in Section \ref{parameters_2dmodel}. The red line is with non-uniform $D_z$ in $z$ while the orange corresponds to  $D_z=cte$. The blue and purple/dotted lines are analytical solutions derived from a 1D diffusion model \citep{dubrulle95} respectively without wind and with uniform wind. All profiles are normalized to unity in the midplane to allow comparison.}
\label{fig_profile2D1}
\end{figure}  

\subsubsection{Parameters of the 2D model}
\label{parameters_2dmodel}
For a given Stokes number St, the 2D model described in \ref{toy_model} contains exactly six free parameters, three associated with the geometry of the outflow ($i$, $\delta$ and $\Omega_w$), two associated with the turbulent diffusivities ($D_x$ and $D_z$)  and one related to the amplitude of the zonal flow. To match with the simulations, we choose $i=40^{\circ}$, $\delta = 0.7 H$ and $\Omega_w=6 \times 10^{-3} L_x/(\delta \sqrt{2\pi})$. The latter is calculated such that the horizontal average $\overline{v_z}(z)$ matches with the 1D prescription of \ref{wind1d}. The amplitude of the zonal flow $A_x$ is fixed to 0.5.  The two diffusion coefficients remain to be estimated from simulations. We remind  that their definition in the 2D formalism are:
\begin{equation}
D_x(z)= \int  \dfrac{\langle \int  \delta \rho_d \, \delta v_{x_d}   \, dy \rangle}{\dfrac{\partial}{dx} \langle   \int  \rho_d \,dy     \rangle} dx
\end{equation}
\begin{equation}
D_z(z)= \int  \dfrac{\left \langle \int  \delta \rho_d \, \delta v_{z_d}  \,  dy\right \rangle}{\left \langle   \int  \rho \,dy \right     \rangle  \dfrac{\partial}{dz} \left (  \dfrac{ \left \langle   \int  \rho_d \,dy\right\rangle }{\left \langle   \int  \rho \,dy \right\rangle}  \right) } dx
\end{equation}
The numerical calculation for $\beta=10^3$ gives $D_x \simeq 2.5 \times 10^{-4}$ and $D_z \simeq 3 \times 10^{-4}$ in the midplane regions within $z\pm 1 H$. However we point out that these coefficients are difficult to evaluate further in the disc atmosphere. 

\subsubsection{2D solutions and comparison with simulations}

Once the dust velocities and parameters of the model are fixed, we compute numerically the 2D equilibrium solutions of (\ref{eq_model2D}). The numerical methods to obtain these solutions are detailed in Appendix \ref{appendixC}.  Figure \ref{fig_rhod2D} shows two examples of solutions computed for $\text{St}=0.001$ and $\text{St}=0.00025$.  The top panels correspond to a uniform wind in $x$ with $\delta=\infty$ while the bottom panels correspond to  a wind  distributed non-uniformly in $x$ with $\delta=0.7H$. Both have non-zero inclination $i=40^{\circ}$.  In the first case, the dust forms two layers off the midplane, located  around  $z=2H$, exactly as in the 1D case (see red line in Fig.~\ref{fig_profile1D_wind} for comparison). Note that the dust distribution have a radial dependence due to the gas density modulation in $x$.  In the second case however, the dust remains confined in the midplane. The strong collimated wind associated with the plume locally depletes the dust and forces some material to float around $z=2H$, but most of the grains fall back into the disc once they leave the plume region. A large scale poloidal circulation, similar to that described in Section \ref{circulation} is then sustained in the disc atmopshere. One critical parameter that controls the rate of mass entering the plume and therefore the amount of dust accumulating at large $z$  is the radial diffusion coefficient. Indeed strong radial diffusion will allow a rapid replenishment of the depleted region (plume) and  thus massive dust outflow. On contrary,  small $D_x$ will tend to reduce the mass loading.  The simulations are  rather in a regime of inefficient wind since the diffusion timescale $H^2/D_x\approx 4000\, \Omega^{-1} $ is much longer than the outflow timescale $\Omega_w^{-1} \approx 200\, \Omega^{-1}$. Note that in the simulations (Fig.~\ref{fig_circulation}) the outflow region was not that depleted because a large scale poloidal roll was trapping the dust inside. This roll has not been included in our model as its effect is probably not dominant in the vertical settling process. \\

To check that our 2D model is compatible with simulations, we compare in Fig.~\ref{fig_profile2D1} the horizontally averaged density profiles $\overline{\rho_d}$ of the 2D solutions (red dashed lines and orange lines) with those obtained in the shearing box (green lines). For both Stokes numbers  $\text{St}=0.001$ and  $\text{St}=0.00025$ our model is able to accurately reproduce the shape of the vertical profiles. For $\text{St}=0.00025$, the result is independent on whether the diffusion coefficients are uniform or not. However this is not necessarily true for $\text{St}=0.001$ for which the 2D model with uniform $D_z$ fails to reproduce the simulations data. We also superimpose in Fig.~\ref{fig_profile2D1} the profiles obtained with the 1D Dubrulle's model (blue lines) and the 1D model with wind (purple dotted line). Clearly the 2D models are better to explain the simulations. This indicates that the dust poloidal circulation induced by the zonal flows is crucial to model the dust vertical equilibrium.

\subsection{Conclusions on settling models}
In the regime of $\beta \gtrsim 10^4$ and for $\text{St} \gg 0.001$, we found that the 1D model of \citet{dubrulle95} correctly predicts the shape of the density profile and the ratio $H_d/H$. However when the magnetization is stronger or the particles size smaller, the 1D model fails to reproduce the simulations profiles. We found that the wind dynamics and its dependence on $x$ has a crucial effect on dust distribution and needs to be modelled in a 2D framework. The dependence of diffusion coefficients on $z$ seems also to matter but does not fundamentally change the physical processes involved. 
\section{Discussion and link with observations}
\label{sec_obs}
\subsection{Comparison with other MHD simulations}
Our results suggest that the dust is highly sedimented in presence of ambipolar diffusion, regardless the magnetization. In particular for particles size larger than the millimetre, we found that the ratio  $H_d/H$ is smaller than 0.1 and the vertical diffusion coefficient less than $3\times 10^{-4}$.  This corresponds to values of the Schmidt number $S_c=\alpha \Omega H^2/ D_{z}$ between 10 for $\beta=10^5$ and 200 for $\beta=10^3$. Hence, the gas transport is always much larger than the vertical dust transport. This in contrast with ideal MRI turbulence simulations (either zero or non-zero net vertical flux) for which typical Schmidt number is never larger than 10 \citep{johansen06,fromang06}. We can interpret this result as a consequence of enhanced Maxwell to Reynolds stress ratio at large $B_z$ (see Section \ref{turbulence_properties}). In fact, in non-ideal MHD flows, angular momentum is mainly extracted through coronal winds and laminar Mawell stress, while dust is more sensitive to turbulent gas motions.  Note that the large $S_c$ found in our simulations, as well as the the small $H_d/H$ and $\tau_{corr} \Omega\simeq 0.5$, seem incompatible with the recent non-ideal MHD simulations of \citet{zhu15} with ambipolar diffusion. Indeed in their case, they  found $S_c\lesssim 1$ and diffusion coefficients 10 to 20 times larger than ours. The main difference between their simulations and ours is the vertical and radial box size (they used $L_x=4H, L_z=6H.$ while we use a larger domain with $L_x=8H, L_z=12H$). We ran few simulations by varying both $L_x$ and $L_z$ and found that the discrepancy is mainly due to the vertical extent $L_z$. The reason is that the wind properties, in particular its strength and mass loss rate, depend strongly on the the vertical extent of the box \citep{fromang13}. We checked that for a box of size  $L_z=6H$, the vertical r.m.s  turbulent fluctuations are stronger by a factor almost 2 and the wind by a factor 3-4,  compared to the case $L_z=12H$. This leads to a much higher diffusion coefficient and a ratio $H_d/H$ three time larger. We emphasize that the convergence of the outflow properties with $L_z$ is probably not reached even for  $L_z=12H$. The ratio $H_d/H$ may therefore be slightly over-estimated by our simulations. To avoid such bias, it is necessary  to model the wind in a global way and characterize the dust settling in global simulations, which will be the object of a future paper. 
\begin{figure}
\centering
\includegraphics[width=\columnwidth]{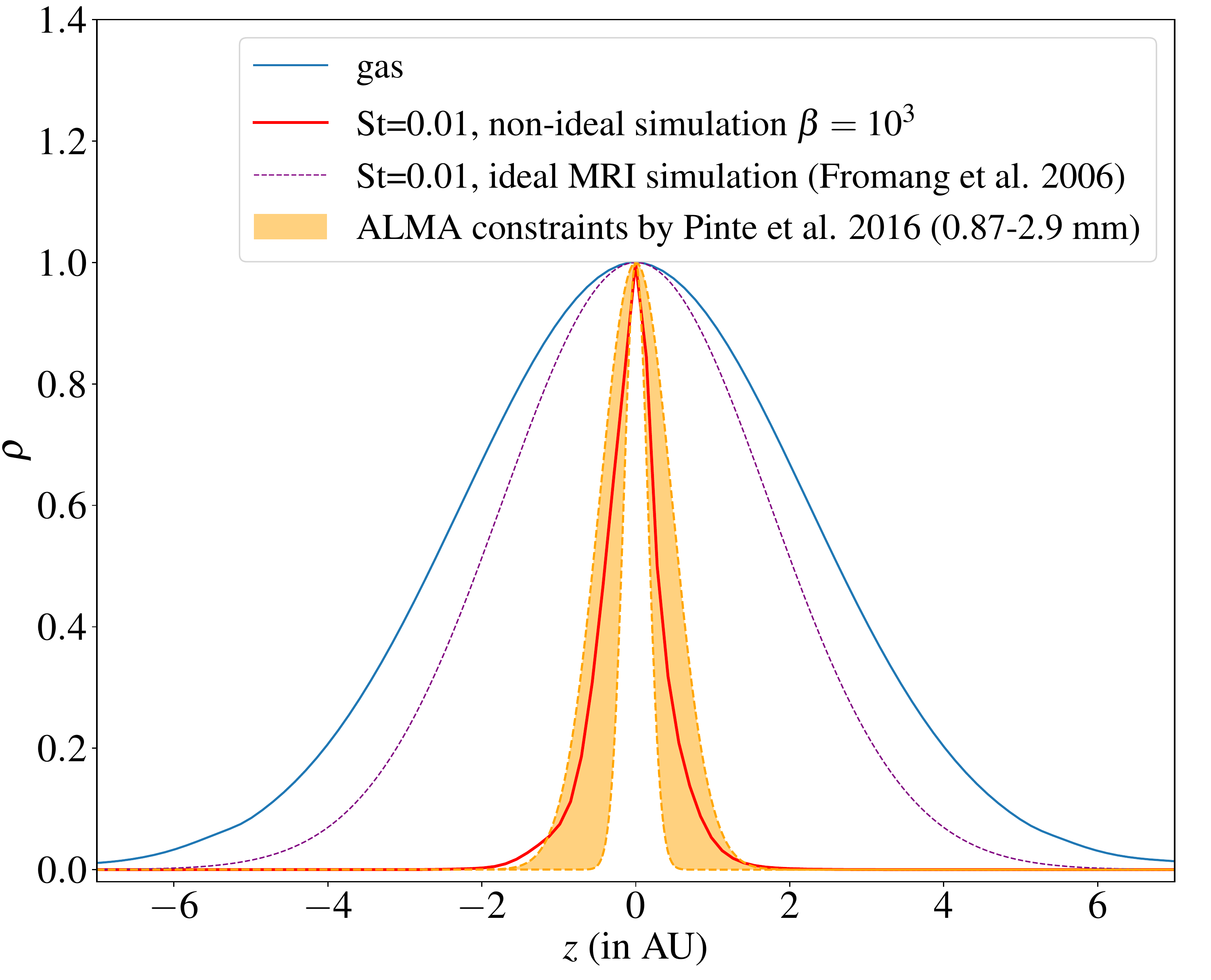}
 \caption{Vertical density profile of the millimetre dust simulated at 30 AU ($\text{St}\simeq 0.025$) and comparison with ALMA observations. The red profile corresponds to our simulation with $\beta=10^3$ and ambipolar diffusion ($\text{Am}=1)$. The purple profile is that obtained in the ideal (zero net flux) MRI simulation of \citet{fromang06} for the same Stokes number.  The orange area corresponds to the range of profiles compatible with the 0.87-2.9 mm dust continuum emission measured by ALMA \citep{pinte16}. To help the comparison, the density profiles are all normalized with $\rho_0=1$; the gas profile is represented by a blue line. The x-axis unit is in AU, the conversion is about $1H \simeq 2.5$ AU at $R=30$ AU.}
\label{fig_obs}
\end{figure}
\subsection{Vertical settling and comparison with ALMA  observations}
We compare here the dust scaleheight $H_d$ measured in our simulations with that inferred from observations. We discuss also about the limitations of such comparison.  Most of the constraints available today on $H_d$ come from the sub-millimetre ALMA observations of T-Tauri discs like HL-Tau \citep{pinte16}. These constraints, based on the measure of the rings contrast, suggest that sub-millimetre particles are contained in a geometrically thin disc with scale height between  0.7 and 2 AU at 100 AU.  If we assume that the ratio $H_d/H$ does not vary strongly with $R$ and consider disc aspect ratio $H/R \simeq 0.1$ at $R=100 \text{AU}$ \citep[suggested by][for HL Tau]{pinte16},  we find  $H_d/H \simeq 0.07 - 0.2$. \\

Using either the MMSN model or current estimation of disc surface density profiles, it is possible to show that sub-millimetre dust particles measured by ALMA  lies into the range $\text{St}=0.005-0.01$ at 30 AU (see Section \ref{conversion}). To obtain this range, we have  considered that most of the emission in the band 6-7 of ALMA (with wavelength $\lambda \simeq 1$ mm) is due to particles of size $a \simeq \lambda/(2\pi)\simeq 160\, \mu$m \citep[][]{kataoka17}. For such St, our simulations with ambipolar diffusion indicate that the dust is contained within  $H_d \simeq 0.11 - 0.19 H$, depending on the magnetization of the disc. This scaleheight is then compatible with observation. To facilitate the comparison, we superimpose in Fig.~\ref{fig_obs} the normalized density profile in $z$ obtained numerically at $\beta=10^3$ , $\text{St}=0.01$ (red line) and a surface (in orange) covering the range of Gaussian profiles estimated by the observations of \citet{pinte16}. Note that the profiles obtained for smaller $\beta$ lies only marginally within the error bar. 
We stress that the scaleheights considered here are 4 to 10 times smaller than those found in  ideal MRI simulations \citep{fromang06b}. In particular, the typical vertical diffusion coefficient in this case ($\simeq 5.5 \times 10^{-3}$) is 20 times larger than those obtained in our simulations with ambipolar diffusion. All these results are indications that the dynamics of discs like HL-Tau, in regions beyond 1 AU, is strongly affected by non-ideal MHD processes such as ambipolar diffusion. \\

There are however many caveats associated with such comparison. First the uncertainties on the gas surface density in the outer regions can be more important than those estimated. The current MMSN models and the observations can differ by one order of magnitude at most. We have also assumed that the dust to gas scaleheight ratio $H_d/H$ does not vary with radius, which is a crude approximation. In reality, it may encounter large variations between 30 AU and 100 AU, either due to radial variation of turbulence strength or local change in the gas to dust ratio \citep{pinte16}. The estimation of $H_d$ from the rings contrast is also largely arguable,  since it depends on the dust opacities and radiative transfer used in the synthesized model of the disc.  High resolution imaging of edge-on discs is probably a better and promising avenue to measure directly the dust scaleheight.  Finally, although disc emission at $\lambda=1$mm peaks around $a \simeq 150\, \mu$m, there is a wide range of particles size from $50\, \mu$m to $500\, \mu$m that also contributes, to a lesser extent, to such emission.   Since particles are not distributed uniformly in size in real discs, there is potentially a bias in our estimation of the Stokes number at 30 AU. This last issue however should not change drastically the conclusions of the present work.\\

{Note that ongoing ALMA observations of edge-on discs (in particular in HH30, Menard et al., in prep), combined with previous HST surveys, are going to provide accurate measure of both millimetre and micrometer dust thickness. We achieved a preliminary confrontation between our model and these measures and found that the vertical extent of both dust components fit with the observations when using $\beta=10^3$.  A forthcoming paper will be dedicated to the details of such comparison.}

\subsection{Rings and comparison with  ALMA observations}

The Table 1 of \citet{pinte16} summarizes the properties of the rings/gaps in HL Tau estimated from the millimetre emission. In particular information about the gap width and depth is given. The width associated with the gap at 30 AU is about 10 AU, although it varies moderately from one gap to another. In our simulations, the typical width  of the sub-millimter dust rings (St $\simeq 0.01$),  averaged over the different structures, is respectively $0.4H$,  $1. 7 H$  and $\gtrsim 3.5H$  for $\beta=10^5, 10^4$ and  $10^3$.  Given that $H\simeq 2. 5$ AU in our case, there is again a good agreement between the numerical result and the observation for $\beta=10^3$. We stress however that the boundary conditions and the lack of global radial structure in shearing-boxes might influence the distance between the rings.  Note also that other discs than HL-Tau, like Elias 24, exhibit much wider rings or gaps with size of few tenth of AU.  To explain such systems,  our simulations are probably not relevant anymore  ;  the presence of an embedded planet with a mass comparable to that of Jupiter might be required to form such gaps \citep{dipierro18}. 

As for the gap depth, which measures the ratio between the average dust density in the ring and the density at the minimum, direct comparison is much less obvious and has to be taken with caution.   Indeed, due to the instrumental resolution ($\simeq 3-5$ AU for HL-Tau) the contrast between the rings cannot be measured accurately. At best, a lower bound for the gaps depth  can be estimated, which is around 20 for HL-Tau. In the simulations, this quantity is also extremely difficult to estimate since it drastically changes between the different gaps. In the case $\beta=10^3$, the density contrast in the region lying between the two rings (see Fig.~\ref{fig_rings} and $\text{St}=0.01$)  is 35. However in the wider region across the radial boundary, this ratio is huge, of order $10^6$, since the dust is completely depleted there. Morevoer, we remind that the dust density within the rings has not necessarily reached a steady value in simulations. 

\section{Conclusions}
In summary, we characterised the dust dynamics in the outer regions of magnetised accretion disks with ambipolar diffusion. By using 3D MHD stratified shearing box simulations, we showed that centimetre to millimetre dust particles are highly sedimented in the midplane with scaleheight $H_d/H \lesssim 0.1-0.2$. This result fits particularly well the observational constraint of \citet{pinte16} (ALMA) in case of a large enough magnetization $\beta \simeq 10^3$. The vertical settling associated with particles of intermediate size (mm-cm) can be described through the  1D diffusion model of \citet{dubrulle95} which nicely reproduces the density profiles and the ratio $H_d/H$. We also found that the dust particles in this range accumulate within the pressure maxima and form large scale axisymmetric structures (rings in a global view) whose properties depend strongly on the magnetization. For $\beta \simeq 10^3$,  the rings separation (or gaps width) is of order 3-4 $H$, which is in agreement with measures in HL-Tau \citep{pinte16}. We checked that the presence of a mean pressure gradient in the box does not change drastically these results, although our simulations are not enough resolved to trigger the streaming instability and thus determine its impact on the dust distribution. 

For small Stokes number (sub-millimetre to micron size particles), we showed on contrary that the dust is less sedimented in the vertical direction, with $H_d/H\approx 1$ and forms much fainter rings. In case of a strong magnetization ($\beta=10^3$), particles are subject to a large scale poloidal circulation driven by the zonal flows and the winds associated with the ambipolar MHD flow.  
Given their small size, particles in the midplane can diffuse toward density minima where the magnetic activity is high. Once they enter such regions, they are lifted by powerful gas plumes and reach the base of the ionized layer. Due to gravitational settling and inclination of the plumes, they fall back onto the disc. Such circulation makes the 1D diffusion model of \citet{dubrulle95} inappropriate to explain the dust settling in this regime. A 2D model taking into account the radial structure of the gas flow is necessary to reproduce the vertical dust profiles of the simulations. One crucial element of this model is the non-uniformity of the wind in the radial direction. Indeed, models assuming a uniform outflow (inclined or not)  lead to "floating" dust layers which are not seen in the present simulations. \\

We discuss finally  the implications of this work on accretion and planet formation scenario in protoplanetary discs. Since they are cold and dense, such objects are inevitably subject to non-ideal MHD effects when threaded by a magnetic field. Our simulations suggest that both accretion efficiency (up to $\alpha \simeq 10^{-2}$) and dust scaleheight in discs dominated by ambipolar diffusion are compatible with observations. The turbulent r.m.s fluctuations obtained in these simulations are also in agreement with the dispersion velocity in CO measured recently by \citet{Flaherty17} who predict an upper limit $v_{\text{turb}}<0.05 c_s$ in the midplane.  Non-ideal MHD effects are therefore a possible avenue to reconcile the accretion theory with the observations. A next step is to perform global simulations, including a more realistic dust size distribution and the three non-ideal effects.  In particular the role of the Hall effect \citep{lesur14,bai14} on the dust is still unknown and has to be investigated. 

This work is also directly relevant for planet formation theory and may be important to understand the growth of grains to  pebbles ("radial-drift" barrier) and  from planetesimals to planets ("fragmentation" barrier)  \citep{birnstiel16}. Indeed, the confinement of solids in pressure maxima (here the rings) is known to be a very efficient way to overcome these barriers. In addition to stop their radial migration and accelerate their growth, such trapping regions also reduce their relative velocities, which prevents grain fragmentation \citep{gonzalez17}.  The high dust to gas ratio in these regions can ultimately lead to gravitational runaway of the solids. Non-ideal MHD effects appear to be an alternative to other trapping mechanisms such as vortices, whose stability and origin remain unclear \citep{lesur09}. It may be also more generic and efficient at collecting dust particles than the streaming instability  (and related self-induced traps) whose excitation requires very specific conditions. The high dust to gas ratio in the midplane due to the strong sedimentation induced by MHD zonal flows could also enhance the so-called “pebble accretion” \citep{ormel10,lambrechts12}, and therefore reduce the growth timescale of massive cores.  
\label{sec_conclusions}

\begin{acknowledgements}
This work acknowledges funding from The French ANR under contracts ANR-17-ERC2-0007 (MHDiscs) and ANR-16-CE31-0013 (Planet-forming-discs). Part of this work was performed using HPC ressources from GENCI-IDRIS under the allocation A0020402231 and using the Froggy platform of the CIMENT infrastructure (https://ciment.ujf-grenoble.fr), which is supported by the Rhône-Alpes region (GRANT CPER07-13 CIRA), the OSUG@2020 labex (reference ANR10 LABX56) and the Equip@Meso project (reference ANR-10-EQPX-29-01) of the programme Investissements d'Avenir supervised by the Agence Nationale pour la Recherche.
\end{acknowledgements}



\bibliographystyle{aa}
\bibliography{refs} 

\begin{thebibliography}{72}
\expandafter\ifx\csname natexlab\endcsname\relax\def\natexlab#1{#1}\fi

\bibitem[{{ALMA Partnership} {et~al.}(2015){ALMA Partnership}, {Brogan},
  {P{\'e}rez}, {Hunter}, {Dent}, {Hales}, {Hills}, {Corder}, {Fomalont},
  {Vlahakis}, {Asaki}, {Barkats}, {Hirota}, {Hodge}, {Impellizzeri}, {Kneissl},
  {Liuzzo}, {Lucas}, {Marcelino}, {Matsushita}, {Nakanishi}, {Phillips},
  {Richards}, {Toledo}, {Aladro}, {Broguiere}, {Cortes}, {Cortes}, {Espada},
  {Galarza}, {Garcia-Appadoo}, {Guzman-Ramirez}, {Humphreys}, {Jung}, {Kameno},
  {Laing}, {Leon}, {Marconi}, {Mignano}, {Nikolic}, {Nyman}, {Radiszcz},
  {Remijan}, {Rod{\'o}n}, {Sawada}, {Takahashi}, {Tilanus}, {Vila Vilaro},
  {Watson}, {Wiklind}, {Akiyama}, {Chapillon}, {de Gregorio-Monsalvo}, {Di
  Francesco}, {Gueth}, {Kawamura}, {Lee}, {Nguyen Luong}, {Mangum}, {Pietu},
  {Sanhueza}, {Saigo}, {Takakuwa}, {Ubach}, {van Kempen}, {Wootten},
  {Castro-Carrizo}, {Francke}, {Gallardo}, {Garcia}, {Gonzalez}, {Hill},
  {Kaminski}, {Kurono}, {Liu}, {Lopez}, {Morales}, {Plarre}, {Schieven},
  {Testi}, {Videla}, {Villard}, {Andreani}, {Hibbard}, \& {Tatematsu}}]{alma15}
{ALMA Partnership}, {Brogan}, C.~L., {P{\'e}rez}, L.~M., {et~al.} 2015, \apjl,
  808, L3

\bibitem[{{Andrews} {et~al.}(2016){Andrews}, {Wilner}, {Zhu}, {Birnstiel},
  {Carpenter}, {P{\'e}rez}, {Bai}, {{\"O}berg}, {Hughes}, {Isella}, \&
  {Ricci}}]{andrews16}
{Andrews}, S.~M., {Wilner}, D.~J., {Zhu}, Z., {et~al.} 2016, ApJl, 820, L40

\bibitem[{{Bai}(2013)}]{bai13b}
{Bai}, X.-N. 2013, ApJ, 772, 96

\bibitem[{{Bai}(2014)}]{bai14}
{Bai}, X.-N. 2014, ApJ, 791, 137

\bibitem[{{Bai}(2015)}]{bai15}
{Bai}, X.-N. 2015, ApJ, 798, 84

\bibitem[{{Bai} \& {Stone}(2013)}]{bai13}
{Bai}, X.-N. \& {Stone}, J.~M. 2013, ApJ, 767, 30

\bibitem[{{Balbus} \& {Hawley}(1991)}]{balbus91}
{Balbus}, S.~A. \& {Hawley}, J.~F. 1991, ApJ, 376, 214

\bibitem[{{Balsara} {et~al.}(2009){Balsara}, {Tilley}, {Rettig}, \&
  {Brittain}}]{balsara09}
{Balsara}, D.~S., {Tilley}, D.~A., {Rettig}, T., \& {Brittain}, S.~D. 2009,
  MNRAS, 397, 24

\bibitem[{{B{\'e}thune} {et~al.}(2016){B{\'e}thune}, {Lesur}, \&
  {Ferreira}}]{bethune16}
{B{\'e}thune}, W., {Lesur}, G., \& {Ferreira}, J. 2016, AAp, 589, A87

\bibitem[{{B{\'e}thune} {et~al.}(2017){B{\'e}thune}, {Lesur}, \&
  {Ferreira}}]{bethune17}
{B{\'e}thune}, W., {Lesur}, G., \& {Ferreira}, J. 2017, AAp, 600, A75

\bibitem[{{Birnstiel} {et~al.}(2016){Birnstiel}, {Fang}, \&
  {Johansen}}]{birnstiel16}
{Birnstiel}, T., {Fang}, M., \& {Johansen}, A. 2016, SSR, 205, 41

\bibitem[{{Carballido} {et~al.}(2006){Carballido}, {Fromang}, \&
  {Papaloizou}}]{carbadillo06}
{Carballido}, A., {Fromang}, S., \& {Papaloizou}, J. 2006, MNRAS, 373, 1633

\bibitem[{{Chiang} \& {Youdin}(2010)}]{chiang10}
{Chiang}, E. \& {Youdin}, A.~N. 2010, Annual Review of Earth and Planetary
  Sciences, 38, 493

\bibitem[{{Dipierro} {et~al.}(2018){Dipierro}, {Ricci}, {P{\'e}rez}, {Lodato},
  {Alexander}, {Laibe}, {Andrews}, {Carpenter}, {Chandler}, {Greaves}, {Hall},
  {Henning}, {Kwon}, {Linz}, {Mundy}, {Sargent}, {Tazzari}, {Testi}, \&
  {Wilner}}]{dipierro18}
{Dipierro}, G., {Ricci}, L., {P{\'e}rez}, L., {et~al.} 2018, MNRAS, 475, 5296

\bibitem[{{Dubrulle} {et~al.}(1995){Dubrulle}, {Morfill}, \&
  {Sterzik}}]{dubrulle95}
{Dubrulle}, B., {Morfill}, G., \& {Sterzik}, M. 1995, ICARUS, 114, 237

\bibitem[{{Dullemond} \& {Dominik}(2004)}]{dullemond04}
{Dullemond}, C.~P. \& {Dominik}, C. 2004, AAp, 421, 1075

\bibitem[{{Dullemond} \& {Penzlin}(2018)}]{dullemond18}
{Dullemond}, C.~P. \& {Penzlin}, A.~B.~T. 2018, AAp, 609, A50

\bibitem[{{Flaherty} {et~al.}(2017){Flaherty}, {Hughes}, {Rose}, {Simon}, {Qi},
  {Andrews}, {K{\'o}sp{\'a}l}, {Wilner}, {Chiang}, {Armitage}, \&
  {Bai}}]{Flaherty17}
{Flaherty}, K.~M., {Hughes}, A.~M., {Rose}, S.~C., {et~al.} 2017, ApJ, 843, 150

\bibitem[{{Fleming} {et~al.}(2000){Fleming}, {Stone}, \& {Hawley}}]{fleming00}
{Fleming}, T.~P., {Stone}, J.~M., \& {Hawley}, J.~F. 2000, ApJ, 530, 464

\bibitem[{{Flock} {et~al.}(2011){Flock}, {Dzyurkevich}, {Klahr}, {Turner}, \&
  {Henning}}]{flock11}
{Flock}, M., {Dzyurkevich}, N., {Klahr}, H., {Turner}, N.~J., \& {Henning}, T.
  2011, \apj, 735, 122

\bibitem[{{Flock} {et~al.}(2013){Flock}, {Fromang}, {Gonz{\'a}lez}, \&
  {Commer{\c c}on}}]{flock13}
{Flock}, M., {Fromang}, S., {Gonz{\'a}lez}, M., \& {Commer{\c c}on}, B. 2013,
  AAp, 560, A43

\bibitem[{{Flock} {et~al.}(2015){Flock}, {Ruge}, {Dzyurkevich}, {Henning},
  {Klahr}, \& {Wolf}}]{flock15}
{Flock}, M., {Ruge}, J.~P., {Dzyurkevich}, N., {et~al.} 2015, AAp, 574, A68

\bibitem[{{Fromang} {et~al.}(2006){Fromang}, {Hennebelle}, \&
  {Teyssier}}]{fromang06}
{Fromang}, S., {Hennebelle}, P., \& {Teyssier}, R. 2006, AA, 457, 371

\bibitem[{{Fromang} {et~al.}(2013){Fromang}, {Latter}, {Lesur}, \&
  {Ogilvie}}]{fromang13}
{Fromang}, S., {Latter}, H., {Lesur}, G., \& {Ogilvie}, G.~I. 2013, \aap, 552,
  A71

\bibitem[{{Fromang} \& {Papaloizou}(2006)}]{fromang06b}
{Fromang}, S. \& {Papaloizou}, J. 2006, AAp, 452, 751

\bibitem[{{Gammie}(1996)}]{gammie96}
{Gammie}, C.~F. 1996, ApJ, 457, 355

\bibitem[{{Garufi} {et~al.}(2017){Garufi}, {Benisty}, {Stolker}, {Avenhaus},
  {de Boer}, {Pohl}, {Quanz}, {Dominik}, {Ginski}, {Thalmann}, {van Boekel},
  {Boccaletti}, {Henning}, {Janson}, {Salter}, {Schmid}, {Sissa}, {Langlois},
  {Beuzit}, {Chauvin}, {Mouillet}, {Augereau}, {Bazzon}, {Biller}, {Bonnefoy},
  {Buenzli}, {Cheetham}, {Daemgen}, {Desidera}, {Engler}, {Feldt}, {Girard},
  {Gratton}, {Hagelberg}, {Keller}, {Keppler}, {Kenworthy}, {Kral}, {Lopez},
  {Maire}, {Menard}, {Mesa}, {Messina}, {Meyer}, {Milli}, {Min}, {Muller},
  {Olofsson}, {Pawellek}, {Pinte}, {Szulagyi}, {Vigan}, {Wahhaj}, {Waters}, \&
  {Zurlo}}]{garufi17}
{Garufi}, A., {Benisty}, M., {Stolker}, T., {et~al.} 2017, The Messenger, 169,
  32

\bibitem[{{Goldreich} \& {Lynden-Bell}(1965)}]{goldreich65}
{Goldreich}, P. \& {Lynden-Bell}, D. 1965, MNRAS, 130, 125

\bibitem[{{Gonzalez} {et~al.}(2017){Gonzalez}, {Laibe}, \&
  {Maddison}}]{gonzalez17}
{Gonzalez}, J.-F., {Laibe}, G., \& {Maddison}, S.~T. 2017, MNRAS, 467, 1984

\bibitem[{{Gressel} \& {Pessah}(2015)}]{gressel15}
{Gressel}, O. \& {Pessah}, M.~E. 2015, ApJ, 810, 59

\bibitem[{{Gressel} {et~al.}(2015){Gressel}, {Turner}, {Nelson}, \&
  {McNally}}]{gressel15b}
{Gressel}, O., {Turner}, N.~J., {Nelson}, R.~P., \& {McNally}, C.~P. 2015, ApJ,
  801, 84

\bibitem[{{Hawley} {et~al.}(1995){Hawley}, {Gammie}, \& {Balbus}}]{hawley95}
{Hawley}, J.~F., {Gammie}, C.~F., \& {Balbus}, S.~A. 1995, ApJ, 440, 742

\bibitem[{{Hayashi}(1981)}]{hayashi81}
{Hayashi}, C. 1981, Progress of Theoretical Physics Supplement, 70, 35

\bibitem[{{Heinemann} \& {Papaloizou}(2009)}]{heinemann09}
{Heinemann}, T. \& {Papaloizou}, J.~C.~B. 2009, MNRS, 397, 64

\bibitem[{{Isella} {et~al.}(2016){Isella}, {Guidi}, {Testi}, {Liu}, {Li}, {Li},
  {Weaver}, {Boehler}, {Carperter}, {De Gregorio-Monsalvo}, {Manara}, {Natta},
  {P{\'e}rez}, {Ricci}, {Sargent}, {Tazzari}, \& {Turner}}]{isella16}
{Isella}, A., {Guidi}, G., {Testi}, L., {et~al.} 2016, Physical Review Letters,
  117, 251101

\bibitem[{{Jacquet} {et~al.}(2011){Jacquet}, {Balbus}, \& {Latter}}]{jacquet11}
{Jacquet}, E., {Balbus}, S., \& {Latter}, H. 2011, MNRAS, 415, 3591

\bibitem[{{Johansen} \& {Klahr}(2005)}]{johansen05}
{Johansen}, A. \& {Klahr}, H. 2005, ApJ, 634, 1353

\bibitem[{{Johansen} {et~al.}(2006){Johansen}, {Klahr}, \& {Mee}}]{johansen06}
{Johansen}, A., {Klahr}, H., \& {Mee}, A.~J. 2006, MNRAS, 370, L71

\bibitem[{{Johansen} {et~al.}(2009){Johansen}, {Youdin}, \&
  {Klahr}}]{johansen09}
{Johansen}, A., {Youdin}, A., \& {Klahr}, H. 2009, ApJ, 697, 1269

\bibitem[{{Kataoka} {et~al.}(2017){Kataoka}, {Tsukagoshi}, {Pohl}, {Muto},
  {Nagai}, {Stephens}, {Tomisaka}, \& {Momose}}]{kataoka17}
{Kataoka}, A., {Tsukagoshi}, T., {Pohl}, A., {et~al.} 2017, ApJ, 844, L5

\bibitem[{{Kunz} \& {Lesur}(2013)}]{kunz13}
{Kunz}, M.~W. \& {Lesur}, G. 2013, MNRAS, 434, 2295

\bibitem[{{Laibe} \& {Price}(2011)}]{laibe11}
{Laibe}, G. \& {Price}, D.~J. 2011, \mnras, 418, 1491

\bibitem[{{Laibe} \& {Price}(2012)}]{laibe12}
{Laibe}, G. \& {Price}, D.~J. 2012, MNRAS, 420, 2345

\bibitem[{{Lambrechts} \& {Johansen}(2012)}]{lambrechts12}
{Lambrechts}, M. \& {Johansen}, A. 2012, AAp, 544, A32

\bibitem[{{Lesur} {et~al.}(2014){Lesur}, {Kunz}, \& {Fromang}}]{lesur14}
{Lesur}, G., {Kunz}, M.~W., \& {Fromang}, S. 2014, AAp, 566, A56

\bibitem[{{Lesur} \& {Papaloizou}(2009)}]{lesur09}
{Lesur}, G. \& {Papaloizou}, J.~C.~B. 2009, AAp, 498, 1

\bibitem[{{Lighthill}(1952)}]{lighthill52}
{Lighthill}, M.~J. 1952, Proceedings of the Royal Society of London Series A,
  211, 564

\bibitem[{{Mignone} {et~al.}(2007){Mignone}, {Bodo}, {Massaglia}, {Matsakos},
  {Tesileanu}, {Zanni}, \& {Ferrari}}]{mignone07}
{Mignone}, A., {Bodo}, G., {Massaglia}, S., {et~al.} 2007, ApJS, 170, 228

\bibitem[{{Miyake} {et~al.}(2016){Miyake}, {Suzuki}, \& {Inutsuka}}]{miyake16}
{Miyake}, T., {Suzuki}, T.~K., \& {Inutsuka}, S.-i. 2016, ApJ, 821, 3

\bibitem[{{Morfill}(1985)}]{morfill85}
{Morfill}, G.~E. 1985, in Birth and the Infancy of Stars, ed. R.~{Lucas},
  A.~{Omont}, \& R.~{Stora}

\bibitem[{{Okuzumi} {et~al.}(2016){Okuzumi}, {Momose}, {Sirono}, {Kobayashi},
  \& {Tanaka}}]{okuzumi16}
{Okuzumi}, S., {Momose}, M., {Sirono}, S.-i., {Kobayashi}, H., \& {Tanaka}, H.
  2016, ApJ, 821, 82

\bibitem[{{Ormel} \& {Klahr}(2010)}]{ormel10}
{Ormel}, C.~W. \& {Klahr}, H.~H. 2010, AAp, 520, A43

\bibitem[{{Perez-Becker} \& {Chiang}(2011)}]{perez11}
{Perez-Becker}, D. \& {Chiang}, E. 2011, ApJ, 735, 8

\bibitem[{{Pinte} {et~al.}(2016){Pinte}, {Dent}, {M{\'e}nard}, {Hales}, {Hill},
  {Cortes}, \& {de Gregorio-Monsalvo}}]{pinte16}
{Pinte}, C., {Dent}, W.~R.~F., {M{\'e}nard}, F., {et~al.} 2016, \apj, 816, 25

\bibitem[{{Ruge} {et~al.}(2016){Ruge}, {Flock}, {Wolf}, {Dzyurkevich},
  {Fromang}, {Henning}, {Klahr}, \& {Meheut}}]{ruge16}
{Ruge}, J.~P., {Flock}, M., {Wolf}, S., {et~al.} 2016, AAp, 590, A17

\bibitem[{{Sano} \& {Stone}(2002)}]{sano02}
{Sano}, T. \& {Stone}, J.~M. 2002, \apj, 577, 534

\bibitem[{{Schr{\"a}pler} \& {Henning}(2004)}]{Schrapler04}
{Schr{\"a}pler}, R. \& {Henning}, T. 2004, ApJ, 614, 960

\bibitem[{{Shakura} \& {Sunyaev}(1973)}]{shakura73}
{Shakura}, N.~I. \& {Sunyaev}, R.~A. 1973, A{\&}A, 24, 337

\bibitem[{{Simon} \& {Armitage}(2014)}]{simon14}
{Simon}, J.~B. \& {Armitage}, P.~J. 2014, ApJ, 784, 15

\bibitem[{{Simon} {et~al.}(2013){Simon}, {Bai}, {Armitage}, {Stone}, \&
  {Beckwith}}]{simon13b}
{Simon}, J.~B., {Bai}, X.-N., {Armitage}, P.~J., {Stone}, J.~M., \& {Beckwith},
  K. 2013, \apj, 775, 73

\bibitem[{{Simon} {et~al.}(2015){Simon}, {Lesur}, {Kunz}, \&
  {Armitage}}]{simon15}
{Simon}, J.~B., {Lesur}, G., {Kunz}, M.~W., \& {Armitage}, P.~J. 2015, \mnras,
  454, 1117

\bibitem[{{Takahashi} \& {Inutsuka}(2014)}]{takahashi14}
{Takahashi}, S.~Z. \& {Inutsuka}, S.-i. 2014, \apj, 794, 55

\bibitem[{{Venuti} {et~al.}(2014){Venuti}, {Bouvier}, {Flaccomio}, {Alencar},
  {Irwin}, {Stauffer}, {Cody}, {Teixeira}, {Sousa}, {Micela}, {Cuillandre}, \&
  {Peres}}]{venuti14}
{Venuti}, L., {Bouvier}, J., {Flaccomio}, E., {et~al.} 2014, AAp, 570, A82

\bibitem[{{Wardle}(2007)}]{wardle07}
{Wardle}, M. 2007, \apss, 311, 35

\bibitem[{{Wardle} \& {Salmeron}(2012)}]{wardle12}
{Wardle}, M. \& {Salmeron}, R. 2012, \mnras, 422, 2737

\bibitem[{{Weidenschilling}(1977)}]{weiden77}
{Weidenschilling}, S.~J. 1977, MNRAS, 180, 57

\bibitem[{{Williams} \& {McPartland}(2016)}]{williams16}
{Williams}, J.~P. \& {McPartland}, C. 2016, ApJ, 830, 32

\bibitem[{{Wolff} {et~al.}(2017){Wolff}, {Perrin}, {Stapelfeldt}, {Duchene},
  {Menard}, {Padgett}, {Pinte}, {Pueyo}, \& {Fischer}}]{wolff17}
{Wolff}, S.~G., {Perrin}, M.~D., {Stapelfeldt}, K., {et~al.} 2017, ArXiv
  e-prints

\bibitem[{{W{\"u}nsch} {et~al.}(2005){W{\"u}nsch}, {Klahr}, \&
  {R{\'o}{\.z}yczka}}]{wunsch05}
{W{\"u}nsch}, R., {Klahr}, H., \& {R{\'o}{\.z}yczka}, M. 2005, MNRAS, 362, 361

\bibitem[{{Youdin} \& {Johansen}(2007)}]{youdin07}
{Youdin}, A. \& {Johansen}, A. 2007, ApJ, 662, 613

\bibitem[{{Youdin} \& {Goodman}(2005)}]{youdin05}
{Youdin}, A.~N. \& {Goodman}, J. 2005, ApJ, 620, 459

\bibitem[{{Zhu} {et~al.}(2015){Zhu}, {Stone}, \& {Bai}}]{zhu15}
{Zhu}, Z., {Stone}, J.~M., \& {Bai}, X.-N. 2015, ApJ, 801, 81

\end{thebibliography}

\appendix
\section{Tests of dust implementation in PLUTO}
\label{appendixA}
\subsection{Numerical methods}
The numerical integration of a given dust specie  is similar to that of the gas, except that the pressure is not computed. We use a HLL Riemann solver to compute the density and momentum flux at cells interfaces. A standard monotonized central flux limiter is used to bound the spatial derivatives. The drag force is treated as a source term in the right hand side of the Runge Kutta solver.  The time step is adapted to take into account the large drag force encountered by small particles and ensure the stability of the scheme. The upper bound for the time step is taken from \citet{laibe12} (cf their section 3.2): 
\begin{equation}
\Delta t <\min_k \left( \dfrac{\rho  \tau_{s_k}}{\rho+\rho_{d_k}}\right), 
\end{equation}
where the subscript k labels the different particle species. This prescription, derived from a Von Neumann analysis, guarantees the stability of the simple explicit Euler-scheme. It is then robust but perhaps not optimal for Runge-Kutta solvers. Finally,  in order to avoid strong numerical diffusion,  we implement a version of the FARGO algorithm for the dust components, which treats the shear  advection separetely from the other flux and source terms.   
\begin{figure}
\centering
\includegraphics[width=1.04\columnwidth]{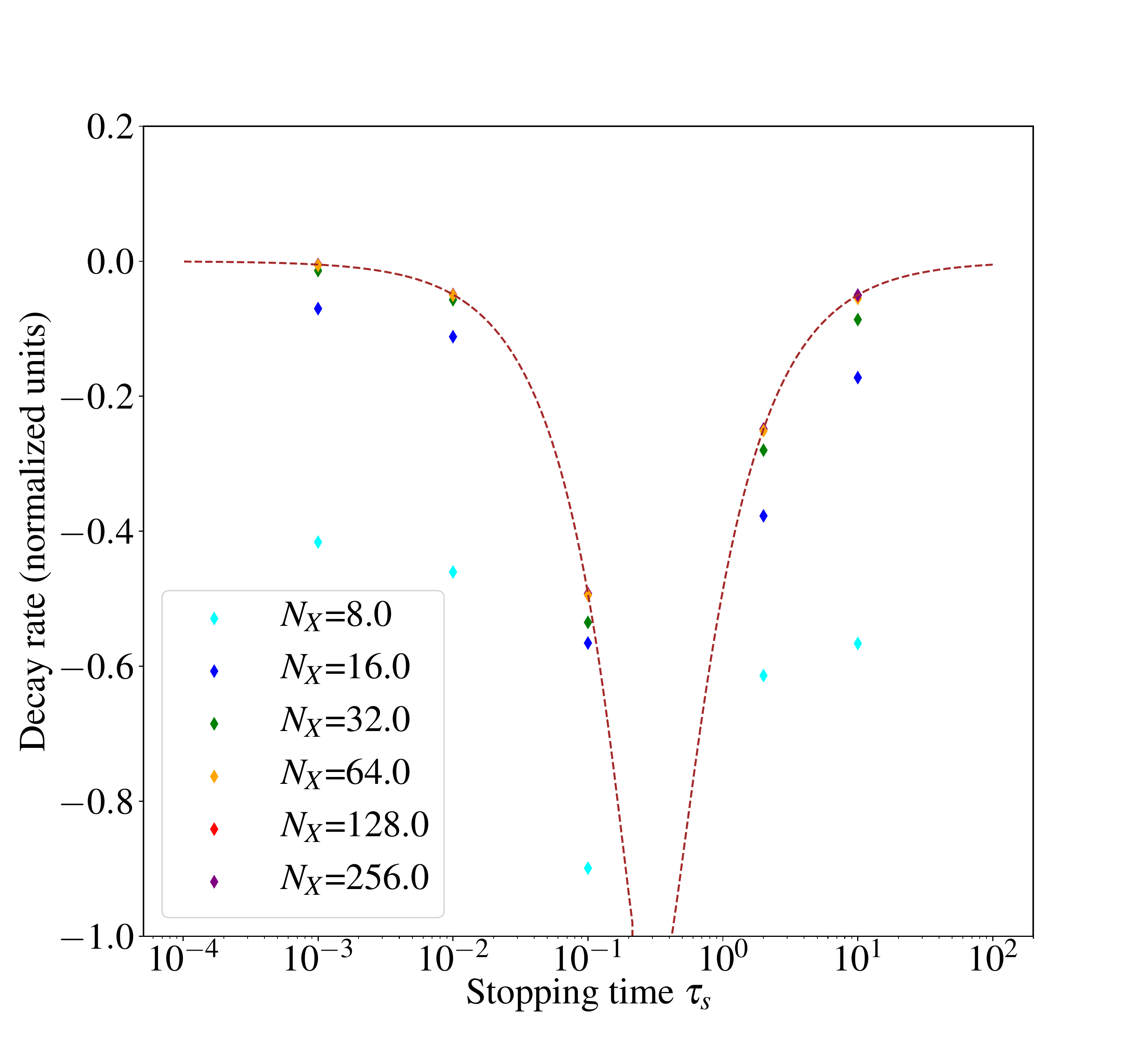}
 \caption{Decay rates of 1D sound waves in a gas/dust mixture  as a function of the stopping time. The dashed/red curve is the analytical solution whereas the diamond markers are the rates measured from PLUTO simulations for different resolution in $x$. The background dust to gas ratio is $\chi=1$ and the wave number is $k_x=2\pi/L_x$. }
\label{fig_soudwave}
\end{figure}
\begin{figure*}
\centering
\includegraphics[width=\textwidth]{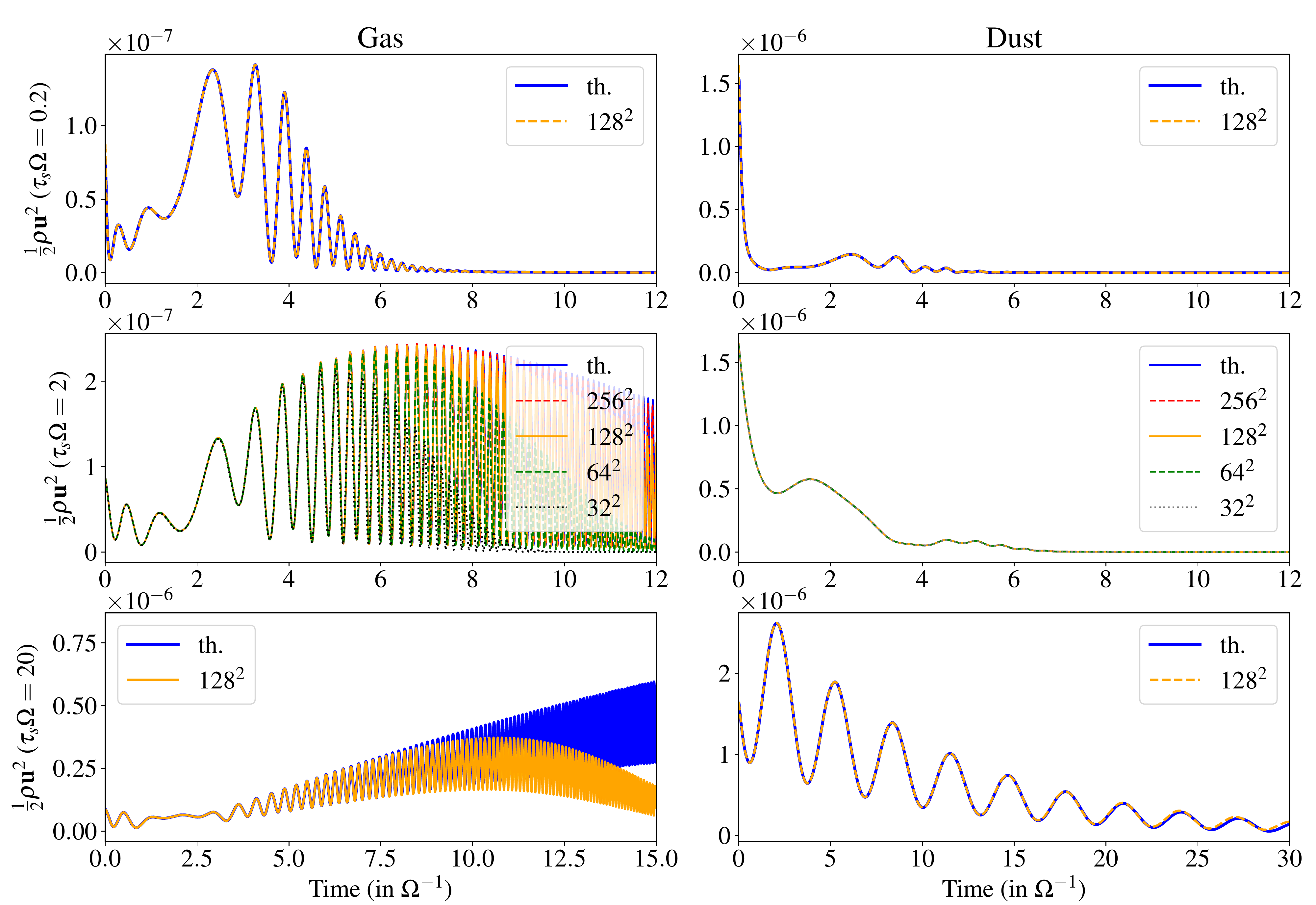}
 \caption{Evolution of shearing waves kinetic energy in a gas/dust mixture with $\chi=0.4$  and for three different stopping times. From top to bottom, $\tau_s \Omega=0.2$, 2, and 20. The right panels correspond to the gas component while the right panels correspond to the dust component. The blue lines are the semi-analytical evolution while other lines are the evolution computed numerically with PLUTO.}
\label{fig_naxi}
\end{figure*}

\subsection{Sound waves}
To check our implementation, we first perform a series of tests involving the propagation of simple 
axisymmetric sound waves in a dust-gas mixture along the radial direction.  As a first step, we neglect the Coriolis force (no rotation) and the shear.  Assuming a single dust specie of size $a$ and perturbations $\delta \rho$, $\delta \rho_d$, $\delta v_{x}$ and $\delta v_{x_d}\propto \exp(i k_x-\omega t)$, the equations governing the linear sound waves in the $x$ direction are: 
\begin{equation}
 -i\omega \delta \rho + ik_x \rho\, \delta v_x = 0,
\end{equation}
\begin{equation}
 -i\omega \delta \rho_d + ik_x \rho_d\, \delta v_{x_d}  = 0,
\end{equation}
\begin{equation}
 \left(-i\omega +\dfrac{\chi}{\tau_s}\right) \, \delta v_x + i \dfrac{k_x c_s^2}{\rho} \delta \rho-\dfrac{\chi \delta v_{x_d}}{\tau_s} = 0,
\end{equation}
\begin{equation}
 \left(-i\omega +\dfrac{1}{\tau_s}\right) \, \delta v_{x _d}-\dfrac{\delta v_{x}}{\tau_s} = 0,
\end{equation}
where $\rho$, $\rho_d$ and $\chi$ are respectively the gas density, dust density and dust to gas ratio  related to the background. By calculating the determinant of this system, and eliminating the trivial mode $\omega=0$  (gas components reduced to 0), we obtain the following dispersion relation:
\begin{equation}
\omega^3 + \dfrac{i \omega^2}{\tau_s} (1+\chi) -k_x^2 c_s^2 \omega - i\dfrac{k_x^2 c_s^2}{\tau_s} = 0.
\end{equation}
The solution is composed of one standing wave and two oscillatory modes with frequencies close to $k_x c_s$ (in the limit of  small $\tau_s$). Because the dust and gas exert a mutual drag,  the oscillatory modes decay in time with a rate that depends on the dust to gas ratio and $\tau_s$.   In Fig.~\ref{fig_soudwave}, we plotted in red/dashed line the theoretical decay rate as  function of $\tau_s$, calculated analytically from the dispersion relation, for a fixed dust to gas ratio $\chi=1$. \\

To check that our PLUTO implementation is correct, we simulate a linear sound wave for different $\tau_s$ between 0.1 and 1000, and for $\chi=1$. The wave corresponds initially to a pure eigenmode of the linear system  with an amplitude of $10^{-3}$.  We choose $k_x=2\pi$ so that the wave fills the box of size $L_x=1$. The diamond points in Fig.~\ref{fig_soudwave} are the  decay rates obtained in the simulations for different $\tau_s$ and numerical resolution (per wavelength). Clearly, there is a perfect agreement with the teoretical prediction, provided that $N_x \gtrsim 64$. However, the difference remains small even for $N_x \simeq 32 $. Below $N_x \simeq 16 $, the decay rate is not reproduced accurately and numerical dissipation starts to be prominent. Note that  the wave never dissipates entirely  and its kinetic energy saturates around a very low value,  regardless of the resolution. This behaviour results  probably from the initial numerical noise that produces spurious modes and pollutes the initial wave. However we emphasize that such noise is completely negligible compared to the initial wave amplitude.

Note that this simple problem has been already tested in SPH codes \citep{laibe11,laibe12}. Unlike SPH implementation, we did not experience any convergence problem when considering stopping times much smaller than the wave period. In particular the rate at which it converges with resolution seems independent on the stopping time.  \citet{laibe12} argued that at small $\tau_s$ the spatial de-phasing between gas and dust is small, of order $c_s \tau_s$,  and must be resolved numerically in order to capture properly the physics of the wave dissipation. Our tests in PLUTO suggest however that the dephasing length does not need to be resolved to obtain correct dissipation. Such length is actually purely geometrical and is not associated with any energy or momentum transfer.  

\subsection{Shearing waves with rotation}

We performed similar tests for 2D non-axisymmetric waves of the form $\exp(i k_x + ik_y-\omega t)$ with $k_y\neq 0$, in the presence of rotation and shear. Since these waves have a wavenumber $k_{x}=k_{x_0}+Sk_yt$ that increases linearly in time, analytical solutions are not straightforward to obtain. Waves are rapidly sheared out and their evolution on long time scales cannot be described through simple decaying trigonometric functions.  To deal with this issue, we solve numerically the linearised problem with a simple Runge Kutta integrator and compare the solutions with those obtained numerically with PLUTO. \\

We consider an initial wave with $k_x=-2\pi/L_x$ and $k_y=2\pi/L_y$ in a box of size $L_x=1$ and $L_y=3$. The amplitudes of the density and velocity fields of the mode are initialized randomly and uniformly between 0 and $10^{-3}$.  We tested three different stopping time by fixing the dust to gas ratio $\chi=0.4$. The equation of state is isothermal and the gas is unmagnetized. 
The results are shown in Fig.~\ref{fig_naxi}, where the blue curves represent the time-evolution of kinetic energy  integrated with our linear solver. The other curves correspond to  PLUTO  simulations with different resolutions.  Our code reproduces quite well the desired solution during the first shearing times but the gas component dissipates at longer times ($t>10\,\Omega^{-1}$) as the wave is strongly sheared out and damped by numerical diffusion. We checked in particular that increasing the numerical resolution makes the evolution of the waves closer to the theoretical prediction. Note that for $\tau_s\Omega=0.2$, the dust component tends to follow the gas oscillations within the first orbits, since $\tau_s$ is smaller than the oscillation period. Dust and gas are well-coupled and the wave is rapidly damped due to the strong drag between the two components.  However for $\tau_s=20\Omega^{-1}$, there is only a weak coupling between gas and dust, since $\tau_s$ is much larger than the gas oscillation period. The dust motion follows its own oscillation with larger period and slowly decays due to the small drag. 

\subsection{Streaming instability}
\begin{figure}
\centering
\includegraphics[width=\columnwidth]{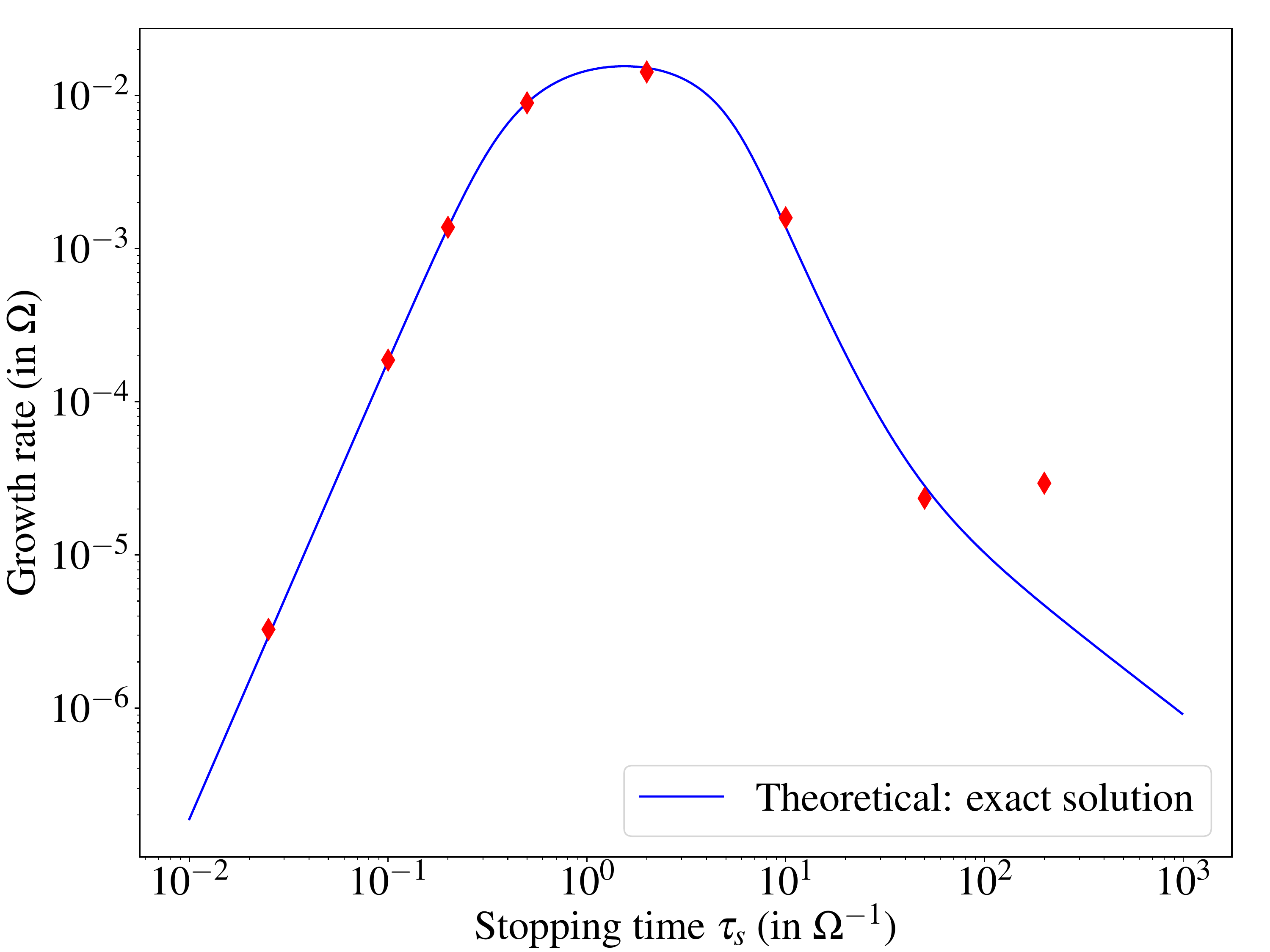}
 \caption{Growth rate of the streaming instability as  function of the stopping time $\tau_s$. The dust to gas ratio is $\chi=0.2$ and the radial normalized acceleration associated with the mean pressure gradient is $\hat{g}_e=0.02$. The blue line is the theoretical growth rate while the red diamond markers are the growth rates computed numerically with PLUTO.}
\label{fig_streaming}
\end{figure}
Finally we tested the streaming instability in the 3D unstratified box with a mean pressure gradient. The properties of this instability and its linear derivation can be found in \citet{youdin05,jacquet11}. The setup is similar to that described in Section \ref{gradP}, but with $\hat{g}_e=0.02$.   The background flow has a dust to gas ratio $\chi= 0.2$ and a mean radial drift given by Eq.~(\ref{eq_radialdrift}).  We first checked the 2D linear unstable axisymmetric modes in $x$ and $z$. The theoretical growth rates and eigenfunctions of these modes are computed semi-analytically by solving the linearised system given in \citet{youdin05}. Figure \ref{fig_streaming} shows the theoretical growth rate as a function of the stopping time $\tau_s$ (blue curve). 

We ran PLUTO simulations for different stopping times (0.025, 0.1, 0.2, 0.5, 2, 10, 50, 200), with initial states composed of a background drift flow plus an unstable eigenmode of amplitude $2\times 10^{-3}$.  The box size is $L_x=1 H, \, L_z=4\pi H$ and the numerical resolution is $N_X=128, N_Z=64$. The initial mode has $k_x=2 \pi/L_x$ and $k_z=2 \pi/L_z$. The red diamond markers in Fig.~\ref{fig_streaming} correspond to the numerical growth rates obtained with PLUTO. We checked that for $\tau_s \lesssim 100$ there is a very good agreement between the theoretical and numerical growth rates, with a relative error less than 2\% in all cases. For $\tau_s = 200 \, \Omega^{-1}$, however we do not reproduce the correct growth rate. Actually we found that this discrepancy is not due to our implementation but to the presence of unstable harmonics with much higher growth rates, probably excited by the initial numerical noise. 
To make sure that our implementation is robust, we also simulate the nonlinear evolution of such mode in a bigger box ($L_x=L_z=6H$) with higher resolution ($N_X=512$, $N_Z=256$) during 1000 orbits. We recover the classical result of \citet{youdin07} that dust concentrates into small-scale clumps with local dust to gas ratio greater than 1. 
\section{Derivation of the 1D advection-diffusion equation of \citet{dubrulle95}}
\label{appendixB}
In this appendix, we derive the 1D advection-diffusion equation of \citet{dubrulle95} (Eq.~\ref{eq_dubrulle}), commonly used to model vertical dust settling in turbulent discs. In particular we aim to clarify the hypothesis behind this equation.
Given the 1D decomposition of a quantity $X$ into a mean component $\overline{X}(z)$ and a fluctuation $\delta X(x,y,z)$, we showed already in Section \ref{dubrulle_model} that the dust mass conservation equation expands to:
\begin{equation}
\dfrac{\partial \overline{\rho_d}}{\partial t}+\dfrac{\partial}{\partial z} \left(\overline{\rho_d}\, \overline{ v_z} + \overline{\delta \rho_d \,\delta v_{z_d}}+ \overline{\rho_d} \, \overline{\Delta v_z}\right)=0,
\label{mass_eq_dust2_a}
\end{equation}
We start by developing the mean drift  $\overline{\Delta v_z}$, which appears in the third flux term. We use the "terminal velocity approximation" \citep{youdin05} which assumes that the grain  instantaneously responds to the gas drag (their inertia is negligible). This approximation is valid only if St $\ll 1$ and if the typical eddies turnover time is larger than $\tau_s$. In that case, the drift velocity at leading order is: 
\begin{equation}
\label{eq_deltaV}
\mathbf{\Delta v} = \tau_s\left( \dfrac{\mathbf{\nabla}{P}}{\rho^\star}-\dfrac{(\nabla\times \mathbf{B})\times\mathbf{B}}{\rho^\star}\right)
 \end{equation}
where $\rho^\star = \rho+\rho_d$ is the total density. Yet, by considering the equation for the center-mass vertical velocity  $v_z^\star=\dfrac{\rho v_z+ \rho_d {v_d}_z}{{\rho^\star}}$, and neglecting the quadratic terms in  $\mathbf{\Delta v}$, we have: 
\begin{equation}
\label{eq_centermass}
\left[\dfrac{\mathbf{\nabla}{P}}{{\rho^\star}}-\dfrac{(\nabla\times \mathbf{B})\times\mathbf{B}}{{\rho^\star}}\right]_{z}=g_z - \frac{\partial{{{v_z^\star}}}}{\partial{t}}-\mathbf{v}^\star\cdot\mathbf{\nabla}
 v_z^\star.
\end{equation}
The  two inertia terms in the right hand side are generally negligible compared to the gravitational term and the gas wind advection. By taking $g_z=-\Omega^2 z$, we deduce from Eqs.~(\ref{eq_deltaV}) and (\ref{eq_centermass}) the mean vertical drift velocity: 
\begin{equation}
\label{eq_vertical_drift}
\overline{\Delta v_z} =- \overline{\tau}_s \Omega^2 z, 
\end{equation}
with 
\begin{equation}
\overline{\tau}_s= \dfrac{a \rho_s}{c_s} \overline{\rho^{-1}}.
\end{equation}
If the gas density fluctuations are small, we have $\overline{\rho^{-1}}\simeq\overline{\rho}^{-1}$. \\

We need then to find a closure relation to express the turbulent correlation term $ \overline{\delta \rho_d \,\delta v_{z_d}}$ in Eq.~(\ref{mass_eq_dust2_a}). 
For that purpose, let us note $\chi={\rho_d}/{\rho} \ll 1$ the dust concentration. 
By combining the mass conservation equation of both gas and dust species,  we obtain
\begin{equation}
\dfrac{\partial \chi}{\partial t} = -\dfrac{1}{\rho}\left(\nabla\cdot \left(\chi\rho \mathbf{v_d}\right)
-  \nabla\cdot \left(\rho\mathbf{v}\right) \chi \right). 
\end{equation}
By using the drift definition $\mathbf{\Delta v} = \mathbf{v_d} - \mathbf{v}$ and simplifying the expression above, the dust to gas ratio obeys the following continuity equation:
\begin{equation}
\rho \left( \dfrac{\partial  \chi}{\partial t}+\mathbf{v} \cdot \nabla  \chi\right) =  -\nabla\cdot \left( \chi\rho \mathbf{\Delta v}\right).
\end{equation}
In the limit of small drifts which corresponds to small Stokes, the dust is a passive scalar and follows the pure advection equation: 
\begin{equation}
\label{eq_advection}
\dfrac{\partial  \chi}{\partial t}+\mathbf{v} \cdot \nabla \chi= 0.
\end{equation}
The horizontal average of the above equation writes:
\begin{equation}
\label{eq_adv_mean}
\dfrac{\partial \overline{ \chi}}{\partial t}+ \overline{v_z} \cdot \dfrac{ \partial \overline{ \chi}}{\partial z}+ \overline{\mathbf{v} \cdot \nabla \delta  \chi}= 0.
\end{equation}
By expanding Eq.~(\ref{eq_advection}), we have also:
\begin{equation}
\label{eq_adv_tot}
\dfrac{\partial  \chi}{\partial t}+\overline{v_z} \cdot \dfrac{ \partial \overline{ \chi}}{\partial z}+ \delta v_z \dfrac{ \partial \overline{ \chi}}{\partial z}+\mathbf{v} \cdot \nabla \delta  \chi = 0.
\end{equation}
Substracting Eq.~(\ref{eq_adv_mean}) with Eq.~(\ref{eq_adv_tot}) gives:
\begin{equation}
\dfrac{\partial \delta  \chi}{\partial t}+ \delta v_z \dfrac{ \partial \overline{ \chi}}{\partial z}+\mathbf{\overline{v}} \cdot \nabla \delta  \chi+\mathbf{\delta v} \cdot \nabla \delta  \chi- \overline{\delta \mathbf{v} \cdot \nabla \delta X} = 0.
\end{equation}
To close the system of equations, one needs to neglect the term $\mathbf{\delta v} \cdot \nabla \delta  \chi- \overline{\mathbf{\delta v} \cdot \nabla \delta  \chi}$, which is justified in the limit of small turbulent correlation time. 
We obtain then: 
\begin{equation}
\dfrac{\partial \delta  \chi}{\partial t}+ \delta v_z \dfrac{ \partial \overline{ \chi}}{\partial z}+\overline{v_z} \cdot \dfrac{\partial \delta  \chi}{\partial z} = 0
\end{equation}
By considering the case without mean wind $\overline{v_z}=0$,  and defining the correlation time $\tau_{\text{corr}}$  of the turbulent eddies such that ${\partial \delta  \chi}/{\partial t}\simeq  \delta \chi /\tau_{\text{corr}}$, we have then 
\begin{equation}
\delta \chi \simeq -\tau_{\text{corr}}\, \delta v_z \dfrac{ \partial \overline{ \chi}}{\partial z}
\end{equation}
We finally use the fact that $\delta \rho / \rho \ll 1$ (anelastic approximation, so that $\delta \chi \simeq \delta \rho_d /\overline{\rho}$) to obtain a closure relation that links the turbulent term with the vertical gradient of $\overline{\chi}$: 
\begin{equation}
\label{eq_fluct_diffusion}
\overline{\delta \rho_d \,\delta v_{z_d}} = -D_z \,\overline{\rho}  \dfrac{\partial}{\partial z} \left(\dfrac{\overline{\rho_d}}{\overline{\rho}}\right), 
\end{equation}
with $D_z \simeq   \overline{v_z^2}  $ $\tau_{\text{corr}}>0$. The turbulent correlation can be then reduced to a diffusive operator with coefficient $D_z$. Using Eq.~(\ref{mass_eq_dust2_a}), (\ref{eq_vertical_drift}) and (\ref{eq_fluct_diffusion}), and neglecting the mean wind term $\overline{\rho_d}\, \overline{ v_z}$ in Eq.~(\ref{mass_eq_dust2_a}), it is straightforward to show that: 
\begin{equation}
\label{eq_dubrulle_a}
\dfrac{\partial\overline{\rho_d} }{\partial t} = \dfrac{\partial}{\partial z} \left(z \Omega^2\, \overline{\tau_s} \,\overline{\rho_d} \right) +\dfrac{\partial}{\partial z}  \left[ D_z \, \overline{\rho}  \dfrac{\partial}{\partial z} \left(\dfrac{\overline{\rho_d}}{\overline{\rho}}\right) \right]
\end{equation}
\section{The 2D advection-diffusion solver}
\label{appendixC}
In Section \ref{2Dmodel}, we study the 2D solutions $\rho_d(x,z)$ that satisfy the following equation:
\begin{equation}
\label{eq_model2D_a}
\dfrac{\partial (\rho_d v_{x_d})}{\partial x} + \dfrac{\partial (\rho_d v_{z_d})}{\partial z} = \dfrac{\partial}{dz} \left[ D_z \, \rho \dfrac{\partial}{\partial z} \left(\dfrac{\rho_d}{\rho}\right)\right] + \dfrac{\partial}{dx} \left[D_x  \dfrac{\partial \rho_d}{\partial x}\right]. 
\end{equation}
where $v_{x_d}$, $v_{z_d}$ and $\rho$ are given 2D fields, and $D_x, D_z$ are given functions of $z$. In this appendix, we describe the methods employed to calculate numerically such solutions.  First, the equation is re-written in the form:  
\begin{equation}
\label{eq_model2D_b}
A \dfrac{\partial^2 \rho_d}{\partial x^2}+B \dfrac{\partial \rho_d}{\partial x} + C \dfrac{\partial^2 \rho_d}{\partial z^2}+D \dfrac{\partial \rho_d}{\partial z} + E \rho_d = 0
\end{equation}with $A,B,C,D$ and $E$ given functions of $x$ and $z$. Then fields are discretised on a 2D grid of size $(n_x,n_z)$ where first and second order partial derivative are  approximated using second-order centred differences.  By defining a vector state $X$ containing the values of $\rho$ at each grid points,  the discretized advection-difffusion equation takes the form of a linear system  $L X = 0$. A possible method to solve such system is to use a singular values decomposition (SVD) and seek for the kernel of $L$. However this method does not guarantee the positivity of the solution $\rho_d(x,z)$ and generates a lot of noise, since the matrix is generally ill-conditioned. Another method we preferred is to add a temporal derivative ${\partial \rho_d}/{\partial t}$ in the left hand side of Eq.~\ref{eq_model2D_a} and solve for the dynamical system (initial value problem). The solution is then obtained by convergence of the system toward a steady state.  To integrate the PDE in time, we use a Runge-Kutta method of order 4 which guarantees  the stability of the numerical scheme. The boundary conditions are periodic in $x$ and with no gradient in $z$. The time-step is fixed and chosen to enforce the Courant–Friedrichs–Lewy condition.  The solver is systematically initialized with the 1D Gaussian solution of \citet{dubrulle95}. Convergence toward a steady state is obtained generally after few thousands of orbits, which corresponds in real time to few minutes with a single processor at resolution $64\times 64$.
\end{document}